\documentclass[12pt,letterpaper,final,peerreview]{IEEEtran}

\newcommand{\TIT}{}

\usepackage{cite}
\usepackage{graphicx,color,epsfig,rotating}
\usepackage{amsfonts,amsmath,amsthm,amssymb,bbm,dsfont,stfloats}
\usepackage{hyperref}
\usepackage{xcolor}\hypersetup{linkbordercolor=green}
\usepackage{algorithm,algorithmic}
\usepackage{subfigure}

\makeatletter
\newcommand{\vast}{\bBigg@{3.5}}
\newcommand{\Vast}{\bBigg@{5}}
\makeatother

\newtheorem{thm}{Theorem}
\newtheorem{prop}{Proposition}
\newtheorem{lem}{Lemma}
\newtheorem{corol}{Corollary}
\newtheorem{conj}{Conjecture}

\newtheorem{rem}{Remark}

\def\p0{{\pmb 0}}

\def\snr{{\small\textsf{SNR}}}

\def\inr{{\small\textsf{INR}}}

\newfont{\bbc}{msbm10 scaled 1100}

\newcommand{\Kc}{{\cal K}}

\newcommand{\Rc}{{\cal R}}

\newcommand{\Sigmam}{\hbox{\boldmath$\Sigma$}}

\def\snr{{\small\textsf{SNR}}}

\def\inr{{\small\textsf{INR}}}

\def\CN{{\small\mathcal{CN}}}

\IEEEoverridecommandlockouts

\begin{document}


\title{Capacity Bounds for \\ the $K$-User Gaussian Interference Channel}


\author{Junyoung Nam
\thanks{The material in this paper was presented in part at the IEEE International Symposium on Information Theory (ISIT), Hong Kong, June 2015.}
\thanks{J. Nam was with the Wireless Communication Division, Electronics and Telecommunications Research Institute (ETRI), Daejeon, Korea. He is now with the department of Wireless Communications and Networks, Fraunhofer Heinrich Hertz Institute (HHI), 10587 Berlin, Germany (e-mail: junyoung.nam@hhi.fraunhofer.de).} }

\maketitle

\begin{abstract}
The capacity region of the $K$-user Gaussian interference channel (GIC) is a long-standing open problem 
and even capacity outer bounds are little known in general. A significant progress on degrees-of-freedom (DoF) analysis, a first-order capacity approximation, for the $K$-user GIC has provided new important insights into the problem of interest in the high signal-to-noise ratio (SNR) limit. However, such capacity approximation has been observed to have some limitations in predicting the capacity at \emph{finite} SNR. In this work, we develop a new upper-bounding technique that utilizes {a new type of genie signal and applies \emph{time sharing} to genie signals at $K$ receivers.}  
Based on this technique, we derive new upper bounds on the sum capacity of the three-user GIC with constant, complex channel coefficients and then generalize to the $K$-user case to {better understand} sum-rate behavior at {finite} SNR. We also provide closed-form expressions of our upper bounds on the capacity of the $K$-user symmetric GIC {easily computable} for \emph{any} $K$. From the perspectives of our results, some sum-rate behavior at finite SNR is in line with the insights given by the known DoF results, while some others are not. In particular, the well-known $K/2$ DoF achievable for almost all constant real channel coefficients turns out to be not embodied as a substantial performance gain over a certain range of the cross-channel coefficient in the $K$-user symmetric real case {especially for \emph{large} $K$}. We further investigate the impact of phase offset between the direct-channel coefficient and the cross-channel coefficients on the {sum-rate upper bound} for the three-user \emph{complex} GIC. As a consequence, we aim to provide new findings that could not be predicted by the prior works on DoF of GICs. 


\end{abstract}

\section{Introduction}
\label{sec:intro}

As the recent emerging wireless networks with a tremendous amount of mutually-interfering links tend to be severely interference-limited, interference management plays a more central role to improve system performance. The classical way to treat interference orthogonalizes the channel access in time, frequency, or even code domain. However, this approach has been known to be suboptimal in general.
The interference channel has been one of the long-standing fundamental problems in network information theory since \cite{Car78}, which finds an optimal way of managing interference and investigates the  fundamental performance limit of all interference management schemes, i.e., the capacity region of this channel.  However, the sum capacity of even the simplest 2-user Gaussian interference channel (GIC)  \cite{Etk08,Ann09,Sha09,Mot09,Etk09,Cha11,Nam12} has not been fully understood, although it was recently shown in \cite{Nam15a} that a relatively small gap between the new upper bounds therein and the time division (or frequency division) lower bound is left in the weak interference regime. The well-known outer bounds on the capacity region of the two-user GIC are the Kramer bound \cite{Kra04} and the Etkin-Tse-Wang (ETW) bound \cite{Etk08}.
The capacity region of the $K$-user GIC with $K\ge 3$ is unknown in general, except for the sum capacity in some special cases including the very strong interference regime \cite{Sri08} for the symmetric\footnote{The special case where all cross-channel coefficients are the same and all users have the same SNR.} GIC and the $K$-user extension of the noisy (very weak) interference regime \cite{Ann09,Sha09}. The notion of strong interference in the two-user case does not naturally extend to even the symmetric three-user case \cite{Sri08}. {For the cyclic $K$-user GIC (Z-interference channel), its capacity region to within a constant gap is studied in \cite{Zho13} based on the well-known (HK) scheme and the ETW bounding approach.}

It is in general quite difficult to obtain {either a constructive lower bound or upper bound on the sum capacity to better understand the more than two-user case. For instance, the HK scheme becomes extremely complicated for $K\ge 3$ even with Gaussian signals and without time sharing. To the best of the author's knowledge, only a few useful upper bounds that help understanding the capacity are known (e.g., \cite{Ann09,Tun11,Tun11b}).} Hence, most of the related works have focused on a simplification of the problem of interest for the $K$-user GIC and restricted our attention to the DoF capacity approximation.
For the DoF characterization, the significant progress has been made mainly owing to interference alignment \cite{Mad08,Cad08}, deterministic channel model \cite{Ave11}, and structured codes \cite{Bre10,Naz11}. {The notion of interference alignment is decoding the sum of interfering signals rather than decoding a part of the individual interference in the HK scheme.} 

\subsection{Prior Works}

There are several generalizations of the above two-user upper bounds to more than two-user cases.
In \cite{Ann09}, the ETW bound was extended to the $K$-user GIC by using a vector genie. However, the resulting \emph{useful genie} bound is not so tight in general. As a consequence, for the three-user symmetric case, the genie-aided upper bound was further tightened by allowing correlation between all additive noise variables and shown to be optimal in the noisy interference case. The Kramer bound was generalized to the three-user GIC in \cite{Tun11} by using the linear minimum mean-squared error (LMMSE) estimation based proof \cite{Kra04} and by following the Sato approach \cite{Sat77} that exploits the fact that the capacity of GIC depends only on the marginal noise distributions so that correlation among Gaussian noises does not affect the capacity. {This bounding technique was further extended to more than three-user cases and some capacity results of {certain classes} of $K$-user GICs were given in \cite{Tun11b}.} A multiple access upper bound was presented in \cite{Cad09} for the sum capacity of the three-user symmetric GIC, where receivers are provided with sufficient side information so as to decode a subset of the users  in the corresponding multiple access channel. This approach was extended to the $K$-user case in \cite{Jos10} but the resulting upper bound has not been evaluated in the literature. Therefore, there have been only few known upper bounds on the capacity for more than three-user GICs. A common framework in the existing bounds for more than two-user cases is based on imposing mutual correlation between noise variables to tighten their bounds. Within this framework, a major difficulty of such bounds is that they involve the numerical optimization of a covariance matrix of jointly Gaussian noise variables, which makes it hard to generalize to the large $K$ case. Even if the generalization is available, it is infeasible to even compute the resulting upper bounds unless $K$ is small.

For the fully connected GIC with more than two users, the interference channels that the existing DoF-based capacity approximation results have considered can be categorized as the following two types: time varying/frequency selective channels and constant (static) channels. Initially, Cadambe and Jafar \cite{Cad08} showed that vector-space interference alignment can achieve $K/2$ DoF for time varying/frequency selective channels. The ergodic interference alignment \cite{Naz12} allows each user to achieve $1/2$ its interference-free ergodic capacity at \emph{any} SNR, but incurring very long delay due to its opportunistic matching of complementary states. Assuming that channel coefficients in each channel use is drawn independently from a continuous random distribution, this type of channels requires sufficiently fast-variation/high-selectivity, which may not be common in practical systems. Hence, we rather focus on the constant GIC in this work. 
For constant channels, the $K/2$ DoF was shown by Motahari \emph{et al.} \cite{Mot09b} to be achievable for {almost all} channel realizations through the use of Diophantine approximation. More recently, Wu \emph{et al.} \cite{Wu15} recovered the same result by developing a general formula based on R{\'e}nyi information dimension. In the complex-valued GIC, phase alignment with asymmetric complex signaling \cite{Cad10a} can be exploited to achieve at least $1.2$ DoF for almost all channel coefficients in the three-user case. In multiple-antenna GIC, vector-space alignment is known in \cite{Bre11} to be feasible for $K$-user symmetric square MIMO GIC if and only if the number of antennas is larger than or equal to $\frac{d(K+1)}{2}$, where $d$ is the number of DoF per user. Meanwhile, the condition of \emph{almost all} channel coefficients in \cite{Mot09b} precludes only a subset of Lebesgue measure zero in reals, $\mathbb{R}$. However, the exceptional cases of measure zero include the set of rational numbers, dense and infinitely many in $\mathbb{R}$. In particular, Etkin and Ordentlich \cite{Etk09b} showed that the DoF of GIC with all rational coefficients is strictly less than $K/2$, thereby concluding that the DoF of GIC with $K\ge 3$ is everywhere discontinuous with respect to channel coefficients. 
Moreover, any irrational number can be approximated by a rational number arbitrarily close to it. This implies that very next to every good channel is bad channel \cite{Nie13}.  
Consequently, the practical implication of the above remarkable DoF results might be limited to some extent.

The {above} discontinuity phenomenon has also been observed in other DoF results that do not exploit the irrationality of channel coefficients. Jafar and Vishwanath \cite{Jaf10} showed that the well-known generalized DoF characterization in \cite{Etk08} for the symmetric two-user GIC naturally extends to the \emph{symmetric (positive) real} $K$-user GIC, with the exception of a singularity when SNR and interference-to-noise ratio (INR) are the same, i.e., the common cross-channel coefficient is $1$, where DoF is only $1$. It was also shown by \cite{Cad10a} that the three-user \emph{complex} GIC has at least $1.2$ DoF for almost all channel coefficients, while just $1$ DoF for a measure-zero subset of channel coefficients satisfying certain phase and amplitude conditions. Furthermore, the symmetric GIC has $1.5$ DoF if all direct-channel coefficients are $1$ and all cross-channel coefficients are $\sqrt{-1}$, which is an exceptional case where the exact capacity of $K/2 \log(1+2\snr)$ is known \cite{Cad08}. 
More recently, it was reported that the discontinuity of DoF characterization with respect to the channel coefficients might be in fact due to the asymptotic analysis in the high SNR limit and that it may not appear  any longer at finite SNR. For example, see \cite{Nie13} for the two-user Gaussian X channel and \cite{Ord14} for the $K$-user symmetric GIC. They provided constant-gap capacity approximations to circumvent the limitation of DoF characterization and to better understand the capacity of X channel and GIC at finite SNR. However, even if the constant gap results hold at any SNR except for an \emph{outage} set of channel coefficients whose measure vanishes exponentially with a target gap $c > 0$ increasing, their constant gaps seem to be {large to date}. In \cite{Ord14}, their scheme approximates the sum capacity of the $K$-user symmetric GIC to within a constant gap of $c+\log(K)+10$ bits up to an outage set of channel coefficients of Lebesgue measure smaller than $2^{-c}$, 
implying that the constant-gap capacity approximation may be weak to appropriately capture the sum-rate behavior at {finite} SNR. 
Therefore, understanding the capacity of the $K$-user GIC at finite SNR still remains {far from the two-user case}. In particular, we have an intriguing open problem as to how much the important \emph{asymptotic} result on  the achievability of $K/2$ DoF for almost all constant channel realizations can be embodied as realistic performance gains outside of the high SNR limit. 

 {In this work, we would like to draw attention to careful interpretation of the DoF results by pointing out that DoF is not necessarily translated into a substantial sum-rate performance gain for practical values (e.g., 10 to 20 dB) of SNR} This is the case with multiple-antenna communications such as multiple-input multiple-output (MIMO) point-to-point/broadcast/multiple-access channels when the multiple-antenna channels are highly correlated (e.g., \cite{Tul05}), where DoF is not fully attainable outside very high SNR. 
Bearing this in mind, we would like to derive useful upper bounds on the capacity of complex-valued GIC with more than two users for SNR of practical interest.

\subsection{Contributions}



We first develop new upper bounds on the sum capacity of the three-user complex GIC, {inspired by the Etkin-type upper bounding approach \cite{Etk08} and the change-of-interference approach \cite{Nam12}. 
The latter is a new genie-aided approach where the noisy interference signals instead of the noisy input signals are provided to the receivers and used to replace those arbitrary interference signals with  independent and identically distributed (i.i.d.) Gaussian signals.}
The known genie-aided bounds were shown to be tight in the noisy interference regime. New ideas in deriving our bounds in this work are the use of genie signals in a different way and a combination of the two bounding approaches {in conjunction with time sharing on the genie signals at the receivers}. The resulting upper bounds are shown to be tighter than the existing bounds over a certain range of channel coefficients. The three-user upper bounds are particularly designed to be amenable to the extension to the general $K$-user case. In particular, we do not involve any auxiliary random {vector} (e.g., vector genie) or optimization of a noise covariance matrix in contrast to the aforementioned framework used in \cite{Ann09,Cad09,Tun11,Jos10}. Even if the resulting $K$-user upper bounds have a relatively low computational complexity, the complexity still becomes prohibitively large even in the symmetric case for $K$ large. To overcome this difficulty, we further provide \emph{closed-form} expressions of our upper bounds for the $K$-user GIC that is a continuous function  for large $K$, whose domain is $\mathbb{R}$, whereby we can investigate the sum-rate behavior for {any} $K$ and any real-valued channel coefficient  irrespectively of whether channel coefficients are irrational or rational. {To this end, a key step is to identify and exploit an intrinsic structure in our upper bounds, which lends themselves to canceling out some pairs of differential entropies.} This is because our bounds intentionally avoid the use of auxiliary random {vectors} and the optimization of a noise covariance matrix. 
The analytical upper bounds have no discontinuous point in the large $K$ symmetric real case. This points out that the exact capacity behavior may not show a large fluctuation due to the irrationality of channel coefficients {for any $K$} in this symmetric real case. {To be fair, the same observation can be found for the three-user case, e.g., in \cite{Tun11}. Another benefit of the proposed analytical bounds is that they are amenable to an affine approximation in the large $K$ and high SNR regime, inspired by \cite{Sha01}. We show that as $K$ grows, the performance benefit promised by the well-known $K/2$ DoF results may not be realized even at high SNR over a certain range (around $g^2=1$) of channel coefficients for the  $K$-user symmetric \emph{positive real} case. 
}

The second part of this work is devoted to the study of our sum-rate upper bounds for GICs with \emph{complex-valued} channel coefficients. {For symmetric complex GICs, where the phases of the cross-channel coefficients are the same but allowed to be different from those of the direct-channel, our study is motivated by the well-known example in \cite{Cad08} where the sum capacity is $\frac{K}{2}\log(1+2P)$ when the common direct- and cross-channel coefficients are $1$ and $\sqrt{-1}$, respectively. In sharp contrast, the sum capacity becomes just $\log(1+KP)$ when the channel coefficients are all $1$. Hence it would be interesting to trace the trajectory of our upper bounds between the two extreme points, which implies that there could be a room} for performance gain achievable by sophisticated schemes including interference alignment and structured codes when the phases of direct and cross-channel coefficients are sufficiently different. 
Then, the following intriguing question naturally arises: Can we always do better by exploiting the phase difference of the direct and cross-channel coefficients? 
To answer this question, the symmetric case is not appropriate because the cross-channel coefficients are already aligned in this case. 

Accordingly, we introduce a ``semi-symmetric" GIC,  {where complex cross-channel coefficients for each user are allowed to be different in contrast to the symmetric case but all users are restricted to experience the same SNR and INR. This more general semi-symmetric GIC includes the above symmetric real and complex GICs as special cases.} For the three-user case, we find that there are ``good" and ``bad" conditions for {potential} performance benefits achievable by sophisticated schemes, yielding a relevant conjecture on certain conditions for such good and bad phase offsets among the direct-channel coefficient and two different cross-channel coefficients. {Interestingly, the bad conditions coincide with those of the DoF result from \cite[Thm.  3]{Cad10a} in the high SNR limit.} Therefore, it turns out that a large phase difference between the direct and cross-channel coefficients does not suffice to achieve a substantial performance gain. This suggests that the good conditions on the phase offset {may deserve} attention to design sophisticated interference management schemes. 


\subsection{The Organization of the Paper}

The remainder of this paper is organized as follows. Section \ref{sec:CM} describes the channel model of the $K$-user GIC that we study. In Section \ref{sec:3IC}, we derive new upper bounds on the sum capacity of the three-user GIC. Section \ref{sec:KIC} generalizes the three-user bounds to the $K$-user GIC, provides closed-form formulas of our upper bounds for the $K$-user symmetric case, and also studies sum-rate behavior of the $K$-user real GIC. In Section \ref{sec:CGIC}, we investigate sum-rate behavior of the $K$-user complex GIC by introducing the semi-symmetric case.  We conclude this work in Section \ref{sec:con}.


\emph{Notations}: We use $X$ for a random variable and $X^n$ for a random sequence. Also, $\sigma^2_X$ denotes the variance of $X$.  For $c\in \mathbb{C}$, let $\mathfrak{R}(c)$ denote the real part of $c$, and $\mathcal{CN}(0,1)$ denotes the zero-mean circularly symmetric complex Gaussian distribution.

\section{Channel Model}
\label{sec:CM}

The $K$-user complex GIC with constant channel coefficients can be defined by
\begin{align} \label{eq:A-1a}
  Y_k&=\sum_{i=1}^K h_{ki}X_i+Z_k 
\end{align}
where $X_k \in \mathbb{C}$ is the channel input for user $k$, subject to an average power constraint $P_k$, $h_{ki} \in \mathbb{C}$ is the channel coefficient from transmitter $i$ to receiver $k$ in which $h_{kk}=1$ for $k=1,2,\cdots,K$, and $Z_k \sim \mathcal{CN}(0,1)$ is drawn from the circularly symmetric complex Gaussian noise process that is i.i.d over time and independent of the channel inputs. The channel coefficients remain constant during the transmission period and are known to all transmitters and receivers. Since every $K$-user complex GIC can be transformed to the standard form in (\ref{eq:A-1a}) with the same capacity region, taking only the normalized direct-channel coefficients into account involves no loss of generality.

Let $M_1, \ldots, M_K$ be independent, uniformly distributed messages over $[1:2^{nR_1}], \ldots, [1:2^{nR_K}]$, and let $X^n_1 \in \boldsymbol{\mathcal{X}}_1,\ldots, X^n_K \in \boldsymbol{\mathcal{X}}_K$ and  $Y^n_1 \in \boldsymbol{\mathcal{Y}}_1, \ldots, Y^n_K \in \boldsymbol{\mathcal{Y}}_K$ be the random sequences induced by encoders $\mathrm{enc}_k: [1:2^{nR_k}]\rightarrow \boldsymbol{\mathcal{X}}_k$ and the channel, respectively, where $\boldsymbol{\mathcal{X}}_1, \ldots, \boldsymbol{\mathcal{X}}_K$ are the codebooks with $|\boldsymbol{\mathcal{X}}_k|=2^{nR_k}$, where the channel input $X_k$ satisfies the average power constraints of $P_k$ such that $||X_k^n||^2\le nP_k, \forall k$.
A rate tuple $(R_1, \ldots,R_K)$ is achievable if there exists a sequence of {$(2^{nR_1},\ldots,2^{nR_K}, n)$} codes with $\lim_{n\rightarrow \infty} P_e^{(n)}=0$, where $P_e^{(n)}$ is the average decoding error probability. The capacity region is the closure of the set of all achievable rate tuple $(R_1, \ldots,R_K)$. The correlation among the Gaussian noises is irrelevant since the capacity region of GIC only depends on the marginal distributions of $Z_k$, i.e., $P_{Y_k|X_1,\ldots,X_K}$, for all $k$.

Throughout this paper, the subscript $_G$ indicates $X_k=X_{kG} \sim \mathcal{CN}(0,P_k)$ so that all other variables including $X_k$ become complex Gaussian distributed. For instance, $Y_{kG}=\sum_{i=1}^K h_{ki}X_{iG}+Z_k$ for all $k$. {The user indices must be understood as $\mathrm{modulo}\; K$ such that $X_{K+1} = X_1$. For the symmetric GIC, SNR and INR are defined as $\snr=P$ and $\inr=|g|^2P$, respectively.}



\section{Three-User Gaussian Interference Channel}
\label{sec:3IC}

The standard channel model of the three-user GIC can be given by 
\begin{align} \label{eq:A-1}
  Y_1&=X_1+h_{12}X_2+h_{13}X_3+Z_1 \nonumber \\ 
  Y_2&=X_2+h_{23}X_3+h_{21}X_1+Z_2 \nonumber \\ 
  Y_3&=X_3+h_{31}X_1+h_{32}X_2+Z_3 .
\end{align}
In order to derive useful upper bounds on the sum capacity of the three-user GIC which are amenable to the more general $K$-user case, we utilize the Etkin-type and the change-of-interference bounding approach in a separate or joint fashion. In this section, we will provide three sum-rate upper bounds, in which the first bound is given by {generalizing the change-of-interference approach in \cite{Nam12}, \cite[Thm. 3]{Nam15a}, {a time-sharing parameter with cardinality 2 was used on genie signals,} for the there-user case.} For the second upper bound, the {Etkin-type genie signals} are used by constructing a new genie-aided channel in conjunction with the \emph{conditional} worst additive noise lemma \cite{Nam12} (see also \cite{Nam15a}), which is a conditional version of the worst additive noise lemma \cite{Dig01}. The last upper bound is to jointly make use of the above two bounding approaches.

\subsection{Change-of-Interference Bound}

The Z channel upper bound in \cite[Thm. 1]{Kra04} was naturally extended in \cite{Tun11,Zho13} for the three-user case as follows:  
\ifdefined\TIT
\begin{align}  \label{eq:3ub-1}
    R_1 +R_2+R_3 \le \frac{1}{2} &\Big\{ I(X_{1G}; Y_{1G} |  X_{2G},X_{3G}) +I(X_{1G}; Y_{1G} |  X_{3G}) \nonumber \\
    &\ +I(X_{2G}; Y_{2G} |  X_{3G},X_{1G}) +I(X_{2G}; Y_{2G} |  X_{1G}) \nonumber \\
    &\ +I(X_{3G}; Y_{3G} |  X_{1G},X_{2G}) +I(X_{3G}; Y_{3G} |  X_{2G})\Big\}
\end{align} 
\else
\begin{align}  \label
    R_1 +R_2&+R_3 \le \frac{1}{2} \big\{ I(X_{1G}; Y_{1G} |  X_{2G},X_{3G}) \nonumber \\
    &\ +I(X_{1G}; Y_{1G} |  X_{3G}) +I(X_{2G}; Y_{2G} |  X_{3G},X_{1G}) \nonumber \\
    &\ +I(X_{2G}; Y_{2G} |  X_{1G}) +I(X_{3G}; Y_{3G} |  X_{1G},X_{2G}) \nonumber \\
    &\  +I(X_{3G}; Y_{3G} |  X_{2G})\big\}
\end{align} 
\fi
for channel coefficients satisfying $|h_{12}|^2\le 1, |h_{23}|^2\le 1$, and $|h_{31}|^2\le 1$.
Permuting the user indices, we obtain $3!$ such bounds in total. 
The first upper bound that we derive is given by modifying (\ref{eq:3ub-1}) with the change-of-interference genie-aided approach in \cite{Nam12}, \cite[Thm. 3]{Nam15a}. Let $U_k, k=1,2,3,$ denote {genie} signals, which we also call change-of-interference variables, defined as
\begin{align} \label{eq:3IC-2}
  U_{1} = h_{12}X_2+h_{13}X_3+W_1 \nonumber\\
  U_{2} = h_{23}X_3+h_{21}X_1+W_2 \nonumber\\
  U_{3} = h_{31}X_1+h_{32}X_2+W_3 
\end{align} 
where the additive noise $W_k$ is distributed as $\mathcal{CN}(0,\sigma_{W_k}^2)$ with $\sigma_{W_k}^2\le 1$, correlated to $Z_k$ with correlation coefficient $\rho_{W_k}$ (i.e., $\mathbb{E}[Z_k W_k^*]=\rho_{W_k}\sigma_{W_k}$) but independent of everything else, for $k=1,2,3$.

Conditioned on the change-of-interference variable $U^n_1$ for the case of user $1$, the arbitrary random sequence $h_{12}X_2^n+h_{13}X_3^n$ (interference signal to user 1) is replaced with the i.i.d. Gaussian random sequence $W^n_1$, which is the main role of the change-of-interference variables.  
Replacing certain $X_2, X_3,$ and  $X_1$ in the side information terms of (\ref{eq:3ub-1}) with $U_1, U_2,$ and  $U_3$, respectively, we can get the following result.

\begin{thm} \label{thm-3ub1}
The sum capacity of the three-user complex GIC is upper-bounded by
\begin{align}  \label{eq:3ub-1c}
    R_1 +R_2+R_3 \le \frac{1}{2} &\Big\{ I(X_{1G}; Y_{1G} |  X_{3G},U_{1G}) +I(X_{1G}; Y_{1G} |  X_{3G}) \nonumber \\
    &\ +I(X_{2G}; Y_{2G} |  X_{1G},U_{2G}) +I(X_{2G}; Y_{2G} |  X_{1G}) \nonumber \\
    &\ +I(X_{3G}; Y_{3G} |  X_{2G},U_{3G})  +I(X_{3G}; Y_{3G} |  X_{2G}) \nonumber \\
    &\ +I(U_{1G};Y_{1G}+\tilde{V}_{W_3}|X_{3G}) +I(U_{2G};Y_{2G}+\tilde{V}_{W_1}|X_{1G})  \nonumber \\
    &\ +I(U_{3G};Y_{3G}+\tilde{V}_{W_2}|X_{2G}) \Big\}    
\end{align} 
for all channel coefficients and $\{W_1,W_2,W_3\}$ satisfying 
\begin{subequations}
\begin{align}  
 &|h_{12}|^2\le 1,\ |h_{23}|^2\le 1,\ |h_{31}|^2\le 1 \label{eq:3ub-30a} \\
 &\sigma^2_{V_{W_1}}\ge |h_{12}|^2 \sigma^2_{Z_2-W_2}  \label{eq:3ub-30b}\\
 &\sigma^2_{V_{W_2}}\ge |h_{23}|^2 \sigma^2_{Z_3-W_3}  \label{eq:3ub-30c}\\ 
 &\sigma^2_{V_{W_3}}\ge |h_{31}|^2 \sigma^2_{Z_1-W_1}  \label{eq:3ub-30d}
\end{align} 
\end{subequations}
where $V_{W_k} {\sim \CN(0,\sigma^2_{W_k|\; Z_k-W_k})}$ for all $k$ and  
\begin{align} 
   \tilde{V}_{W_1}&=\sqrt{|h_{12}|^{-2}-\sigma_{V_{W_1}}^{-2}\sigma_{Z_2- W_2}^2}\;V_{W_1}\nonumber \\ 
   \tilde{V}_{W_2}&=\sqrt{|h_{23}|^{-2}-\sigma_{V_{W_2}}^{-2}\sigma_{Z_3- W_3}^2}\;V_{W_2}\nonumber \\ 
   \tilde{V}_{W_3}&=\sqrt{|h_{31}|^{-2}-\sigma_{V_{W_3}}^{-2}\sigma_{Z_1- W_1}^2}\;V_{W_3}. \nonumber 
\end{align}
Permuting the user indices (i.e., changing the order of the users), we obtain $3!$ such bounds in total. 
\end{thm}

\begin{IEEEproof}
Refer to Appendix \ref{proof-1}. 
\end{IEEEproof}

{
\begin{rem} \label{rem-1}
Comparing the bounds in Theorem \ref{thm-3ub1} and the bound in (\ref{eq:3ub-1}), we can see that the more general side information $U_1, U_2,$ and  $U_3$ (noisy interference) than $X_2, X_3,$ and  $X_1$ (noiseless interference) can tighten upper bounds at the cost of the penalty terms $I(U_{1G};Y_{1G}+\tilde{V}_{W_3}|X_{3G}) +I(U_{2G};Y_{2G}+\tilde{V}_{W_1}|X_{1G}) +I(U_{3G};Y_{3G}+\tilde{V}_{W_2}|X_{2G}) $ in (\ref{eq:3ub-1c}). Hence the bound in Theorem \ref{thm-3ub1} improves upon (\ref{eq:3ub-1}) at a certain range of channel coefficients but also degrades due to the penalty terms and the constraints in (\ref{eq:3ub-30b}) -- (\ref{eq:3ub-30d}) at some other range, as will be shown later in Fig. \ref{fig-1}. 
\end{rem}
}

\subsection{Etkin-Type Bound}

The second sum-rate upper bound to be derived in the following is inspired by the Etkin-type genie-aided approach \cite{Etk08,Sha09,Mot09,Ann09} for the two-user GIC. A generalization  of this approach for more than two-user cases is given by \cite{Ann09} in the standard form of $\sum_{i=1}^K R_i \le \sum_{i=1}^K I(X_{iG}; Y_{iG}, S_{iG})$. However, this type of genie-aided bound is tight only in the noisy interference regime, where cross-channel coefficients are very weak and transmission power should be restricted, and becomes quickly loose by construction, i.e., even quite larger than the interference-free upper bound as cross-channel coefficients get close to $1$.
It is non-trivial to design a \emph{different} genie-aided channel from the standard form, $\sum_{i=1}^K I(X_{iG}; Y_{iG}, S_{iG})$, which should yield a new genie-aided upper bound useful for the moderately weak interference regime rather than the noisy interference regime. In order to construct such a new form of genie-aided channel, we first define the genie signals for the three-user case as
\begin{align} \label{eq:3IC-1}
  S_{1} = h_{31}X_{1}+h_{32}X_{2}+N_1 \nonumber\\
  S_{2} = h_{12}X_{2}+h_{13}X_{3}+N_2 \nonumber\\
  S_{3} = h_{23}X_{3}+h_{21}X_{1}+N_3 
\end{align} 
where $N_k$ is distributed as $\mathcal{CN}(0,\sigma_{N_k}^2)$ with $\sigma_{N_k}^2\le 1$,  correlated to $Z_k$ with correlation coefficient $\rho_{N_k}$ (i.e., $\mathbb{E}[Z_kN_k^*]=\rho_{N_k}\sigma_{N_k}$) but independent of everything else, for $k=1,2,3$.  

With the above definitions of genie signals and additive Gaussian noises, the following result presents a new type of genie-aided upper bound.

\begin{thm} \label{thm-3ub2}
The sum capacity of the three-user complex GIC is upper-bounded by
\ifdefined\TIT
\begin{align}  
    R_1 +R_2+R_3 \le &\; I(X_{1G}; Y_{1G} ) +I(X_{2G}; Y_{2G}, S_{2G} |X_{1G}) +I(X_{3G}; Y_{3G} |  X_{1G},X_{2G}) \label{eq:3ub-2a} 
\end{align}
for all $N_2$ satisfying 
{ \begin{align}  
    |h_{13}|^2\le \sigma^2_{V_{N_2}} \le 1   \ &\text{ or } \ |h_{23}|^2 \le \sigma^2_{V'_{N_2}} \le 1
    \label{eq:3ub-2d} 
\end{align}
where 
\begin{subequations}  
\begin{align} 
  V_{N_1}&\sim \CN(0,\sigma^2_{N_1|\; Z_1-h_{12}h_{32}^{-1}N_1}) \label{eq:3IC-141}\\
  V_{N_2}&\sim \CN(0,\sigma^2_{N_2|\; Z_2-h_{23}h_{13}^{-1}N_2}) \label{eq:3IC-142}\\
  V_{N_3}&\sim \CN(0,\sigma^2_{N_3|\; Z_3-h_{31}h_{21}^{-1}N_3}) \label{eq:3IC-143}
\end{align} 
\end{subequations}  
and
\begin{subequations}  
\begin{align} 
  V'_{N_1}&\sim \CN(0,\sigma^2_{Z_1|\; N_1-h_{32}h_{12}^{-1}Z_1}) \label{eq:3IC-14ba}\\
  V'_{N_2}&\sim \CN(0,\sigma^2_{Z_2|\; N_2-h_{13}h_{23}^{-1}Z_2}) \label{eq:3IC-14bb} \\
  V'_{N_3}&\sim \CN(0,\sigma^2_{Z_3|\; N_3-h_{21}h_{31}^{-1}Z_3}). \label{eq:3IC-14bc}
\end{align} 
\end{subequations}  
}
 \else
\begin{align}  
    R_1 +R_2+R_3 \le &\; I(X_{1G}; Y_{1G} ) +I(X_{2G}; Y_{2G}, S_{2G} |X_{1G}) \nonumber\\
   &+I(X_{3G}; Y_{3G} |  X_{1G},X_{2G}) \label{eq:3ub-2a}   
\end{align}
where (\ref{eq:3ub-2a}) is valid for $|h_{13}|^2 \le\sigma^2_{V_{N_2}}$. 
\fi
Permuting the user indices, we obtain $3!$ such bounds in total. 
\end{thm}

\begin{IEEEproof}
Refer to Appendix \ref{proof-2}.
\end{IEEEproof}

It follows from (\ref{eq:3IC-7}) that, for the first condition in (\ref{eq:3ub-2d}), we can rewrite (\ref{eq:3ub-2a}) as
\begin{align} \label{eq:3IC-25}
   R_1 +R_2+R_3 & \le \log \left(1+\frac{P_1}{|h_{12}|^2P_2+|h_{13}|^2P_3+1}\right)  +\log \left(\frac{|h_{12}|^2P_2+|h_{13}|^2P_3+\sigma_{N_2}^2}{1+|h_{23}h_{13}^{-1}|^2\sigma_{N_2}^2-2\mathfrak{R}\big\{|h_{23}h_{13}^{-1}|^2\rho_{N_2}\sigma_{N_2}\big\}}\right) \nonumber \\ 
   &\ \ \ +\log \left(\frac{P_2+|h_{23}|^2P_3+1-\frac{\left|h_{12}^*P_2+h_{23}h_{13}^*P_3+\rho_{N_2}\sigma_{N_2}\right|^2}{|h_{12}|^2P_2+|h_{13}|^2P_3+\sigma_{N_2}^2}}{|h_{13}|^2P_3+\sigma_{V_{N_2}}^2}\right)  +\log \big(1+P_3\big)
\end{align}
where 
\begin{align} \label{eq:3IC-25b}
  \sigma_{V_{N_2}}^2=\sigma_{N_2}^2-\frac{\left|\rho_{N_2}\sigma_{N_2}-h_{23}h_{13}^{-1}\sigma_{N_2}^2\right|^2}{1+|h_{23}h_{13}^{-1}|^2\sigma_{N_2}^2-2\mathfrak{R}\big\{|h_{23}h_{13}^{-1}|^2\rho_{N_2}\sigma_{N_2}\big\}}.
\end{align}
For the second condition in (\ref{eq:3ub-2d}), we can also get the following expression:
\begin{align} \label{eq:3IC-27}
   R_1 +R_2+R_3 & \le \log \left(1+\frac{P_1}{|h_{12}|^2P_2+|h_{13}|^2P_3+1}\right)  +\log \left(\frac{|h_{12}|^2P_2+|h_{13}|^2P_3+\sigma_{N_2}^2}{\sigma_{N_2}^2+|h_{13}h_{23}^{-1}|^2-2\mathfrak{R}\big\{|h_{13}h_{23}^{-1}|^2\rho_{N_2}\sigma_{N_2}\big\}}\right) \nonumber \\ 
   &\ \ \ +\log \left(\frac{P_2+|h_{23}|^2P_3+1-\frac{\left|h_{12}^*P_2+h_{23}h_{13}^*P_3+\rho_{N_2}\sigma_{N_2}\right|^2}{|h_{12}|^2P_2+|h_{13}|^2P_3+\sigma_{N_2}^2}}{|h_{23}|^2P_3+\sigma_{V'_{N_2}}^2}\right)  +\log \big(1+P_3\big)
\end{align}
where 
\begin{align} \label{eq:3IC-27b}
  \sigma_{V'_{N_2}}^2=1-\frac{\left|\rho_{N_2}\sigma_{N_2}-h_{13}h_{23}^{-1}\right|^2}{\rho_{N_2}^2+|h_{23}h_{13}^{-1}|^2-2\mathfrak{R}\big\{|h_{13}h_{23}^{-1}|^2\rho_{N_2}\sigma_{N_2}\big\}}.
\end{align}
Therefore, the upper bound in Theorem \ref{thm-3ub2} is given by the minimum of (\ref{eq:3IC-25}) and (\ref{eq:3IC-27}) over all parameters $\rho_{N_2}$ and $\sigma_{N_2}$ satisfying (\ref{eq:3ub-2d}). For the symmetric case, (\ref{eq:3IC-25}) and (\ref{eq:3IC-27}) are equivalent. Interchanging the user indices, we have additional $(3!-1)$ bounds as well.

\begin{rem}
The standard genie-aided channel is different from our genie-aided channel in (\ref{eq:3ub-2a}) where only a single receiver (receiver 2) is provided with the corresponding genie signal ($S_2^n$), apart from the condition on $X_1^n$.
In general, the most difficult part to find a single-letter expression for genie-aided upper bounds is how to handle the negative non-Gaussian entropy terms, as well addressed in \cite{Tel07}. To this end, the key step in the proof of Theorem \ref{thm-3ub2} was to carefully design the genie signal $S_2^n$ and the additional side information $X_1^n$ so as to apply Lemma \ref{lem-6} to $-h(S_2^n|X^n_2)-h(Y^n_2| X_1^n,X^n_2,S_2^n)$ in (\ref{eq:3IC-3b}), thus replacing $h(S_2^n|X^n_2)$ with the Gaussian entropy $nh(Z_2-h_{23}h_{13}^{-1}N_2)$. 
\end{rem}

For the special case where the cross-channel coefficients are all unity, it is well known \cite{Jaf10} that the time division scheme achieves the sum capacity. We can easily show that the upper bound coincides with the time division lower bound in this case. Letting $N_2=Z_2$ for (\ref{eq:3ub-2a}), we get $h(Z_2-h_{23}h_{13}^{-1}N_2) =h(Y_{2G}|X_{1G},S_{2G})=0$. Then, (\ref{eq:3IC-25}) reduces to $R_\text{sum} \le \frac{1}{6}\log(1+3P)$ for the symmetric real case. Therefore, our upper bound is tight for this special case. 

{Notice that the generalized Z-channel bound in \cite{She12} has the same mutual information terms as Theorem \ref{thm-3ub2}. However the constraints are different from each other, and $N_i$ in our bound are not restricted to have the same marginal probability as $Z_i$, which leads to different bounds in general.}

\subsection{Hybrid Genie-Aided Bound}

We point out that the previous upper bounds in Theorems \ref{thm-3ub1} and \ref{thm-3ub2} are restricted to {the ``mixed" interference channel. The mixed (i.e., weaker and stronger interference signals than the intended signal) interference channel is defined such that at least one of the amplitudes of cross-channel coefficients $|h_{ki}|, k\neq i,$ should be less than or equal to $1$, as shown in (\ref{eq:3ub-30a}) and (\ref{eq:3ub-2d}). Unlike the mixed interference regime in the two-user case \cite{Sha09}, our mixed interference scenario includes the weak interference regime as well.}  
Furthermore, the first bound {in Theorem \ref{thm-3ub1}} in fact comes from the existing two-user bounds and hence rather loose in the three-user real GIC, while the second bound in {Theorem \ref{thm-3ub2}} is outperformed by the first one when there is even small phase offset between direct link and cross link in the complex GIC, as will be shown later in subsection \ref{sec:CGIC-A}. Therefore, we need the third bound which is valid irrespectively of channel coefficients.

{
Inspired by \cite{Nam15a}, we first introduce a time-sharing operation with respect to side information at the three receivers. 
Let $Q$ denote a time sharing random variable. In order to conduct time sharing on the genie signals $S_{k}$ and $U_{k}$ with $|Q|=3$, we define a new genie signal $T_{k}^n$ as 
\begin{align} \label{eq:IS-2}
  T_{k}^n = \left\{ \begin{array}{ll}
  0 &  \text{if } Q=0\\
  S_{k}^n &  \text{if } Q=1\\
  U_{k}^n &  \text{if } Q=2
  \end{array} \right.
\end{align}
for $k=1,2,3.$ The order of $0, S_{k}, U_{k}$ with the equal probability does not change a resulting capacity bound.  
The random sequences $T_{k}^n$ are conditionally independent given $Q$. 
Using Fano's inequality and letting $\mathrm{Pr}(Q=0)=\mathrm{Pr}(Q=1)=\mathrm{Pr}(Q=2)=1/3$, we can write 
\begin{align} \label{eq:ob-7c}
  n(R_1+R_2+R_3-3\epsilon_n) &\le I({X}_{1}^n;{Y}_{1}^n) +I(X_{2}^n;Y_{2}^n) +I(X_{3}^n;Y_{3}^n) \nonumber \\
  &\le I({X}_{1}^n;{Y}_{1}^n,T_{1}^n) +I(X_{2}^n;Y_{2}^n,T_{2}^n) +I(X_{3}^n;Y_{3}^n,T_{3}^n)  \nonumber \\
  &\overset{(a)}{\le} I({X}_{1}^n;{Y}_{1}^n,T_{1}^n|Q) +I(X_{2}^n;Y_{2}^n,T_{2}^n|Q) +I(X_{3}^n;Y_{3}^n,T_{3}^n|Q)   \nonumber \\
  &= \frac{1}{3}\sum_{k=1}^3\Big\{I(X_{k}^n;Y_{k}^n) +I(X_{k}^n;Y_{k}^n,S_k^n) +I({X}_{k}^n;{Y}_{k}^n,U_{k}^n) \Big\} \nonumber \\
  &\overset{(b)}{\le} \frac{1}{3}\sum_{k=1}^3\Big\{I(X_{k}^n;Y_{k}^n) +I(X_{k}^n;Y_{k}^n,S_k^n) +I({X}_{k}^n;{Y}_{k}^n|U_{k}^n) \Big\} \nonumber \\
  &\le \frac{1}{3}\sum_{k=1}^3\Big\{I(X_{k}^n;Y_{k}^n) +I(X_{k}^n;Y_{k}^n,S_k^n|X_{k-1}^n) +I({X}_{k}^n;{Y}_{k}^n|U_{k}^n) \Big\} 
\end{align}
where $(a)$ follows from the independence between $X_k^n$ and $Q$, and $(b)$ is from the independence between $X_k^n$ and $U_k^n$ all by definition. 
Starting from (\ref{eq:ob-7c}), we can get the third upper bound for the three-user case in the following hybrid fashion.  
}

\begin{thm} \label{thm-3ub3}
The sum capacity of the three-user complex GIC is upper-bounded by
\begin{align}  \label{eq:3ub-3}
    R_1 +R_2+R_3 \le \frac{1}{3} &\Big\{ I(X_{1G}; Y_{1G} ) +I(X_{2G}; Y_{2G}, S_{2G} |X_{1G}) +I(X_{3G}; Y_{3G} |  U_{3G}) \nonumber \\
    &\ +I(X_{2G}; Y_{2G} ) +I(X_{3G}; Y_{3G}, S_{3G} |X_{2G}) +I(X_{1G}; Y_{1G} |   U_{1G}) \nonumber \\
    &\ +I(X_{3G}; Y_{3G} ) +I(X_{1G}; Y_{1G}, S_{1G} |X_{3G}) +I(X_{2G}; Y_{2G} |   U_{2G})  \Big\} \nonumber \\    
    +&  \min \big\{I_0, I_1 \big\}
\end{align}
where 
\begin{align}  
  I_0 = \frac{1}{3}\Big\{ I(U_{1G};Y_{1G}+\tilde{V}_{N_3})+ I(U_{2G};Y_{2G}+\tilde{V}_{N_1})+ I(U_{3G};Y_{3G}+\tilde{V}_{N_2})\Big\} \label{eq:3ub-3b}
\end{align}
with the set of noise terms $(N_1,N_2,N_3,W_1,W_2,W_3)$ satisfying 
\begin{subequations}  \label{eq:3ub-4}
\begin{align} 
  &\sigma^2_{V_{W_1}}\ge \sigma^2_{N_2}, \ \sigma^2_{V_{N_2}}\ge |h_{13}|^2 \sigma^2_{Z_3-W_3} \label{eq:3ub-4a} \\
  &\sigma^2_{V_{W_2}}\ge \sigma^2_{N_3}, \ \sigma^2_{V_{N_3}}\ge |h_{21}|^2 \sigma^2_{Z_1-W_1} \label{eq:3ub-4b} \\
  &\sigma^2_{V_{W_3}}\ge \sigma^2_{N_1} , \ \sigma^2_{V_{N_1}}\ge |h_{32}|^2 \sigma^2_{Z_2-W_2} \label{eq:3ub-4c}
\end{align}
\end{subequations}
where
\begin{align}  
  \tilde{V}_{N_1} &= \sqrt{{|h_{32}|^{-2}-\sigma_{V_{N_1}}^{-2}\sigma_{Z_2- W_2}^2}}\;V_{N_1} \nonumber \\
  \tilde{V}_{N_2} &=\sqrt{{|h_{13}|^{-2}-\sigma_{V_{N_2}}^{-2}\sigma_{Z_3- W_3}^2}}\;V_{N_2} \nonumber \\
  \tilde{V}_{N_3} &=\sqrt{{|h_{21}|^{-2}-\sigma_{V_{N_3}}^{-2}\sigma_{Z_1- W_1}^2}}\;V_{N_3} \nonumber 
\end{align}
and ${V_{N_k}}$ are given in (\ref{eq:3IC-141}) -- (\ref{eq:3IC-143}), and  
\begin{align}  
  I_1 = \frac{1}{3}\Big\{ I(U_{1G};Y_{1G}+\tilde{V}_{N_3}')+ I(U_{2G};Y_{2G}+\tilde{V}_{N_1}')+ I(U_{3G};Y_{3G}+\tilde{V}_{N_2}')\Big\} \label{eq:3ub-3c}
\end{align}
with $(N_1,N_2,N_3,W_1,W_2,W_3)$ satisfying
\begin{subequations} \label{eq:3ub-4z}
\begin{align}  
  &\sigma^2_{V_{W_1}}\ge \sigma^2_{W_1}, \ \sigma^2_{V'_{N_2}}\ge |h_{23}|^2 \sigma^2_{Z_3-W_3} \label{eq:3ub-4d} \\
  &\sigma^2_{V_{W_2}}\ge \sigma^2_{W_2}, \ \sigma^2_{V'_{N_3}}\ge |h_{31}|^2 \sigma^2_{Z_1-W_1} \label{eq:3ub-4e} \\
  &\sigma^2_{V_{W_3}}\ge \sigma^2_{W_3} , \ \sigma^2_{V'_{N_1}}\ge |h_{12}|^2 \sigma^2_{Z_2-W_2} \label{eq:3ub-4f}
\end{align}
\end{subequations}
where
\begin{align}  
  \tilde{V}_{N_1}' &= \sqrt{{|h_{12}|^{-2}-\sigma_{V_{N_1}'}^{-2}\sigma_{Z_2- W_2}^2}}\;V_{N_1}' \nonumber \\
  \tilde{V}_{N_2}' &=\sqrt{{|h_{23}|^{-2}-\sigma_{V_{N_2}'}^{-2}\sigma_{Z_3- W_3}^2}}\;V_{N_2}' \nonumber \\
  \tilde{V}_{N_3}' &=\sqrt{{|h_{31}|^{-2}-\sigma_{V_{N_3}'}^{-2}\sigma_{Z_1- W_1}^2}}\;V_{N_3}'. \nonumber 
\end{align}
and ${V_{N_k}'}$ are given in (\ref{eq:3IC-14ba}) -- (\ref{eq:3IC-14bc}). 
Permuting the user indices, we obtain $3!$ such bounds in total.
\end{thm}

\begin{figure}
\hspace{-7mm}
  \center 
  \includegraphics[scale=1]{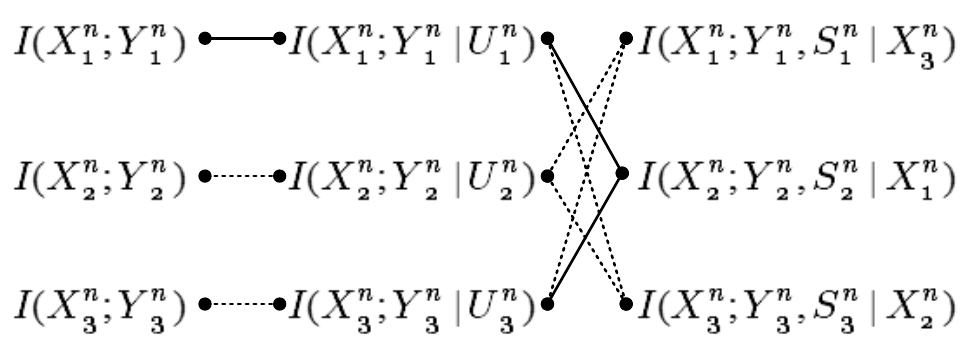}
  \caption{Graphical representation of the relation between mutual informations in Theorem \ref{thm-3ub3}. {The solid lines correspond to the chain of differential entropies $\alpha_{1X}\rightarrow\alpha_{2X}\rightarrow\alpha_{3X}$ given by appropriately pairing of positive and negative entropies in (\ref{eq:3IC-8}), where $X=A, B$.}}\label{fig-0}
\end{figure}

\begin{IEEEproof}
We can bound $R_1$ as
\begin{align}  \label{eq:3IC-9}
   n(R_1-\epsilon_n) \le &\; I(X^n_1;Y^n_1|U^n_1) \nonumber \\
   =&\;h(Y^n_1|U^n_1)-h(Y^n_1|X^n_1,U^n_1) \nonumber \\
   =&\; h(Y^n_1|U^n_1)-h(h_{12}X_2^n+h_{13}X_3^n+Z_1^n|h_{12}X_2^n+h_{13}X_3^n+W_1^n) \nonumber \\
   =&\;  h(Y^n_1|U^n_1) - h(h_{12}X^n_2+h_{13}X_3^n+V^n_{W_1}) +h(U^n_1) -nh(Z_1-W_1) \nonumber \\
   =&\;  h(X^n_1+Z_1^n-W_1^n|U^n_1) - h(h_{12}X^n_2+h_{13}X_3^n+V^n_{W_1}) +h(U^n_1) -nh(Z_1-W_1).
\end{align}
and also $R_2$ as
\ifdefined\TIT
\begin{align}  
  n(R_2-\epsilon_n) \;&\le I(X^n_2;Y^n_2,S_2^n| X_1^n) \nonumber \\ 
  &= I(X^n_2;S_2^n| X_1^n) +I(X^n_2;Y^n_2| X_1^n,S_2^n) \nonumber \\ 
  &= h(S_2^n) -h(S_2^n|X^n_2) +h(Y^n_2| X_1^n,S_2^n) -h(Y^n_2| X_1^n,X^n_2,S_2^n) \nonumber \\ 
  &= h(h_{12}X_{2}^n+h_{13}X_{3}^n+N_2^n)-h(h_{13}X_{3}^n+N_2^n) \nonumber \\ 
  &\ \ \ \ \ +h(X_2^n+h_{23}X_3^n+Z_2^n|S_2^n)  -h(h_{23}X_3^n+Z_2^n|h_{13}X_{3}^n+N_2^n)  \label{eq:3IC-3c} \\  
  & \overset{(a)}{=} h(h_{12}X_{2}^n+h_{13}X_{3}^n+N_2^n)-h(Z_2^n-h_{23}h_{13}^{-1}N_2^n)  \nonumber \\ &\ \ \ \ \ +h(X_2^n+h_{23}X_3^n+Z_2^n|S_2^n)  -h(h_{13}X_3^n+V_{N_2}^n) \label{eq:3IC-3b}  \\ 
  &\overset{(b)}{\le} h(S_{2}^n)-nh(Z_2-h_{23}h_{13}^{-1}N_2) +nh(X_{2G}+h_{23}X_{3G}+Z_2|S_{2G})  -h(h_{13}X_3^n+V_{N_2}^n)  \label{eq:3IC-3}
 \end{align}
\else
\begin{align}  \label{eq:3IC-3}
&\hspace{-3mm}  n(R_2-\epsilon_n) \le I(X^n_2;Y^n_2,S_2^n| X_1^n) \nonumber \\ 
  &\hspace{-3mm}= I(X^n_2;S_2^n| X_1^n) +I(X^n_2;Y^n_2| X_1^n,S_2^n) \nonumber \\ 
  &\hspace{-3mm}= h(S_2^n) -h(S_2^n|X^n_2) +h(Y^n_2| X_1^n,S_2^n) -h(Y^n_2| X_1^n,X^n_2,S_2^n) \nonumber \\ 
  &\hspace{-3mm}= h(h_{12}X_{2}^n+h_{13}X_{3}^n+N_2^n)-h(h_{13}X_{3}^n+N_2^n) \nonumber \\ 
  &\hspace{-3mm} +h(X_2^n+h_{23}X_3^n+Z_2^n|S_2^n)  -h(h_{23}X_3^n+Z_2^n|h_{13}X_{3}^n+N_2^n)  \nonumber \\
  &\hspace{-3mm}\overset{(a)}{=} h(h_{12}X_{2}^n+h_{13}X_{3}^n+N_2^n)-h(Z_2^n-h_{23}h_{13}^{-1}N_2^n)  \nonumber \\ &\ +h(X_2^n+h_{23}X_3^n+Z_2^n|S_2^n)  -h(h_{13}X_3^n+V_{N_2}^n) \nonumber  \\ 
  &\hspace{-3mm}\overset{(b)}{\le} h(S_{2}^n)-nh(Z_2-h_{23}h_{13}^{-1}N_2)\nonumber \\ &  +nh(X_{2G}+h_{23}X_{3G}+Z_2|S_{2G})  -h(h_{13}X_3^n+V_{N_2}^n)  
 \end{align}
\fi
where $(a)$ follows from applying Lemma \ref{lem-6} in Appendix \ref{proof-1} to $h(h_{23}X_3^n+Z_2^n|h_{13}X_{3}^n+N_2^n)$, and $(b)$ follows from \cite[Lemma 1]{Tho87}. 

Using (\ref{eq:3IC-9}) and (\ref{eq:3IC-3}), we can write
\ifdefined\TIT
\begin{align}  
  I(X^n_1; Y^n_1) &\le nh(Y_{1G})-\underbrace{h(h_{12}X_2^n+h_{13}X_3^n+Z_1^n)}_{\triangleq \;\alpha_{1B}}  \nonumber \\
  I(X^n_1;Y^n_1|U^n_1) &= \underbrace{h(X^n_1+Z_1^n-W_1^n|U^n_1)}_{\triangleq \;\alpha_{9B}} - \underbrace{h(h_{12}X^n_2+h_{13}X_3^n+V^n_{W_1})}_{\triangleq \;\alpha_{2B}} \nonumber \\ & \ \ \ +\underbrace{h(h_{12}X_2^n+h_{13}X_3^n+W_1^n)}_{\triangleq \;\alpha_{1A}} -nh(Z_1-W_1)  \nonumber \\
  I(X^n_2;Y^n_2,S_2^n| X_1^n) &\le  \underbrace{h(h_{12}X_{2}^n+h_{13}X_{3}^n+N_2^n)}_{\triangleq \;\alpha_{2A}}-nh(Z_2-h_{23}h_{13}^{-1}N_2)  \nonumber \\ &\ \ \ +nh(X_{2G}+h_{23}X_{3G}+Z_2|S_{2G})  -\underbrace{h(h_{13}X_3^n+V_{N_2}^n)}_{\triangleq \;\alpha_{3B}} \nonumber \\
I(X^n_3; Y^n_3) &\le nh(Y_{3G})-\underbrace{h(h_{31}X_1^n+h_{32}X_2^n+Z_3^n)}_{\triangleq \;\alpha_{4B}}  \nonumber \\
  I(X^n_3;Y^n_3|U^n_3) &= \underbrace{h(X^n_3+Z_3^n-W_3^n|U^n_3)}_{\triangleq \;\alpha_{3A}} - \underbrace{h(h_{31}X^n_1+h_{32}X_2^n+V^n_{W_3})}_{\triangleq \;\alpha_{5B}} \nonumber \\ & \ \ \ +\underbrace{h(h_{31}X_1^n+h_{32}X_2^n+W_3^n)}_{\triangleq \;\alpha_{4A}} -nh(Z_3-W_3) \nonumber 
\end{align}
\begin{align}  \label{eq:3IC-8}    
  I(X^n_1;Y^n_1,S_1^n| X_3^n) &\le  \underbrace{h(h_{31}X^n_1+h_{32}X_2^n+N_1^n)}_{\triangleq \;\alpha_{5A}}-nh(Z_1-h_{12}h_{32}^{-1}N_1)  \nonumber \\ &\ \ \ +nh(X_{1G}+h_{12}X_{2G}+Z_1|S_{1G})  -\underbrace{h(h_{32}X_2^n+V_{N_1}^n)}_{\triangleq \;\alpha_{6B}} \nonumber \\
  I(X^n_2; Y^n_2) &\le nh(Y_{2G})-\underbrace{h(h_{23}X_3^n+h_{21}X_1^n+Z_2^n)}_{\triangleq \;\alpha_{7B}}  \nonumber \\
  I(X^n_2;Y^n_2|U^n_2) &= \underbrace{h(X^n_2+Z_2^n-W_2^n|U^n_2)}_{\triangleq \;\alpha_{6B}} - \underbrace{h(h_{23}X_{3}^n+h_{21}X_{1}^n+V^n_{W_2})}_{\triangleq \;\alpha_{8B}} \nonumber \\ & \ \ \ +\underbrace{h(h_{23}X_3^n+h_{21}X_1^n+W_2^n)}_{\triangleq \;\alpha_{7A}} -nh(Z_2-W_2)  \nonumber \\
  I(X^n_3;Y^n_3,S_3^n| X_2^n) &\le  \underbrace{h(h_{23}X_{3}^n+h_{21}X_{1}^n+N_3^n)}_{\triangleq \;\alpha_{8A}}-nh(Z_3-h_{31}h_{21}^{-1}N_3)  \nonumber \\ &\ \ \ +nh(X_{3G}+h_{31}X_{1G}+Z_3|S_{3G})  -\underbrace{h(h_{21}X_1^n+V_{N_3}^n)}_{\triangleq \;\alpha_{9A}}. 
\end{align}

\else

\begin{align}  
  &I(X^n_1; Y^n_1) \le nh(Y_{1G})-\underbrace{h(h_{12}X_2^n+h_{13}X_3^n+Z_1^n)}_{\triangleq \;\alpha_{1B}}  \nonumber \\
  &I(X^n_1;Y^n_1|U^n_1) = \underbrace{h(X^n_1+Z_1^n-W_1^n|U^n_1)}_{\triangleq \;\alpha_{9B}} \nonumber \\
  &\ \ \ - \underbrace{h(h_{12}X^n_2+h_{13}X_3^n+V^n_{W_1})}_{\triangleq \;\alpha_{2B}} \nonumber \\ & \ \ \ +\underbrace{h(h_{12}X_2^n+h_{13}X_3^n+W_1^n)}_{\triangleq \;\alpha_{1A}} -nh(Z_1-W_1)  \nonumber \\ 
  &I(X^n_2;Y^n_2,S_2^n| X_1^n) \le  \underbrace{h(h_{12}X_{2}^n+h_{13}X_{3}^n+N_2^n)}_{\triangleq \;\alpha_{2A}} \nonumber \\ &\ \ \ -nh(Z_2-h_{23}h_{13}^{-1}N_2)  \nonumber \\ &\ \ \ +nh(X_{2G}+h_{23}X_{3G}+Z_2|S_{2G})  -\underbrace{h(h_{13}X_3^n+V_{N_2}^n)}_{\triangleq \;\alpha_{3B}} \nonumber \\
  &I(X^n_3; Y^n_3) \le nh(Y_{3G})-\underbrace{h(h_{31}X_1^n+h_{32}X_2^n+Z_3^n)}_{\triangleq \;\alpha_{4B}}  \nonumber 
\end{align}
\begin{align}  
  &I(X^n_3;Y^n_3|U^n_3) = \underbrace{h(X^n_3+Z_3^n-W_3^n|U^n_3)}_{\triangleq \;\alpha_{3A}}\nonumber \\ &\ \ \ - \underbrace{h(h_{31}X^n_1+h_{32}X_2^n+V^n_{W_3})}_{\triangleq \;\alpha_{5B}} \nonumber \\ & \ \ \ +\underbrace{h(h_{31}X_1^n+h_{32}X_2^n+W_3^n)}_{\triangleq \;\alpha_{4A}} -nh(Z_3-W_3) \nonumber \\
  &I(X^n_1;Y^n_1,S_1^n| X_3^n) \le  \underbrace{h(h_{31}X^n_1+h_{32}X_2^n+N_1^n)}_{\triangleq \;\alpha_{5A}}\nonumber \\ &\ \ \ -nh(Z_1-h_{12}h_{32}^{-1}N_1)  \nonumber \\ &\ \ \ +nh(X_{1G}+h_{12}X_{2G}+Z_1|S_{1G})  -\underbrace{h(h_{32}X_2^n+V_{N_1}^n)}_{\triangleq \;\alpha_{6B}} \nonumber \\
  &I(X^n_2; Y^n_2) \le nh(Y_{2G})-\underbrace{h(h_{23}X_3^n+h_{21}X_1^n+Z_2^n)}_{\triangleq \;\alpha_{7B}}  \nonumber \\
  &I(X^n_2;Y^n_2|U^n_2) = \underbrace{h(X^n_2+Z_2^n-W_2^n|U^n_2)}_{\triangleq \;\alpha_{6B}}\nonumber \\ &\ \ \ - \underbrace{h(h_{23}X_{3}^n+h_{21}X_{1}^n+V^n_{W_2})}_{\triangleq \;\alpha_{8B}} \nonumber \\ & \ \ \ +\underbrace{h(h_{23}X_3^n+h_{21}X_1^n+W_2^n)}_{\triangleq \;\alpha_{7A}} -nh(Z_2-W_2)  \nonumber \\
  &I(X^n_3;Y^n_3,S_3^n| X_2^n) \le  \underbrace{h(h_{23}X_{3}^n+h_{21}X_{1}^n+N_3^n)}_{\triangleq \;\alpha_{8A}}\nonumber \\ &\ \ \ -nh(h_{21}h_{31}^{-1}Z_2-N_2)  \nonumber \\ &\ \ \ +nh(X_{3G}+h_{31}X_{1G}+Z_3|S_{3G})  -\underbrace{h(h_{21}X_1^n+V_{N_3}^n)}_{\triangleq \;\alpha_{9A}}.  \label{eq:3IC-8} 
\end{align}
\fi

Using the worst additive noise lemma and its conditional version in \cite[Lemma 3]{Nam12}, we can upper-bound the multi-letter expressions of $\alpha_{1A}-\alpha_{1B}$, $\alpha_{2A}-\alpha_{2B}$, and $\alpha_{3A}-\alpha_{3B}$ as  
\ifdefined\TIT
\begin{align}  
   \frac{1}{n}\big(\alpha_{1A}-\alpha_{1B}\big) & \le h(h_{12}X_{2G}+h_{13}X_{3G}+W_1) -h(h_{12}X_{2G}+h_{13}X_{3G}+Z_1)  \label{eq:3IC-11a} \\
   \frac{1}{n}\big(\alpha_{2A}-\alpha_{2B}\big) & \le h(h_{12}X_{2G}+h_{13}X_{3G}+N_2) -h(h_{12}X_{2G}+h_{13}X_{3G}+V_{W_1})   \label{eq:3IC-11b} \\
   \frac{1}{n}\big(\alpha_{3A}-\alpha_{3B}\big) & \le h(X_{3G}+Z_3-W_3|U_{3G}) -  h(X_{3G}+Z_3-W_3+\tilde{V}_{N_2}|U_{3G}) -\log|h_{13}|^2 \label{eq:3IC-11c}
\end{align} 
\else
\begin{align}  
   \frac{1}{n}\big(\alpha_{1A}-\alpha_{1B}\big) & \le h(h_{12}X_{2G}+h_{13}X_{3G}+W_1) \nonumber \\ & \hspace{-15mm} -h(h_{12}X_{2G}+h_{13}X_{3G}+Z_1)  \label{eq:3IC-11a} \\
   \frac{1}{n}\big(\alpha_{2A}-\alpha_{2B}\big) & \le h(h_{12}X_{2G}+h_{13}X_{3G}+N_2) \nonumber \\ & \hspace{-15mm} -h(h_{12}X_{2G}+h_{13}X_{3G}+V_{W_1})   \label{eq:3IC-11b} \\
   \frac{1}{n}\big(\alpha_{3A}-\alpha_{3B}\big) & \le h(X_{3G}+Z_3-W_3|U_{3G}) \nonumber \\ & \hspace{-15mm} -  h(X_{3G}+Z_3-W_3+\tilde{V}_{N_2}|U_{3G}) -\log|h_{13}|^2 \label{eq:3IC-11c}
\end{align} 
\fi
where the last bound is valid for the conditions in (\ref{eq:3ub-4a}).
Similar to (\ref{eq:3IC-12c}), the second term in the right-hand side of (\ref{eq:3IC-11c}) can be expressed as
\ifdefined\TIT
\begin{align}  
   h&(X_{3G}+Z_3-W_3+\tilde{V}_{N_2}|U_{3G}) \nonumber \\
   &= h(X_{3G}+Z_3-W_3+\tilde{V}_{N_2}|U_{3G}) -h(X_{3G}+h_{13}^{-1}V_{N_2})+h(X_{3G}+h_{13}^{-1}V_{N_2}) \nonumber \\
   &= h(X_{3G}+Z_3-W_3+\tilde{V}_{N_2}|U_{3G}) -h(X_{3G}+Z_3-W_3+\tilde{V}_{N_2})+h(X_{3G}+h_{13}^{-1}V_{N_2}) \nonumber \\
   &= h(Y_{3G}+\tilde{V}_{N_2}|U_{3G}) -h(Y_{3G}+\tilde{V}_{N_2})+h(X_{3G}+h_{13}^{-1}V_{N_2}) \nonumber \\
   &= -I(U_{3G};Y_{3G}+\tilde{V}_{N_2}) +h(X_{3G}+h_{13}^{-1}V_{N_2}) \label{eq:3IC-12}
\end{align} 
\else 
\begin{align}  
   h&(X_{3G}+Z_3-W_3+\tilde{V}_{N_2}|U_{3G}) \nonumber \\
   &= h(X_{3G}+Z_3-W_3+\tilde{V}_{N_2}|U_{3G}) -h(X_{3G}+h_{13}^{-1}V_{N_2}) \nonumber \\
   &\ \ +h(X_{3G}+h_{13}^{-1}V_{N_2}) \nonumber 
\end{align} 
\begin{align}  
   &= h(X_{3G}+Z_3-W_3+\tilde{V}_{N_2}|U_{3G}) \nonumber \\
   &\ \ -h(X_{3G}+Z_3-W_3+\tilde{V}_{N_2})+h(X_{3G}+h_{13}^{-1}V_{N_2}) \nonumber \\
   &= h(Y_{3G}+\tilde{V}_{N_2}|U_{3G}) -h(Y_{3G}+\tilde{V}_{N_2})\nonumber \\
   &\ \ +h(X_{3G}+h_{13}^{-1}V_{N_2}) \nonumber \\
   &= -I(U_{3G};Y_{3G}+\tilde{V}_{N_2}) +h(X_{3G}+h_{13}^{-1}V_{N_2}) \label{eq:3IC-12}
\end{align} 
\fi
since $\sigma^2_{\tilde{V}_{N_2}}\ge 0$ due to the second condition in (\ref{eq:3ub-4a}). 

Plugging (\ref{eq:3IC-12}) into (\ref{eq:3IC-11c}), we get
\begin{align}  \label{eq:3IC-13}
   \frac{1}{n}\big(\alpha_{3A}-\alpha_{3B}\big) & \le h(X_{3G}+Z_3-W_3|U_{3G})\nonumber \\
   &\hspace{-10mm} -h(h_{13}X_{3G}+V_{N_2}) +I(U_{3G};Y_{3G}+\tilde{V}_{N_2}). 
\end{align} 
In the above inequality, $I(U_{3G};Y_{3G}+\tilde{V}_{N_2})$ is a penalty term due to the conditional worst additive noise lemma. 
Repeating the same procedure to $\alpha_{4A}-\alpha_{4B}$ through $\alpha_{9A}-\alpha_{9B}$, we arrive at (\ref{eq:3ub-3}) for $I_0$ and the  conditions in (\ref{eq:3ub-4a}), (\ref{eq:3ub-4b}), and (\ref{eq:3ub-4c}). 

As for $I_1$ and the other conditions in (\ref{eq:3ub-4d}), (\ref{eq:3ub-4e}), and (\ref{eq:3ub-4f}), it suffices to relate $\alpha_{1A}-\alpha_{2B}$ and $\alpha_{2A}-\alpha_{1B}$ instead of $\alpha_{1A}-\alpha_{1B}$ and $\alpha_{2A}-\alpha_{2 B}$, to replace the negative entropy terms $-nh(Z_2-h_{23}h_{13}^{-1}N_2)  -h(h_{13}X_3^n+V_{N_2}^n)$ in $I(X^n_2;Y^n_2,S_2^n| X_1^n)$ in (\ref{eq:3IC-8})  with $-nh(N_2-h_{13}h_{23}^{-1}Z_2)  -h(h_{23}X_3^n+V_{N_2}'^n)$ as already done in the proof of Theorem \ref{thm-3ub2}, and to follow the above same steps. This completes the proof.
\end{IEEEproof}

\begin{rem}
Constructing the relation shown in Fig. \ref{fig-0} among mutual information terms to handle the negative non-Gaussian entropies was the key step to derive the upper bound in Theorem \ref{thm-3ub3} by combining Etkin-type and change-of-interference approaches. The resulting upper bound is not restricted to the mixed interference regime any longer owing to the careful matching of $\alpha_{3A}-\alpha_{3B}$ in (\ref{eq:3IC-8}). {Notice that there are several other possible matchings of positive and negative non-Gaussian entropies due to the conditional worst case noise lemma, each of which clearly yields a valid upper bound. Nevertheless, such bounds turn out not so useful (i.e., not tighter than the existing bounds).}
\end{rem}

{
\begin{rem}
On one hand, notice that it is generally possible to find a tighter upper bound by letting all noise terms correlated to each other as in the existing bounds \cite{Ann09,Tun11, She12}. On the other hand, the noises $W_k$ and $N_k$ in our bounds are not restricted to have the same marginal distribution as $Z_k$. This interesting tradeoff will be addressed in the context of the symmetric GIC and numerical results in the following subsections. 
\end{rem}
}

Eventually, our new upper bound on the capacity of the three-user complex GIC is given by the minimum of four upper bounds given in Theorems \ref{thm-3ub1},  \ref{thm-3ub2}, \ref{thm-3ub3}, and the bound in (\ref{eq:3ub-1}). This will be the same case in the $K$-user case. All the upper bounds in this section are useful since they have their own ranges in terms of channel coefficients and SNR, where one of them is tighter than the others. 
Finally, it should be pointed out that the upper bounds in this work may be further tightened by using  techniques in the two-user case \cite[Thms. 4 and 5]{Nam15a}.

\subsection{Symmetric Case: Simplifying Upper Bounds} 
\label{sec:3IC-A}

In this subsection, we focus on the special case of the \emph{symmetric} GIC since it is sometimes useful to provide new insights into understanding the capacity.
The $K$-user symmetric complex  GIC with constant channel coefficients is given by
\begin{align} \label{eq:A-1b}
  Y_k=X_k+g\sum_{i\neq k} X_i+Z_k
\end{align}
where $P_k=P$ for all $k$.  The amplitude and phase of the symmetric cross-channel coefficient $g$ are denoted by $|g|$ and $\phi(g)$, respectively.

For the symmetric case, we evaluate sum-rate bounds of the {symmetric} three-user GIC at finite SNR to see how useful the derived upper bounds are to investigate the sum-rate behavior, compared to the existing bounds.  
We will use the \emph{normalized} symmetric rate for which the sum rate is {normalized} by the number of users and the number of dimension (real/complex) and hence its unit is bits/channel use/user/dimension throughout this work. 
A generalization of the Han-Kobayashi coding scheme \cite{Han81} has a prohibitively large complexity as the number of users increases, since each user should decode a different common message for every subset of non-intended receivers. Despite the overwhelming complexity, the resulting achievable rate seems to be outperformed by interference alignment and structured codes. 
Therefore, throughout this paper, we will use the simple lower bound that is given by the maximum symmetric rate of treating interference as noise, time division (with power control), and simultaneous non-unique decoding schemes.

In the following, we briefly review some existing upper bounds to compare the new upper bound with. First, the Kramer upper bound \cite[Thm. 1]{Kra04} and ETW upper bound \cite{Etk08} on the symmetric capacity of the symmetric two-user GIC can be used as a simple upper bound for the $K$-user case, as shown in \cite{Jaf10}. Deactivating all but any two users, the symmetric capacity for the general $K$-user GIC is upper-bounded by the the two-user case since removing interferers only increases the symmetric rates of the selected users 1 and 2. {Let $C_\text{sym}$ denote the symmetric sum capacity of the symmetric GIC.} The Kramer upper bound on the symmetric capacity of the two-user GIC can be written as \cite[Thm. 1]{Kra04}
\begin{align} \label{eq:Kra}
  C_\text{sym}\le
  \begin{cases}
    \frac{1}{2} \log (1+P)+\frac{1}{2} \log \big(1+\frac{P}{1+|g|^2P }\big), &   |g| < 1 \\
    \frac{1}{2} \log (1+P+|g|^2P ), &  |g| \ge 1 \\
  \end{cases}.
\end{align}
The ETW upper bound is given by \cite{Etk08} 
\begin{align} \label{eq:ETW}
  C_\text{sym}\le  \log \Big(1+|g|^2P +\frac{P}{1+|g|^2P }\Big).
\end{align}


The Kramer bound is extended in \cite{Tun11} to the three-user GIC by using the LMMSE estimation based proof. For the {symmetric} three-user case, the generalized Kramer upper bound can be simplified as follows:
\begin{align} \label{eq:3IC-28a}
  C_\text{sym}\le \;& \log\bigg(P+2|g|^2P+1\bigg)
              +\log\bigg(\frac{P+1}{(|g|^2P+1)({|g|^2P+1-\frac{||g|^2P+\rho|^2}{|g|^2P+1}})}\bigg) \nonumber \\
              & \ \ +\log\Bigg({P+|g|^2P+1-\frac{|g^*(g+1)P+\rho^*|^2}{2|g|^2P+1}}\Bigg) 
\end{align}
for $\rho$ satisfying $\left [ \begin{smallmatrix} g^* & g^*  \end{smallmatrix} \right ] \left [ \begin{smallmatrix} 1 & \rho \\ \rho^* & 1 \end{smallmatrix} \right ]^{-1} \left [ \begin{smallmatrix} g \\ g  \end{smallmatrix} \right ]< 1$, where $\rho$ is the correlation coefficient between $Z_1$ and $Z_2$. For $\rho$ satisfying $\left [ \begin{smallmatrix} g^* & g^*  \end{smallmatrix} \right ] \left [ \begin{smallmatrix} 1 & \rho \\ \rho^* & 1 \end{smallmatrix} \right ]^{-1} \left [ \begin{smallmatrix} g \\ g  \end{smallmatrix} \right ]\ge 1$, we have
\begin{align} \label{eq:3IC-28b}
  C_\text{sym}\le \;&  \log\bigg(\frac{P+2|g|^2P+1}{1-|\rho|^2}\bigg) +\log\bigg(P+|g|^2P+1-\frac{|g^*(g+1)P+\rho^*|^2}{2|g|^2P+1}\bigg) .
\end{align}
{In fact, the above two bounds coincide with each other for the symmetric GIC. Thus we will only consider (\ref{eq:3IC-28b}) in the following. 
Meanwhile, a generalized ETW upper bound for the three-user symmetric GIC is proposed in \cite[Sec. VII-C]{Ann09} using the vector genie $\underline{S}_{iG}$ defined in Table II in \cite{Ann09} and allowing correlation over all noise variables, as follows: 
\begin{align} \label{eq:3IC-30}
    C_\text{sym} \le \sum_{i=1}^3 I(X_{iG}; Y_{iG}, \underline{S}_{iG})
\end{align}
for $(\Sigmam, \mu_1,\mu_2)$ satisfying $\mathrm{Cov}([Z_1 \; g\mu_1W_{11}]^T|W_{12}) -\mathrm{Cov}([g\mu_1W_{11}\; g\mu_2W_{12}]^T) \succeq 0$, 
where $\Sigmam$ is the  covariance matrix of the noise random vector $[Z_1 \;W_{11}  \;W_{12}]^T$, 
$W_{1j}, j=1,2,$ are the additive noises in $\underline{S}_{iG}$, and $\mu_j$ denotes the variance of $W_{1j}$.}

{
For the symmetric case, we can naturally simplify our bounds to avoid the numerical optimization over all possible ranges of the parameters. In this paper, we only simplify the bound in Theorem \ref{thm-3ub3}, which reduces to               
\begin{align} \label{eq:3IC-32}
   C_\text{sym} \le\;& \log \left(\frac{P+2|g|^2P+1}{\sigma_{Z-N}^2}\right)  +\log \left(\frac{2|g|^2P+\sigma_{N}^2}{2|g|^2P+1}\right) +\log \left(\frac{2|g|^2P+\sigma_{W}^2}{2|g|^2P+\sigma_{V_{W}}^2}\right) \nonumber \\ 
   &+\log \left(\frac{P+|g|^2P+1-\frac{|g^*(g+1)P+\rho_N^*\sigma_{N}|^2}{2|g|^2P+\sigma_{N}^2}}{\sigma_{Z-W}^2}\right)  +\log \left(\frac{P+\sigma_{Z-W}^2 -\frac{|\rho_{W}\sigma_{W}-\sigma_{W}^2|^2}{2|g|^2P+\sigma_{W}^2}}{|g|^2P+\sigma_{V_{N}}^2-\frac{|g\rho_{W}\sigma_{W}-g\sigma_{W}^2|^2}{2|g|^2P+\sigma_{W}^2} }\right)  
\end{align}
where the Gaussian noises $Z, W, N$ are irrespective of user index $k$ by symmetry. 

To derive the simplified bound as a special case of (\ref{eq:3IC-32}), we will follow the technique in \cite{Nam15a}, which exploits the fact that the equalities in the worst additive noise lemma in \cite{Dig01} and its conditional version in \cite{Nam12} trivially hold true when $\sigma_Z^2=0$ as well as when $X^n=X_g^n$ (and $U^n=U_g^n$ for the conditional version). Therefore, those lemmas incur no loss in tightness of the resulting bounds if one can let $\sigma_Z^2=0$. Then we have the following result. 

\begin{thm} \label{thm-3ub3b}
The sum capacity of the three-user symmetric GIC is upper-bounded as 
\begin{align} \label{eq:3IC-33}
   C_\text{sym} \le\;&  \min (\textsf{R}_0, \textsf{R}_1) 
\end{align}
where 
\begin{align} 
   \textsf{R}_0 =\;& \min_{\sigma_{N}}\; \textsf{R}  \label{eq:3IC-34} 
\end{align} \vspace{-5mm}
\begin{subequations}
\begin{align} 
   \text{subject to}\ & 0\le(|\rho_W|,|\rho_N|,\sigma_{W},\sigma_{N}) \le 1 \\
   &\sigma_{W}=1 \label{eq:3IC-34d} \\
   &\rho_W=2\sigma_{N}^2-1 \label{eq:3IC-34b} \\
   &\rho_N=4|g|^2\Big((\sigma_{N}^{-1}-\sigma_{N}) -\sqrt{(1-(2|g|)^{-2})\sigma_{N}^{-2} +(1+(2|g|)^{-2})\sigma_{N}^2-2+(2|g|)^{-4}}\Big) \label{eq:3IC-34c}
\end{align}
\end{subequations}
and
\begin{align} 
   \textsf{R}_1 =\;& \min_{\rho_N}\; \textsf{R}  \label{eq:3IC-36} 
\end{align} \vspace{-5mm}
\begin{subequations}
\begin{align} 
   \text{subject to}\ &0\le(|\rho_W|,|\rho_N|,\sigma_{W},\sigma_{N}) \le 1 \\
   &\sigma_{N}=1 \label{eq:3IC-36d} \\
   &\sigma_{W}= |\rho_W|^2 \label{eq:3IC-36c} \\
   &\rho_W=\sqrt{1-\frac{1+\rho_N}{2|g|^2}} \label{eq:3IC-36b} 
\end{align}
\end{subequations}
where 
\begin{align} \label{eq:3IC-38}
  \textsf{R} =\log \left(\frac{P+2|g|^2P+1}{|g|^2\sigma_{Z-N}^2\sigma_{Z-W}^2}\right) +\log \left(P+|g|^2P+1-\frac{|g^*(g+1)P+\rho_N^*\sigma_{N}|^2}{2|g|^2P+\sigma_{N}^2}\right). 
\end{align}
\end{thm}

\begin{IEEEproof}
We first consider (\ref{eq:3IC-32}) under the conditions in (\ref{eq:3ub-4}). For the symmetric GIC, there are four parameters $(\rho_W, \rho_N,\sigma_{W},\sigma_{N})$ and  three inequalities including the two conditions in (\ref{eq:3ub-4a}) and the implicit condition (i.e., $\sigma_{W}\le1$) in (\ref{eq:3IC-11a}) due to the worst additive noise lemma and its conditional version. Based on the argument in \cite[Thm. 6]{Nam15a}, we need to consider the equalities to optimize the parameters.  
Given $\sigma_{N}$, we do the following steps:  
\begin{enumerate}
\item[A1)] Let $\sigma_{W}^{2}=1$ to satisfy $\alpha_{1A}-\alpha_{1B}=0$ in (\ref{eq:3IC-11a}), which is simply (\ref{eq:3IC-34d}).
\item[A2)] Let $\sigma_{V_{W}}^{2}=\sigma_{N}^2$ to satisfy $\alpha_{2A}-\alpha_{2B}=0$ in (\ref{eq:3IC-11b}), which yields (\ref{eq:3IC-34b}) given (\ref{eq:3IC-34d}).
\item[A3)] Let $\sigma_{V_{N}}^{2}=|g|^2\sigma_{V_N}^2$ (i.e., $\sigma^2_{\tilde{V}_{N}}= 0$) to satisfy $\alpha_{3A}-\alpha_{3B}=-n\log |g|^2$ in (\ref{eq:3IC-11c}),  which yields (\ref{eq:3IC-34c}) given (\ref{eq:3IC-34d}) and (\ref{eq:3IC-34b}).
\end{enumerate}
Substituting (\ref{eq:3IC-34d}) -- (\ref{eq:3IC-34c}) into (\ref{eq:3IC-32}), we have $\textsf{R}_0$.

Similarly, we can do the same things for the conditions in (\ref{eq:3ub-4z}) as follows: 
\begin{enumerate}
\item[B1)] Given $\rho_N$, let $\sigma_{N}^{2}=1$ to satisfy $\alpha_{1A}-\alpha_{2B}=0$, which is simply (\ref{eq:3IC-36d}).
\item[B2)] Let $\sigma_{V_{W}}^{2}=\sigma_{W}^2$ to satisfy $\alpha_{2A}-\alpha_{1B}=0$, which yields (\ref{eq:3IC-36c}) given (\ref{eq:3IC-36d}).
\item[B3)] Let $\sigma_{V_{N}}^{2}=|g|^2\sigma_{V_N}^2$ (i.e., $\sigma^2_{\tilde{V}'_{N}}= 0$) to satisfy $h(Y_{3G}|U_{3G}) -h(gY_{3G}+\tilde{V}'_{N}|U_{3G})=-\log |g|^2$ (for more details, see the $I_1$ part in the proof of Theorem \ref{thm-3ub3}),  which yields (\ref{eq:3IC-36b}) given (\ref{eq:3IC-36d}) and (\ref{eq:3IC-36c}).
\end{enumerate}
Substituting (\ref{eq:3IC-36d}) -- (\ref{eq:3IC-36b}) into (\ref{eq:3IC-32}), we have $\textsf{R}_1$.
\end{IEEEproof}

We now show that $\textsf{R}_1$ reduces to the generalized Kramer bound in (\ref{eq:3IC-28b}). It immediately follows from (\ref{eq:3IC-36d}) -- (\ref{eq:3IC-36b}) that ${|g|^2\sigma_{Z-N}^2\sigma_{Z-W}^2}=1-|\rho_N|^2$ in (\ref{eq:3IC-38}), thus yielding that the objective functions in (\ref{eq:3IC-36}) and (\ref{eq:3IC-28b}) are the same by just letting $\rho_N=\rho$. It is easy to check that the constraints of $\textsf{R}_1$ and the generalized Kramer bound can be rewritten as  $\rho\le 2|g|^2-1$. Therefore, the two optimization problems are equivalent. This is not necessarily the case with $\textsf{R}_1$. 
The main difference between $\textsf{R}_0$ and $\textsf{R}_1$ is that the noise $N_k$ in $\textsf{R}_0$ is not restricted to have the same marginal distribution as $Z_k$, i.e., (\ref{eq:3IC-36d}).}

\subsection{Numerical Results}

{In what follows, we focus on capacity bounds for the \emph{symmetric (positive) real} GIC, which has been widely used due to the simplicity of the resulting channel model (e.g., \cite{Jaf10,Ord14}). The behavior of sum-rate bounds for more general complex GICs will be deferred to Sec. \ref{sec:CGIC}.
}

\begin{figure}
\vspace{-5mm}
\hspace{-2mm}
\ifdefined\TIT
  \center \includegraphics[scale=.72]{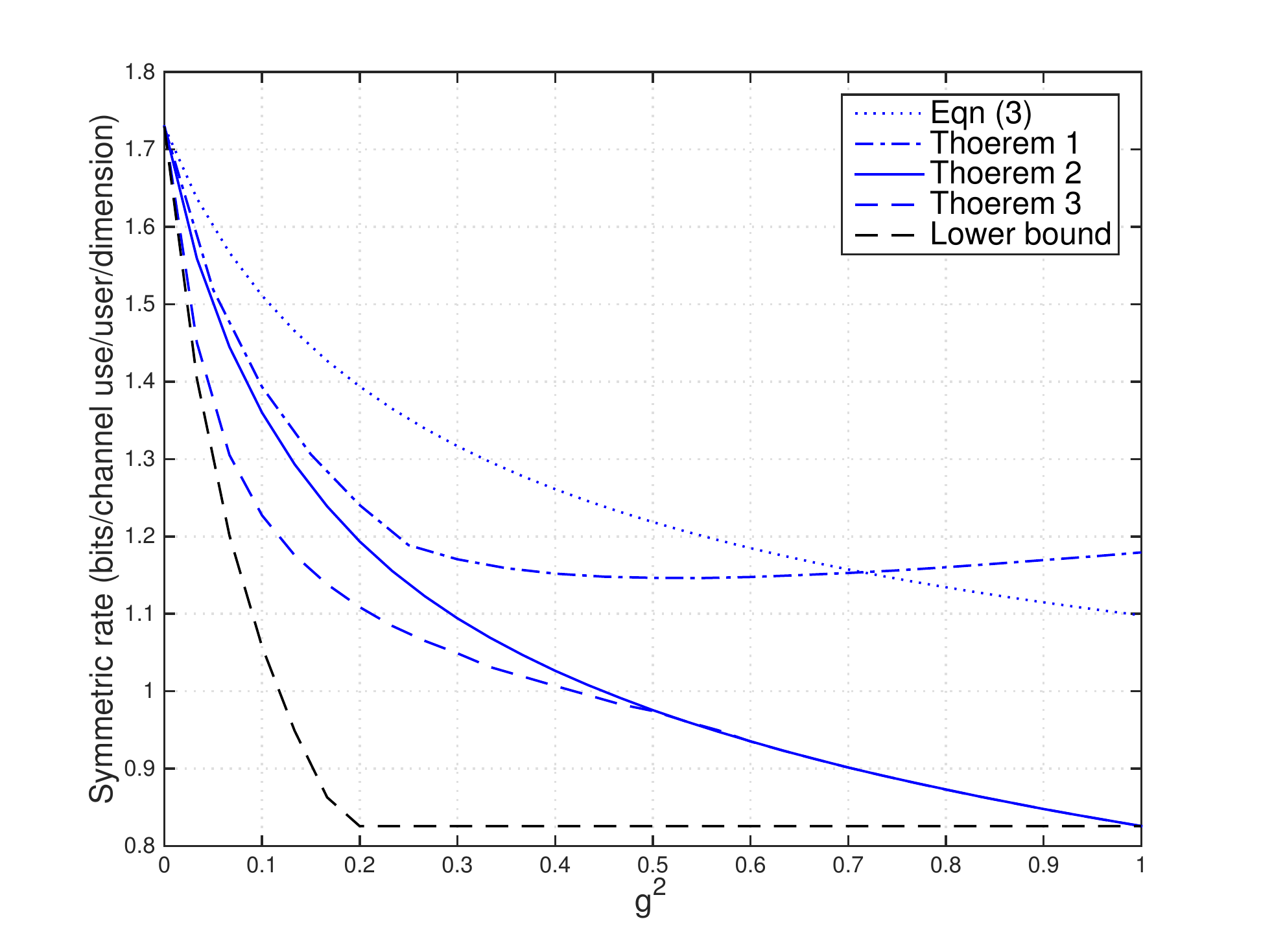}
\else
  \center \includegraphics[scale=.55]{./Fig/3K_fig1_3}
\vspace{-6mm}
\fi
  \caption{Bounds on the sum capacity of three-user symmetric real GIC over $0\le g^2\le 1$ when $P=10$.}\label{fig-1}
\end{figure}

\begin{figure}
\vspace{-5mm}
\hspace{-2mm}
\ifdefined\TIT
  \center \includegraphics[scale=.72]{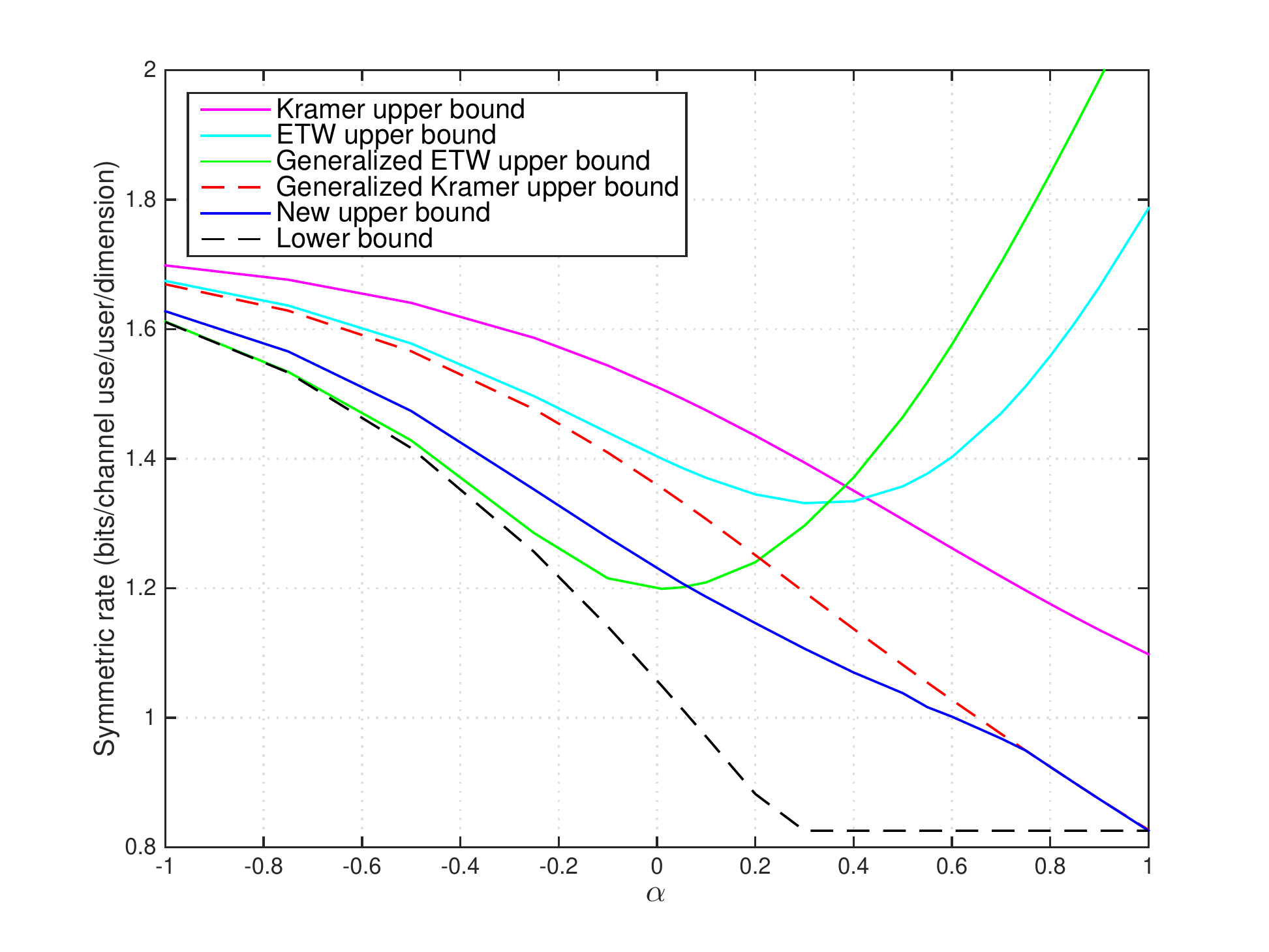}
\else
  \center \includegraphics[scale=.55]{./Fig/K3_alpha_P10_final}
\vspace{-6mm}
\fi
  \caption{Bounds on the sum capacity of three-user symmetric real GIC over $-1\le \alpha\le 1$ for $P=10$. {The new upper bound is given by the minimum of bounds in Theorems 1, 2, 3.}
  }\label{fig-4}
\end{figure}

Fig. \ref{fig-1} depicts the symmetric rates of the four upper bounds derived in the previous section and the lower bound for the three-user symmetric {\emph{real}} GIC. For this symmetric case, the upper bounds in Theorem \ref{thm-3ub1} and (\ref{eq:3ub-1}) reduce to \cite[Thm. 3]{Nam15a} and \cite[Thm. 1]{Kra04}, respectively. {Notice that any of those bounds does not contain the other bounds when complex channel coefficients are taken into account, as will be shown later in Sec. \ref{sec:CGIC}.}


Fig. \ref{fig-4} compares our upper bound with the above two-user upper bounds, the generalized Kramer upper bound, the generalized ETW upper bound, and the simple lower bounds in the previous subsection for the three-user symmetric real GIC over $-1\le \alpha\le 1$ at SNR $=10$ dB, where $\alpha=\frac{\log \inr}{\log \snr}=\frac{\log |g|^2P}{\log P}$.
The new upper bound is shown to be tightest in the medium range of $\alpha$ by nature of the change-of-interference approach in Theorem \ref{thm-3ub3} {because the change-of-interference variables in (\ref{eq:3IC-2}) are more general than $X_1, X_2, X_3$ (i.e., noiseless interference as side information), as mentioned in Remark \ref{rem-1}.
This is consistent with \cite{Nam15a} for the two-user case. Namely, the change-of-interference approach improves upon the Kramer-type bounds in \cite{Kra04,Tun11} at a certain range of channel coefficients.} 

{As mentioned in subsection \ref{sec:3IC-A}, we can see that the proposed bound coincides with the generalized Kramer bound in a certain range of $\alpha$. This is quite interesting as the former does not impose correlation over all noise variables.} The generalized ETW bound is tightest when $\alpha\le 0.07$, i.e., interference is very weak. 

\begin{figure}
\vspace{-5mm}
\hspace{-2mm}
\ifdefined\TIT
  \center \includegraphics[scale=.72]{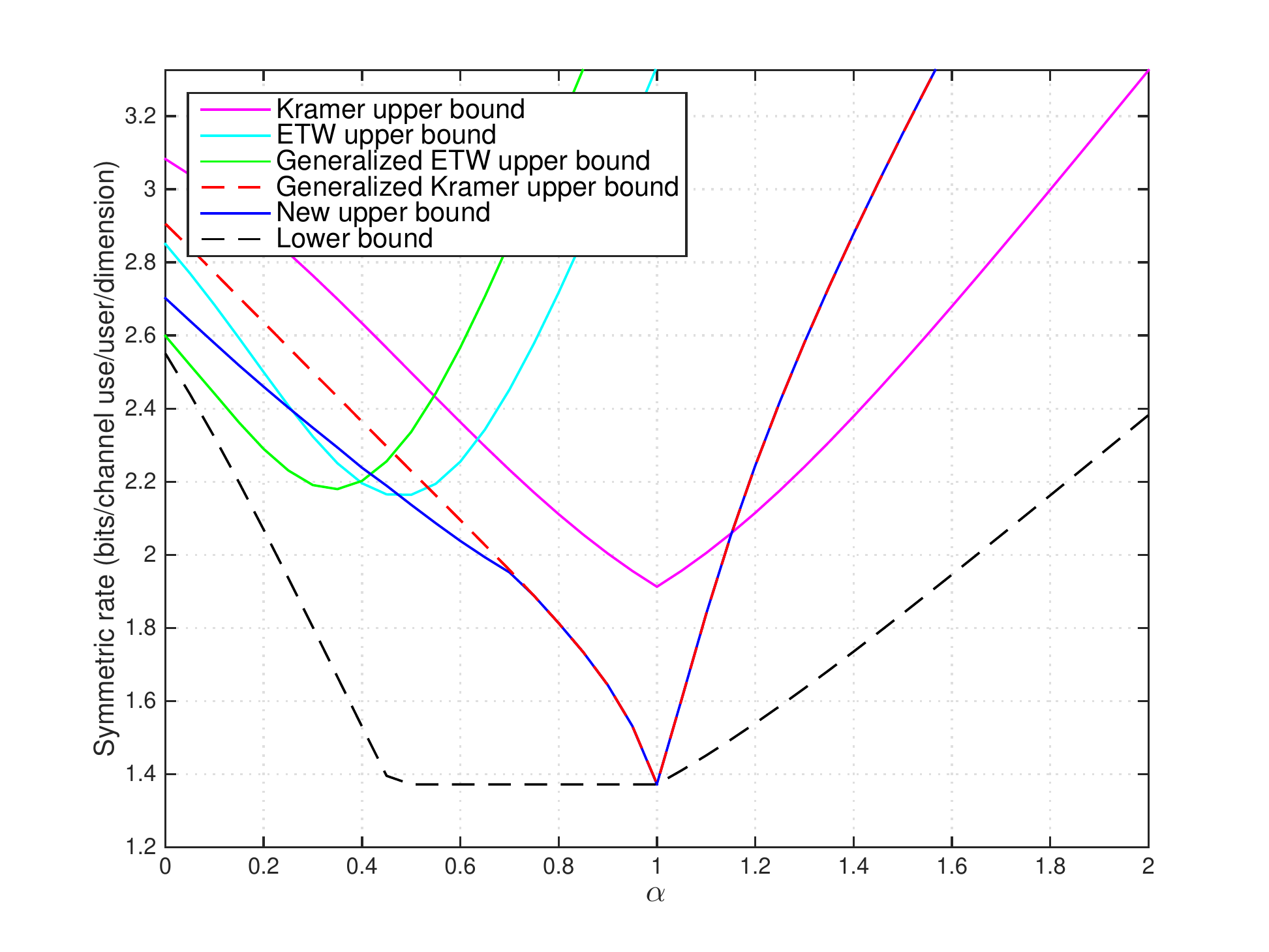}
\else
  \center \includegraphics[scale=.55]{./Fig/K3_alpha_P100_final}
\vspace{-6mm}
\fi
  \caption{Bounds on the sum capacity of three-user real symmetric GIC over $0\le \alpha \le 2$ for $P=100$ (i.e., SNR  $= 20$ dB). 
  }\label{fig-4a}
\end{figure}

Fig. \ref{fig-4a} depicts the same upper and lower bounds over $0\le \alpha\le 2$ at high SNR of $20$ dB. Our upper bound is still tightest in a certain range of $\alpha$ and coincides with the generalized Kramer bound for $\alpha> 0.71$. For this high SNR, the two-user upper bounds are tightest in two different ranges of $\alpha$, respectively.  {A sophisticated lower bound based on lattice interference alignment using the compute-and-forward approach was proposed in \cite{Ord14}. In particular, the readers are encouraged to refer to Fig. 7(a) therein for a tighter lower bound than the simultaneous decoding lower bound plotted in Fig. \ref{fig-4a} for $\alpha>1$. Notice that $g=P^\frac{\alpha-1}{2}$.} Furthermore, the multiple access upper bound in \cite{Cad09} is valid only when $|g|\ge 1$ (i.e., $\alpha\ge 1$) and is shown in \cite[Fig. 3]{Tun11} to coincide with the generalized Kramer bound in some range of $g$ and to be quite loose in the remaining range. One may consider a trivial upper bound obtained by a multiple-access channel (MAC) formed by allowing the receivers to cooperate, which is also shown in \cite[Fig. 3]{Tun11} to be in general loose relative to the other bounds.

From Figs. \ref{fig-4} and \ref{fig-4a}, we observe that the gap between the upper bound and the simple lower bound over a certain range around $g^2=1$ is not significant. 
We will see in Sec. \ref{sec:KIC} that the rate gap around $g^2=1$ is still not significant at least for $K=4$ symmetric real GICs. {Moreover, the range of $\alpha$ over which the new bound is tightest tends to shrink as $P$ increases. Comparing  $\textsf{R}_0$ and $\textsf{R}_1$ (equivalently, the generalized Kramer bound) in Theorem \ref{thm-3ub3b}, we can see that the rate difference of the two bounds vanishes as $P\rightarrow\infty$ irrespectively of $\sigma_N^2$.}



\section{$K$-User Gaussian Interference Channel}
\label{sec:KIC}

In this section, we generalize the upper bounds on the sum capacity of the three-user GIC to the $K$-user case. We first consider the $K$-user symmetric GIC and then the asymmetric case.
A straightforward generalization of Theorem \ref{thm-3ub1} and (\ref{eq:3ub-1}) is skipped due to the space limitation.

\subsection{Symmetric Case}
\label{sec:KIC-A}

In this subsection, we are interested in the symmetric case in (\ref{eq:A-1b}), for which the three-user upper bounds in Theorems \ref{thm-3ub2} and \ref{thm-3ub3} can be naturally extended to the $K$-user case. 
For the $K$-user symmetric case, we rewrite the {genie} random variables in (\ref{eq:3IC-2}) and (\ref{eq:3IC-1}) as
\begin{align} 
  U_{k} &= g\sum_{i\neq k} X_i+W_{k} \nonumber \\
  S_{k } &= g\sum_{i\neq k-1} X_i+N_{k}. \nonumber 
\end{align}
With these definitions, we have the following generalization of Theorem \ref{thm-3ub2} for the symmetric case.


\begin{thm} \label{thm-kub2}
The sum capacity of the $K$-user symmetric complex GIC in the weak interference regime, where $|g|^2\le 1$, is upper-bounded by
\begin{align}  \label{eq:kub-2}
    \sum_{k=1}^K R_k \le &\;  I\big(X_{1G}; Y_{1G}\big) + \sum_{k=2}^{K-1} I\big(X_{k G}; Y_{k G},S_{k G}\big| X_{1G}^{k-1}\big)  +I\big(X_{K G}; Y_{K G}\big| X_{1G}^{K-1}\big)
\end{align}
for all $(N_1,N_2\ldots,N_K)$ satisfying 
\begin{align}  
  \sigma^2_{V_{N_k}} &\ge \sigma^2_{N_{k+1}} \ \text{ for } k=2,\ldots,K-2 \label{eq:Kub-2a} \\
  \sigma^2_{V_{N_{K-1}}} &\ge |g|^2  \label{eq:Kub-2b} 
\end{align}
where 
\begin{align} \label{eq:3IC-14c}
  V_{N_k}&= N_k|\; Z_k-N_k .
\end{align}
\end{thm}

\begin{IEEEproof}
See Appendix \ref{proof-4}.
\end{IEEEproof}


In order to obtain the $K$-user extension of Theorem \ref{thm-3ub3}, we need to generalize the relation of mutual informations for the three-user case in Fig. \ref{fig-0} to the $K$-user GIC. The extension to the four-user case will be given in Fig. \ref{fig-0b} in Appendix \ref{proof-5}. The relation for more than four-user cases can be obtained in the same manner.
We have then the following result.

\begin{thm} \label{thm-kub3}
The sum capacity of the $K$-user symmetric complex GIC is upper-bounded by 
\begin{align}  \label{eq:kub-3}
    \sum_{k=1}^K R_k \le \; & I\big(X_{1G}; Y_{1G}\big) +\sum_{k=2}^{K-1} I\big(X_{k G}; Y_{k G}, S_{k G} \big| X_{1G}^{k-1}\big) \nonumber \\ &\ +I\big(X_{K G}; Y_{K G}\big| U_{K G}\big) +I\big(U_{K G};Y_{K G}+\tilde{V}_{N_{K}}\big)
\end{align}
for all $(N_1,\ldots,N_K,W_1,\ldots,W_K)$ satisfying 
\begin{align}  
  &\sigma^2_{V_{W_{1}}}\ge \sigma^2_{N_{2}} \label{eq:Kub-4a} \\
  &\sigma^2_{V_{N_{k}}}\ge \sigma^2_{N_{k+1}}, \ k=2,\ldots,K-2 \label{eq:Kub-4b} \\
  &\sigma^2_{V_{N_{K-1}}}\ge |g|^2\sigma^2_{Z_{K}-W_{K}} \label{eq:Kub-4c}
\end{align}
where $$\tilde{V}_{N_{K-1}} = \sqrt{{|g|^{-2}-\sigma_{V_{N_{K-1}}}^{-2}\sigma_{Z_{K}- W_{K}}^2}}  V_{N_{K-1}}.$$
\end{thm}

\begin{IEEEproof}
See Appendix \ref{proof-5}.
\end{IEEEproof}

\begin{rem}
{The ``useful genie" upper bound in \cite{Ann09} for the symmetric three-user GIC defines a vector genie and imposes correlation over all additive noise variables. However, this approach makes it hard to compute the resulting outer bounds even in the symmetric four-user case. Specifically, not only we have $\frac{K(K-1)}{2}$ complex correlation coefficients to optimize for the $K$-user symmetric case, but  also matrix inversions and positive semi-definiteness tests of $(K-1)$-dimensional complex matrices. The dimension of the noise covariance matrix increases exponentially with $K$ for the asymmetric case. Similar difficulty arises in generalizing the Kramer bound in \cite[Thm. 1]{Kra04} to the more than two-user case due to the optimization of a positive-definite covariance matrix of jointly Gaussian noises by imposing correlation over all noise variables (e.g., see \cite{Tun11, She12} for the three-user case).  In contrast, our noise variables $N_i$ defined in (\ref{eq:3IC-1}) are correlated only with $Z_i$ having the same user index, respectively. This is the same case with the other noise variables $W_i$ in Theorem \ref{thm-3ub1}. The parameters to be optimized are $(K-1)$ complex correlation coefficients and $(K-1)$ variances. Therefore, we intentionally avoided the use of the above optimization of the noise covariance matrix as well as vector genie in this work.} 
\end{rem}


\subsection{Closed-Form Upper Bounds for the $K$-User Symmetric GIC}
\label{sec:KIC-E}

This subsection is devoted to find appropriate closed-form expressions of our upper bounds that do not involve the optimization of the covariance matrix of additive noise variables, which makes it hard to compute upper bounds on the capacity of the $K$-user GIC unless $K$ is quite small, as mentioned earlier. By exploiting an intrinsic structure of our upper bounds, we will simplify upper bound formulas so as to predict sum-rate behavior of the $K$-user symmetric GIC in the large-number-of-user regime. Notice that the closed-form upper bounds in this subsection are given for the symmetric real GIC but they are straightforwardly extendable to the symmetric complex case. We begin with the following proposition based on Theorem \ref{thm-kub2}.

\begin{prop} \label{prop-1}
For the cross-channel coefficient $|g|< 1$, a closed-form upper bound on the capacity of the $K$-user symmetric GIC is given by
\begin{align} \label{eq:KIC-18}
   C_\text{sym} \le &\;\log \left(1+\frac{P}{(K-1)|g|^2P+1}\right)  +\log \Big(1+{(K-1)P}\Big) \nonumber \\ 
   &+\sum_{k=2}^{K-1}\log \left(1+\frac{|1-g|^2P}{1-|g|^2}\cdot\frac{(k-1)P+1}{kP+1 }\right).
\end{align}
\end{prop}

\begin{IEEEproof}
We can rewrite (\ref{eq:kub-2}) as
\begin{align} \label{eq:KIC-19}
   C_\text{sym} \le  \min_{\Kc_\text{sym}^{(1)}} &\;\vast\{\log \left(1+\frac{P}{(K-1)|g|^2P+1}\right)  +\log \left((K-1)|g|^2P+\sigma_{N_2}^2\right) \nonumber \\ 
   &+\sum_{k=2}^{K-2}\log \left(\frac{k|g|^2P+\sigma_{N_{k+1}}^2}{k|g|^2P+\sigma_{V_{N_{k}}}^2 }\right) +\log \left(\frac{|g|^2P+|g|^2}{|g|^2P+\sigma_{V_{N_{K-1}}}^2 }\right)  -\log |g|^2 \nonumber \\
   &+\sum_{k=2}^{K-1}\log \left(\frac{|1-g|^2P+\sigma^2_{Z_k-N_k}-\frac{|(1-g)g^*P+\rho_{N_k}\sigma_{N_k}-\sigma_{N_k}^2|^2}{(K-k+1)|g|^2P+\sigma_{N_k}^2}}{\sigma^2_{Z_k-N_k} }\right)  \vast\}
\end{align}
where $\Kc_\text{sym}^{(1)}$  is the set of all $(N_2,\ldots,N_{K-1})$ satisfying (\ref{eq:Kub-2a}) and (\ref{eq:Kub-2b}).

Letting $\rho_{N_k}^2=\sigma_{N_k}^2=|g|^2$ for any $k$, we get $\sigma_{V_{N_k}}^2=|g|^{2}$, which naturally satisfies (\ref{eq:Kub-2a}) and (\ref{eq:Kub-2b}) with equality.
Then, the third and fourth terms inside the brace in (\ref{eq:KIC-19}) are canceled out. Substituting these values into (\ref{eq:KIC-19}), we get (\ref{eq:KIC-18}).
\end{IEEEproof}

The next result is a closed-form expression of Theorem \ref{thm-kub3}.

\begin{prop} \label{prop-2}
A closed-form upper bound on the capacity of the $K$-user symmetric GIC is given by
\begin{align} \label{eq:KIC-20}
   C_\text{sym} \le &\;\log \left(1+\frac{P}{(K-1)|g|^2P+1}\right)  +\log \left(1+(K-1)(1+|g|^2)P\right) \nonumber \\ 
   &+\sum_{k=2}^{K-1}\log \left(1+|1-g|^2(1+|g|^2)P\cdot\frac{(k-1)P+\frac{1}{1+|g|^2}}{kP+\frac{1}{1+|g|^2}} \right) .
\end{align}
\end{prop}

\begin{figure}
\vspace{-5mm}
  \center \includegraphics[scale=.72]{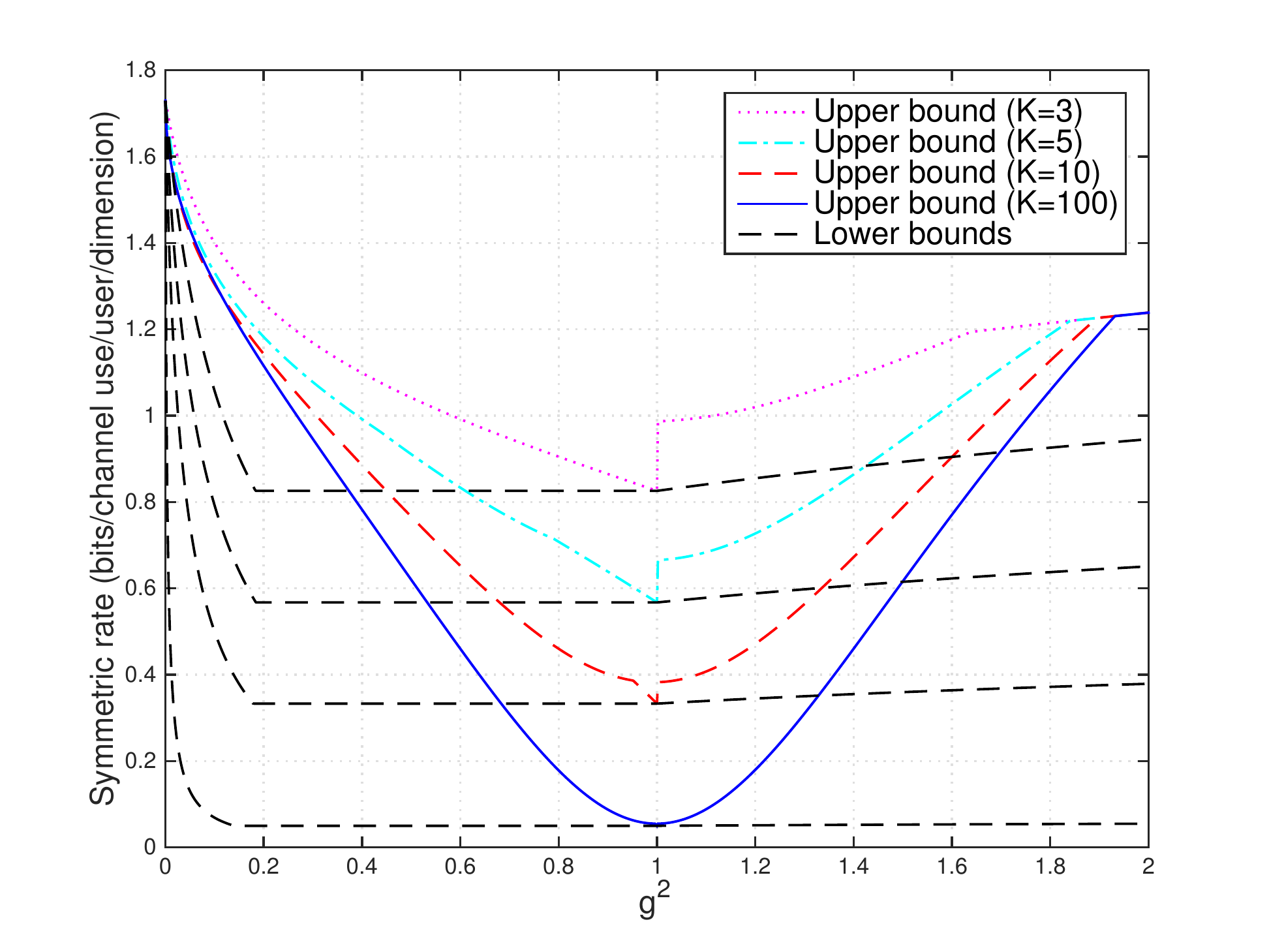}
  \caption{Closed-form upper bounds on the symmetric rate of the $K$-user symmetric GICs at SNR $=10$ dB, where $K=3,5,10,100.$}\label{fig-11}
\end{figure}

\begin{IEEEproof}
Similar to (\ref{eq:KIC-19}), we can rewrite (\ref{eq:kub-3}) as
\begin{align} \label{eq:KIC-15}
   C_\text{sym} \le  \min_{\Kc_\text{sym}^{(2)}} &\;\vast\{\log \left(1+\frac{P}{(K-1)|g|^2P+1}\right)  +\log \left(\frac{(K-1)|g|^2P+\sigma_{N_2}^2}{\sigma^2_{Z_1-W_1} }\right) \nonumber \\ 
   &+\log \left(\frac{(K-1)|g|^2P+\sigma_{W_1}^2}{(K-1)|g|^2P+\sigma_{V_{W_1}}^2 }\right) +\sum_{k=2}^{K-2}\log \left(\frac{k|gㅂP+\sigma_{N_{k+1}}^2}{k|g|^2P+\sigma_{V_{N_{k}}}^2 }\right) 
\nonumber \\ 
   &+\log \left(\frac{P+\sigma^2_{Z_K-W_K}-\frac{(\rho_{W_K}\sigma_{W_K}-\sigma_{W_K}^2)^2}{(K-1)|g|^2P+\sigma_{W_K}^2}}{P+|g|^{-2}\sigma_{V_{N_{K-1}}}^2-\frac{(\rho_{W_K}\sigma_{W_K}-\sigma_{W_K}^2)^2}{(K-1)|g|^2P+\sigma_{W_K}^2} }\right) -\log |g|^2   \nonumber \\
   &+\sum_{k=2}^{K-1}\log \left(\frac{|1-g|^2P+\sigma^2_{Z_k-N_k}-\frac{|(1-g)g^*P+\rho_{N_k}\sigma_{N_k}-\sigma_{N_k}^2|^2}{(K-k+1)|g|^2P+\sigma_{N_k}^2}}{\sigma^2_{Z_k-N_k} }\right)   \vast\}
\end{align}
where $\Kc_\text{sym}^{(2)}$ is the set of all $(W_1,W_K,N_2,\ldots,N_{K-1})$ satisfying (\ref{eq:Kub-4a}), (\ref{eq:Kub-4b}), and  (\ref{eq:Kub-4c}) .

Letting $\rho_{W_k}^2=\sigma_{W_k}^2=\rho_{N_k}^2=\sigma_{N_k}^2={\frac{|g|^2}{1+|g|^2}}\le 1$ for any $k$ by exploiting the symmetry of the noise parameters of the upper bound in Theorem \ref{thm-kub3}, we can get 
\begin{align} \label{eq:KIC-21}
   &\sigma_{V_{W_k}}^2=\sigma_{V_{N_k}}^2=\frac{|g|^2}{1+|g|^2}\nonumber \\
   &\sigma^2_{Z_k-W_k}=\sigma^2_{Z_k-N_k}=\frac{1}{1+|g|^2} 
\end{align}
which satisfy all conditions in (\ref{eq:Kub-4a}), (\ref{eq:Kub-4b}), and  (\ref{eq:Kub-4c}) with equality. Then the third, fourth, and fifth terms inside the brace in (\ref{eq:KIC-15}) are removed. After straightforward manipulation, we have (\ref{eq:KIC-20}).
\end{IEEEproof}

Fig. \ref{fig-11} depicts the closed-from upper bounds given by the minimum of  (\ref{eq:KIC-18}),  (\ref{eq:KIC-20}), and (\ref{eq:Kra}) for different numbers of users at SNR $=20$ dB. Notice that these closed-form upper bounds are rather loose relative to Theorems \ref{thm-kub2} and \ref{thm-kub3} {since the closed forms are special cases of the latter bounds without optimizing the parameters}, e.g., see Fig. \ref{fig-4a} for the three-user case. The closed-form upper bound in (\ref{eq:KIC-20}) can be further tightened for large $K$ in the following way. 

\begin{prop} \label{prop-3}
For the cross-channel coefficient $|g|>1$ satisfying $|g|^{2\gamma}-|g|^2-1\ge 0$, where $\gamma>1$ is a positive real, a closed-form upper bound is given by
\begin{align} \label{eq:KIC-16}
   C_\text{sym} \le &\;\log \left(1+\frac{P}{(K-1)|g|^2P+1}\right)  +\log \left(1+(K-1)|g|^{2\gamma}P\right) \nonumber \\ 
   &+\sum_{k=2}^{K-1}\log \left(1+\frac{|1-g|^2P}{1-|g|^{-2(\gamma-1)}}\cdot\frac{(k-1)P+|g|^{-2\gamma}}{kP+|g|^{-2\gamma} }\right) .
\end{align}
\end{prop}

\begin{IEEEproof}
It suffices to let $\rho_{W_k}^2=\sigma_{W_k}^2=1-|g|^{-2\gamma}$ and $\rho_{N_k}^2=\sigma_{N_k}^2=|g|^{-2(\gamma-1)}$ for any $k$. Then, we get 
\begin{align} \label{eq:KIC-17}
   \sigma^2_{Z_k-W_k}&=\sigma_{V_{N_k}}^2=|g|^{-2\gamma} \nonumber \\
   \sigma^2_{Z_k-N_k}&=1-|g|^{-2(\gamma-1)} \nonumber \\
   \sigma_{V_{W_k}}^2&=1-|g|^{-2\gamma} 
\end{align}
which transforms (\ref{eq:Kub-4a}) into 
\begin{align} 
   |g|^{2\gamma}-|g|^2-1\ge 0 \nonumber
\end{align}
and satisfies (\ref{eq:Kub-4b}) and (\ref{eq:Kub-4c}) with equality. Notice that the above inequality holds true when $|g|>1$. 
Plugging (\ref{eq:KIC-17}) into (\ref{eq:KIC-15}), we obtain (\ref{eq:KIC-16}).
\end{IEEEproof}

\begin{figure}
\vspace{-5mm}
  \center \includegraphics[scale=.72]{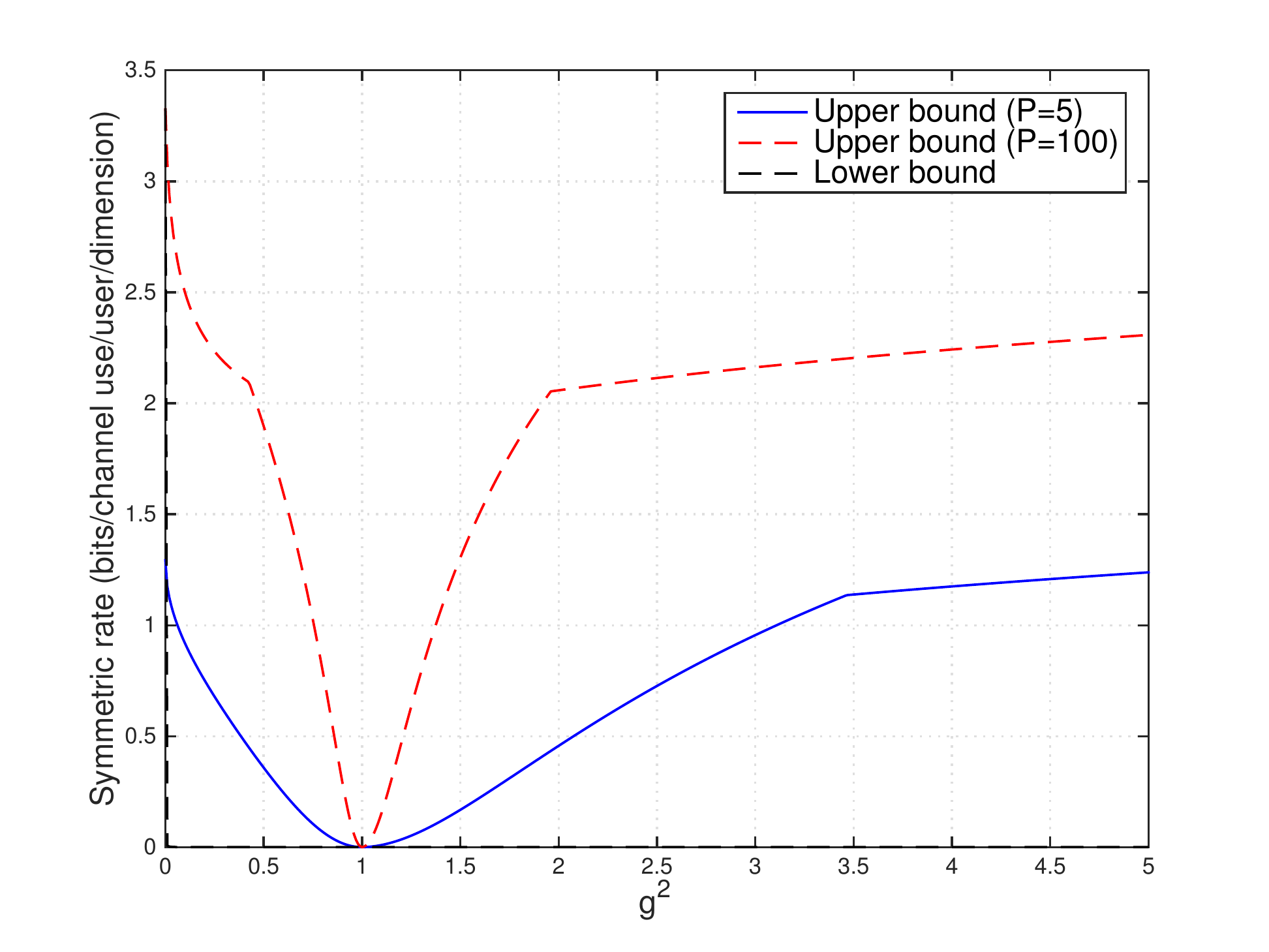}
  \caption{Closed-form upper bounds on the symmetric rate of  the $K$-user symmetric GIC at medium and high SNRs, where $K=10^5$.}\label{fig-12}
\end{figure}

Fig. \ref{fig-12} shows the upper bounds given by the minimum of (\ref{eq:KIC-18}), (\ref{eq:KIC-16}), and (\ref{eq:Kra}) for the symmetric positive real GIC with a very large number of users ($K=10^5$) at medium and high SNRs, where $\gamma$ is numerically found and $\gamma=200$ was used. For $g> 1$, the upper bound in (\ref{eq:KIC-16}) is shown to be tighter than (\ref{eq:KIC-20}). {This figure reveals that the sum-rate upper bound is still far from $K/2$ DoF near $g^2=1$ at realistic SNRs for the large number of user regime. For instance, the symmetric rates of our upper bound and TDM lower bound are $0.0186$ and $0.001$ bit, respectively, whereas that of the Kramer bound representing $K/2$ DoF is $0.8795$ for $g^2=0.9$ and $P=5$.}

\begin{rem}
The closed-form upper bounds were derived {by} exploiting an intrinsic structure\footnote{{This structure is represented by the fact that the proposed bounds are amenable to systematic canceling out some pairs of positive and negative differential entropies.}} of the conditions in (\ref{eq:Kub-2a}), (\ref{eq:Kub-2b}), (\ref{eq:Kub-4a}), (\ref{eq:Kub-4b}), and (\ref{eq:Kub-4c}) for the additional noise variables in our bounds. In fact, the structure was possible by removing the need for the optimization of the noise covariance matrix, i.e., $N_i$ is only correlated with $Z_i$ in our upper bounds and so $W_i$ is. Our closed-form upper bounds in Figs. \ref{fig-11} and \ref{fig-12} have no discontinuous point at $g^2=1$ for large $K$ and practical SNR, where the closed-form bounds are a continuous function. The discontinuous point appears only in the high SNR limit and it ceases to happen at high SNR of practical interest {(e.g., 20 dB)}. Combined with the small $K$ cases in Figs. \ref{fig-4a} and \ref{fig-5}, this result suggests that the  sum capacity of the symmetric real GIC {might} have no drastic change at least around $g^2=1$ {at SNR $=10, 20$ dB} for any $K$. Therefore, this finite SNR analysis is not in line with the known DoF result in \cite{Etk09b} that the DOF of $K$-user GICs is everywhere discontinuous with respect to channel coefficients. 
\end{rem}

{
\subsection{Large $K$ Regime}
\label{sec:KIC-D}

Based on Propositions \ref{prop-2} and \ref{prop-3}, we will focus on the large number of users regime and conduct the large $K$ analysis in this subsection. Inspired by the affine approximation of the capacity of code-division multiple access with random spreading introduced by Shamai and Verd\'u \cite{Sha01}, we first consider the high-SNR approximation of the upper bound on the symmetric rate given by the upper bound in Proposition \ref{prop-2}.  
For large $K$, the symmetric capacity $C(K,P)$ of  $K$-user GICs􏰺 can be approximated at high-SNR by 􏰺the zero-order and first-order terms in the expansion of the capacity as an affine function of $K$ and SNR ($P$) 
\begin{align}  
  C(K,P) = d_K \big(K\log P - K\ell_K\big) +o(1) \nonumber 
\end{align}
where 􏱔􏰾$d_K$ denote the per-user DoF (first-order term) for large $K$ defined by 
\begin{align} \label{eq:Kub-20}
   d_K\triangleq \lim_{K\rightarrow \infty}\lim_{P\rightarrow \infty} \frac{C(K,P)}{K\log P}
\end{align} 
and 􏱦􏰾$\ell_K$ is the power offset (zero-order term) in 3-dB units defined by
 $$\ell_K\triangleq \lim_{K\rightarrow \infty}\lim_{P\rightarrow \infty} \bigg(\log P - \frac{C(K,P)}{d_KK}\bigg).$$ 
Also $o(1)\rightarrow 0$ as $K,P\rightarrow\infty$.  
Let ${\Rc_\text{sym}^\text{ub}}$ denote the minimum of the upper bounds on the symmetric rates given by the upper bound in Propositions \ref{prop-2} and \ref{prop-3}. Then we have the following result. 

\begin{thm} \label{prop-4}
For large $K$ and $P$, the symmetric-rate upper bound ${\Rc_\text{sym}^\text{ub}}$ is characterized as 
\begin{align} \label{eq:Kub-21}
  {\Rc_\text{sym}^\text{ub}} = \log P +\ell_K^* +o(1) 
\end{align}
where $d_K=1$ and the power offset is given by 
\begin{align} \label{eq:Kub-21b}
\ell_K^* = \left\{ \begin{array}{ll}
  -\log\big(|1-g|^2(1+|g|^2)\big)  &  \   |g|^2 \le 1\\
  -\log\big(|1-g|^2\big) &  \ |g|^2 > 1.
\end{array} \right.   
\end{align}

\end{thm}

\begin{IEEEproof}
Taking the limit of ${P\rightarrow \infty}$ on the right-hand side of (\ref{eq:KIC-20}), we have
\begin{align} \label{eq:KIC-23}
    {\Rc_\text{sym}^\text{ub}} 
    &= \log \left((K-1)(1+|g|^2)P\right) +\sum_{k=2}^{K-1}\log \left(|1-g|^2(1+|g|^2)P\cdot\frac{k-1}{k}\right) +O(1) \nonumber \\
   &= \log \left((1+|g|^2)P\right) +\sum_{k=2}^{K-1}\log \left(|1-g|^2(1+|g|^2)P\right) +O(1) \nonumber \\
   &= K\log \left(|1-g|^2(1+|g|^2)P\right) +O(\log P).
\end{align}
Dividing the above equation by $K$ and taking the limit of ${K\rightarrow \infty}$, we get (\ref{eq:Kub-21}) for all values of $|g|^2$. Similarly, we do the same steps to the right-hand side of (\ref{eq:KIC-16}), which yields (\ref{eq:Kub-21}) for $|g|^2 > 1$. Noticing that $\log\big(|1-g|^2(1+|g|^2)\big)\ge\log\big(|1-g|^2\big)$, we obtain the desired result. 
\end{IEEEproof}

\begin{figure*}
\vspace{-2mm}
\center \subfigure[$g^2=1.1$]{\includegraphics[scale=.72]{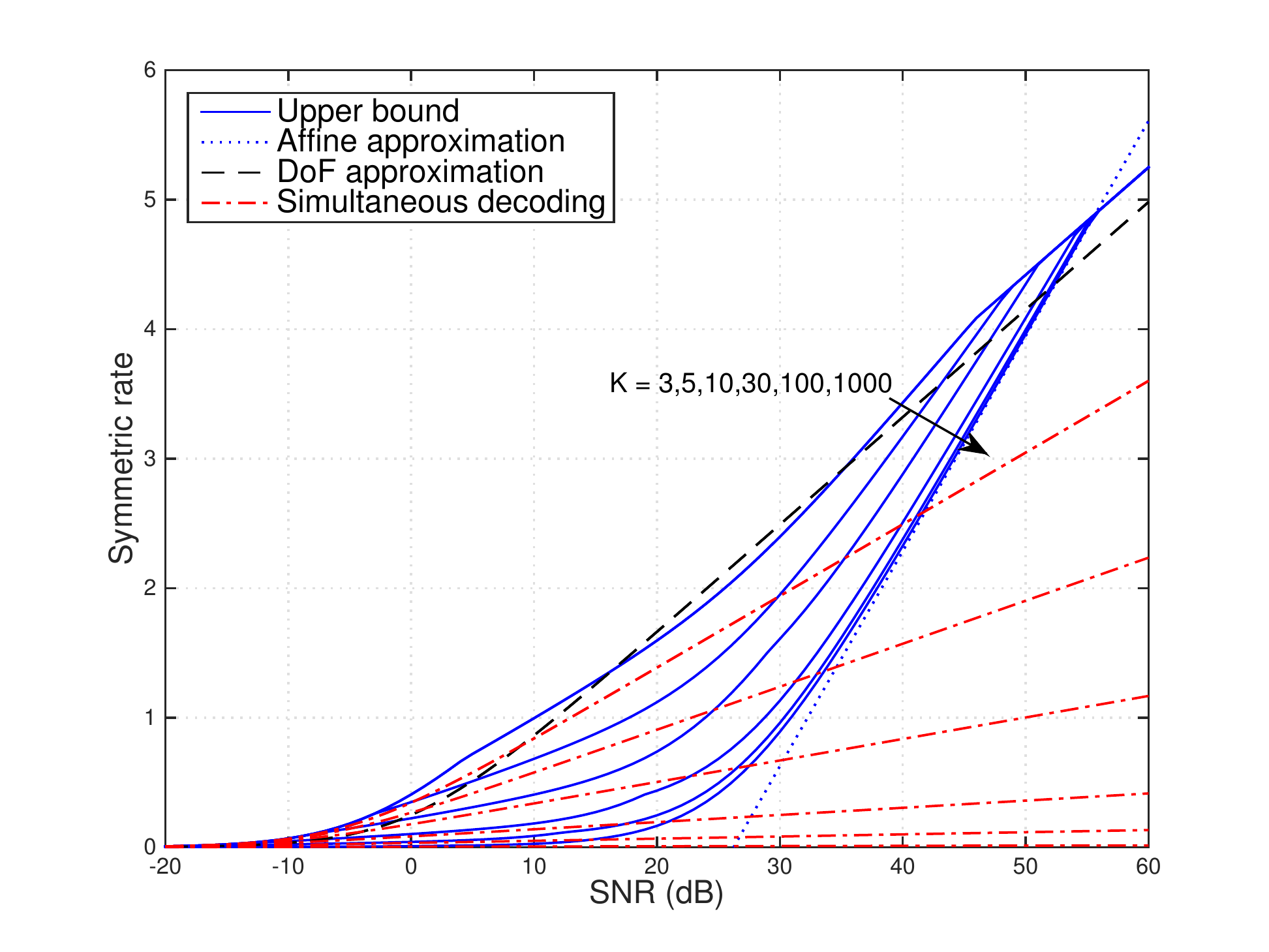}\label{fig-13}} \\
\hspace{1mm}\centerline{\subfigure[$g^2=0.7$]{\includegraphics[width=4.2in]{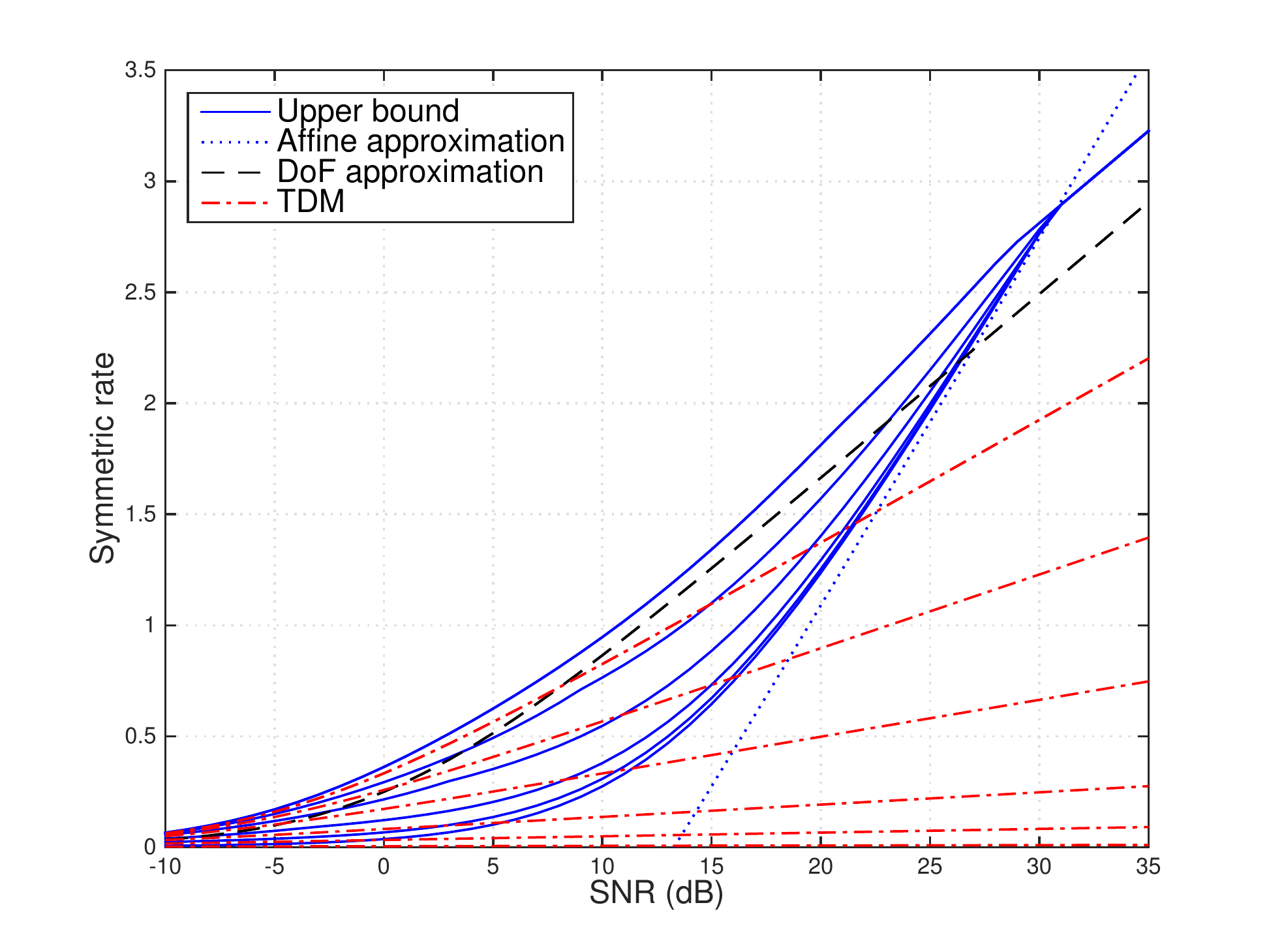}
\label{fig-2a}}
\hfil \hspace{-12mm}
\subfigure[$g^2=1.5$ and $K=1000$]{\includegraphics[width=4.2in]{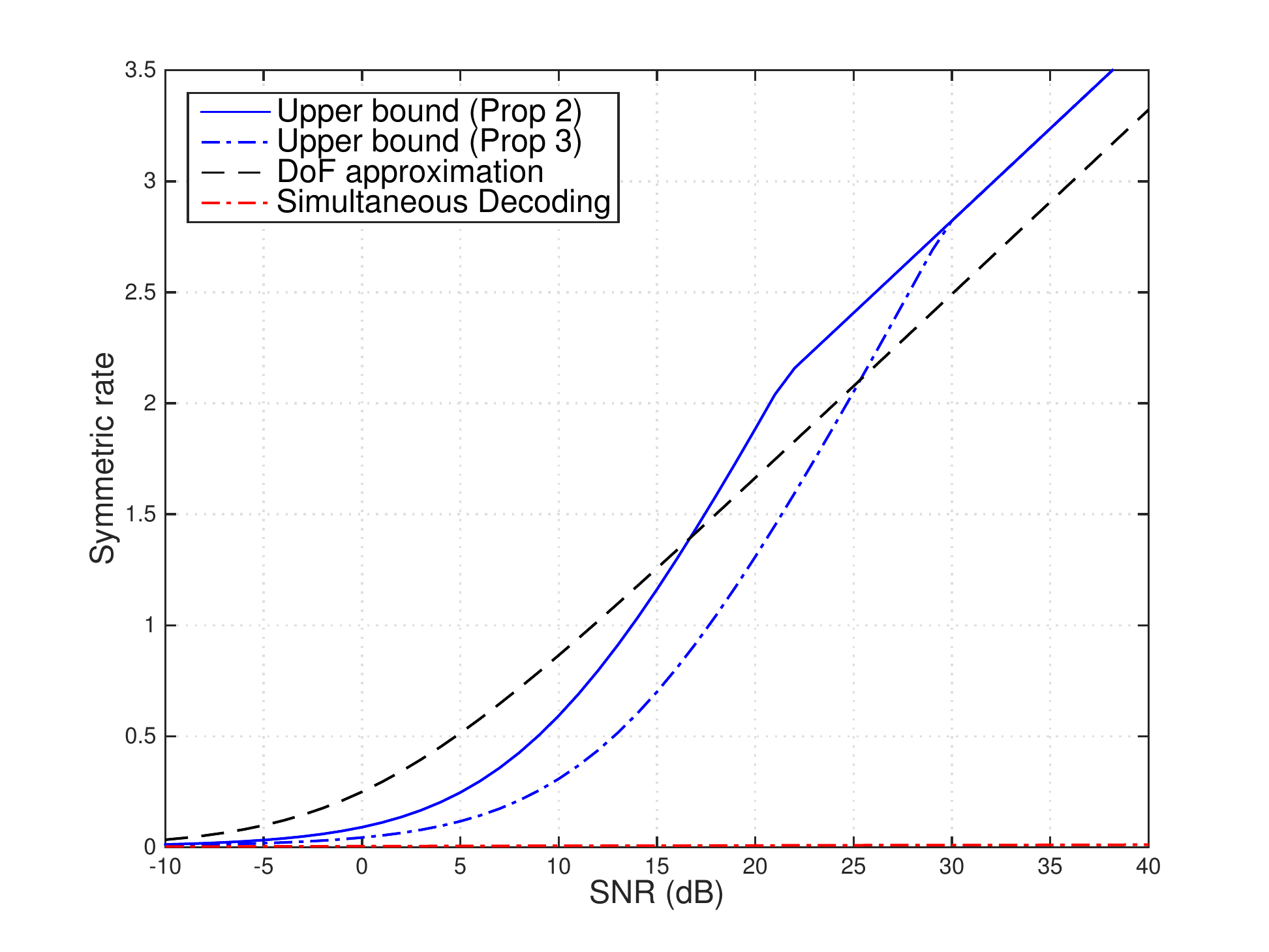}
\label{fig-2b}}}
\vspace{-2mm}
\caption{Bounds on the symmetric rate of the positive real $K$-user GIC for different SNRs and $K$. In (a), the upper bound is given by the minimum of Propositions \ref{prop-2} and \ref{prop-3}. In (b), it is given by Proposition \ref{prop-2}. The red dash-dotted curves that represent simultaneous (non-unique) decoding or TDM lower bound are for $K=3,5,10,30,100,1000$ from top to bottom. In (c), we compare two upper bounds in Propositions \ref{prop-2} and \ref{prop-3} for $K=1000$.} \label{fig-13b}
\vspace{-4mm}
\label{fig2}
\end{figure*}

For large $K$ with \emph{positive real} symmetric GICs, we can characterize the following three regimes of the ratio $\eta \triangleq \lim_{K\rightarrow \infty}\frac{\Rc_\text{sym}}{K\log P}$:
\begin{align} \label{eq:Kub-22}
  \eta = \left\{ \begin{array}{ll}
  0 &  \  \phantom{2\ell_K^*< } \ \snr \le \ell_K^*\\
  1 &  \ \phantom{2}\ell_K^* < \snr\le 2\ell_K^*\\
  \frac{1}{2} &  \ 2\ell_K^*< \snr. \\
  \end{array} \right.
\end{align}
While the first two regimes are suggested by Theorem \ref{prop-4}, 
the third regime $\eta=\frac{1}{2}$ is due to the Kramer bound. 
It is easy to see that at $\snr= 2\ell_K^*\gg 1$, ${\Rc_\text{sym}^\text{ub}}$ meets the DoF approximation  given by $\frac{1}{4}\log (1+\snr)$ for the capacity approximation of the $1/2$ DoF per user. 

Fig. \ref{fig-13b} shows how much the well-known $1/2$ DoF (per user) result could be translated into a real potential\footnote{Notice that the solid curve in Fig. \ref{fig-13} comes from a simplified, closed-form (i.e., not the best bound we can get from our results) upper bound on the achievable rate of any sophisticated interference management scheme.} gain at finite SNR for different $K$. 
In Fig. \ref{fig-13}, we can see the behavior of lower and upper bounds when $K=3$ to $1000$ and $g^2$ is close to $1$, where $\ell_K^*=43$ in dB scales.
Fig. \ref{fig-2a} shows that (\ref{eq:Kub-22}) is valid even when $g^2$ is not so close to $1$. However, a close look at $\ell_K^*$ reveals that $\ell_K^*$ decreases as $g^2$ tends to be far from unity.
While the regime $\eta=0$ is somewhat supported by the simultaneous decoding or TDM lower bound, the regime $\eta=1$ might be fundamental if one finds a matching lower bound in this regime.
Recall that our closed-form upper bound in Proposition \ref{prop-2} has been tightened in Proposition \ref{prop-3}. Hence, it is shown in Fig. \ref{fig-2b} that the regime $\eta=1$ can be shifted to the right by 
$10\log_{10}(1+|g|^2)\approx 4$ dB, which validates (\ref{eq:Kub-21b}).

An important implication of the above results is that for large $K$, the capacity gain promised by $K/2$ DoF may not be achievable if SNR is no more than the power offset (or threshold) $\ell_K^*$. 
In other words, the surprising performance benefit of sophisticated interference management schemes suggested by the DoF results in \cite{Mot09b,Wu15,Jaf10} seems not realized at a realistic SNR {(e.g., 10, 20 dB)} over a certain range (around $g^2=1$) of channel coefficients for the (constant) $K$-user symmetric real case. Such a range of $g^2$ depends on the number of users and the SNR of interference networks. Namely, as SNR deceases and $K$ grows, the range of $g^2$ gets wider according to the rule given by the SNR threshold $\ell_K^*$ in (\ref{eq:Kub-21b}). 
Therefore, the well-known DoF result should be carefully interpreted at finite SNR from the perspective of our results. 

}

\ifdefined\TIT

\subsection{Asymmetric Case}
\label{sec:KIC-B}

{For the $K$-user asymmetric GIC, the upper bounds in Theorems \ref{thm-3ub2} and \ref{thm-3ub3} cannot be naturally generalized. We can  naturally extend the genie variables in (\ref{eq:3IC-2}) to the general $K$-user case as follow: }
\begin{align}  \label{eq:KIC-0a}
    U_k&=\sum_{i\neq k} h_{ki}X_i+W_k \nonumber \\
  S_{k} &= \sum_{i\neq k-1}h_{k-1,i} X_i+N_{k} .
\end{align} 
Then, we present the following asymmetric version of Theorem \ref{thm-kub2} for the $K$-user GIC.

\begin{thm} \label{thm-kub2b}
The sum capacity of the $K$-user complex GIC in the mixed interference regime, where $|h_{1,K)}|^2\le 1$, is upper-bounded by
\begin{align}  \label{eq:kub-2b}
    \sum_{k=1}^K R_k \le &\;   I\big(X_{1G}; Y_{1G}\big)  +I\Big(X_{K G}; Y_{K G}\big| X_{1G}^{K-1}\Big) \nonumber \\ \ \ \ 
    &+ \sum_{k=2}^{K-1} \bigg\{ I\Big(X_{k G}; S_{2 G}\big| X_{1G}^{k-1}\Big) +I\Big(X_{k G}; Y_{k G}\big| X_{1G}^{k-1} ,X_{(k+1)G}^{K)},S_{2 G}\Big) \bigg\}
\end{align}
for all $N_{2}$ satisfying $\sigma^2_{N_{2}}\ge h_{1,K}$, where $N_k$ and $Z_k$ are independent (i.e., $\rho_{N_k}=0$) for all $k$. Permuting the user indices, we obtain $K!$ bounds.
\end{thm}

\begin{IEEEproof}
See Appendix \ref{proof-7}.
\end{IEEEproof}

A major difficulty in the general asymmetric case with more than three users was that we cannot use the technique of (\ref{eq:3IC-3b}) and (\ref{eq:KIC-1c}) in the proofs of Theorems \ref{thm-3ub2} and \ref{thm-kub2}, based on Lemma \ref{lem-6}. As a result, the first mutual information inside the brace in (\ref{eq:kub-2b})
serves as a penalty term for the asymmetric case, compared to the symmetric case in (\ref{eq:kub-2}) of Theorem \ref{thm-kub2}.
Similar to the Theorem \ref{thm-kub2b}, we can obtain the following result on the asymmetric complex GIC. 

\begin{thm} \label{thm-kub3b}
The sum capacity of the $K$-user complex GIC is upper-bounded by
\begin{align}  \label{eq:kub-3b}
    \sum_{k=1}^K R_k \le  &\; \frac{1}{K} \sum_{k=1}^{K} \Bigg\{  I\big(X_{{k,1}G}; Y_{{k,1}G}\big)  +I\big(X_{_{k,K}G}; Y_{{k,K}G}\big| U_{{k,K}G}\big) +I\big(U_{{k,K}G};Y_{{k,K}G}+\tilde{N}_{{k,2}}\big) \nonumber \\ 
    &\ \ \ \ \ \ + \sum_{\ell=2}^{K-1} \bigg\{ I\Big(X_{{k,\ell} G}; S_{{k,2} G}\big| X_{{k,1}G}^{{k,\ell-1}}\Big) +I\Big(X_{{k,\ell} G}; Y_{{k,\ell} G}\big| X_{{k,1}G}^{{k,\ell-1}} ,X_{{k,\ell+1}G}^{{k,K}},S_{{k,2} G}\Big) \bigg\}    \Bigg\}
\end{align}
for all $\{N_k,W_k: k=1,\ldots,K\}$ satisfying 
\begin{align}  
  &\sigma^2_{V_{W_{{k,1}}}}\ge \sigma^2_{N_{{k,2}}} \ge |h_{{k,1},{k,K}}|^2\sigma^2_{Z_{{k,K}}-W_{{k,K}}} \label{eq:Kub-10a} 
\end{align}
where $$\tilde{N}_{{k,2}} = \sqrt{{|h_{{k,1},{k,K}}|^{-2}-\sigma_{N_{{k,2}}}^{-2}\sigma_{Z_{{k,K}}- W_{{k,K}}}^2}}  N_{{k,2}}.$$
Permuting the user indices, we obtain $K!$ bounds. 
\end{thm}

\begin{IEEEproof}
See Appendix \ref{proof-8}.
\end{IEEEproof}

In this case, $I(X_{{k,\ell} G}; S_{{k,2} G}\big| X_{{k,1}G}^{{k,\ell-1}})$ in (\ref{eq:kub-3b}) is a penalty term, compared to the symmetric case in (\ref{eq:kub-3}) of Theorem \ref{thm-kub3}. In the sequel, we provide a special case where we can avoid the penalty terms. 

If the channel coefficients satisfy
$$h_{14}=\frac{h_{13}}{h_{23}} h_{24},\; h_{15}=\frac{h_{13}}{h_{23}} h_{25},\; \cdots, h_{1K}=\frac{h_{13}}{h_{23}} h_{2K}$$
we can rewrite (\ref{eq:KIC-4a}) in Appendix \ref{proof-7} using the same way in (\ref{eq:3IC-3b}) for the three-user case as
\begin{align}  \label{eq:KIC-14}
  I(X^n_2;Y^n_2,S_2^n| X_1^n) &\le  {\textstyle h(\sum_{i=2}^{K}h_{1i}X_i^n+N_2^n)} -{\textstyle h(\sum_{i=3}^{K}h_{1i}X_i^n+N_2^n)}  \nonumber \\ &\ \ \ +nh(Y_{2G}| X_{1G},S_{2G})  -{\textstyle h(\sum_{i=3}^{K}h_{2i}X_i^n+Z_2^n|\sum_{i=3}^{K}h_{1i}X_i^n+N_2^n)} \nonumber \\
  &=  {\textstyle h(\sum_{i=2}^{K}h_{1i}X_i^n+N_2^n)}-{n h(Z_2-h_{23}h_{13}^{-1}N_2)} \nonumber \\ &\ \ \  +nh(Y_{2G}| X_{1G},S_{2G})  -{\textstyle h(\sum_{i=3}^{K}h_{1i}X_i^n+V_{N_2}^n)}
\end{align}
where \begin{align} \label{eq:3IC-14b2}
  V_{N_k}&\sim \CN(0,\sigma^2_{N_k|\; Z_k-h_{k,k+1}h_{k-1,k+1}^{-1}N_k}) .
\end{align}
Repeating the same procedure for all $k=2,\ldots,K$ and all permutations and doing the same things to (\ref{eq:3IC-17b}) in Appendix \ref{proof-8}, we arrive at the following result.

\begin{corol} \label{cor-3}
Given a permutation in terms of user ordering, when channel coefficients satisfy the condition
\begin{align} \label{eq:KIC-25}
 h_{i-1,j}=\frac{h_{i-1,i+1}}{h_{i,i+1}}\; h_{i,j}
\end{align} 
for $i=2,\ldots,K-2, \ j=i+2,\ldots,K,$
the sum-rate upper bounds for the corresponding $K$-user asymmetric GIC are equivalent to the symmetric case in Theorems \ref{thm-kub2} and \ref{thm-kub3}. 
\end{corol}



\section{Sum-Rate Behavior of ``Asymmetric Complex"  Gaussian Interference Channels}
\label{sec:CGIC}

{So far, we have focused on the sum-rate analysis of the $K$-user \emph{symmetric (positive) real} GIC, where the phases of channel coefficients are not taken into account. Some practical implications of the existing DoF results have been revisited.} In this section, we will investigate the sum-rate behavior of the $K$-user \emph{asymmetric complex} GIC at finite SNR, based on our upper bounds derived in the previous sections.
{Our study is motivated by the well-known toy example in \cite{Cad08} where the sum capacity is $\frac{K}{2}\log(1+2P)$ when the common direct- and cross-channel coefficients are $1$ and $\sqrt{-1}$, respectively. In sharp contrast, the sum capacity becomes just $\log(1+KP)$ when the channel coefficients are all $1$. Hence it would be interesting to figure out what happens between the two extreme points.}

\subsection{Symmetric Case}
\label{sec:CGIC-A}

Bearing the somewhat negative result {in subsections \ref{sec:KIC-E} and \ref{sec:KIC-D}} for the symmetric \emph{positive real} case in mind, we first turn to the three-user symmetric \emph{complex} GIC. We restrict our attention to $|g|^2\le 1$ for ease of illustration. Although some existing upper bounds taken into account in this work are shown to be tighter than our upper bound in the very weak interference regime (e.g, $|g|^2\le 0.12$ at SNR $=10$ dB), we consider only the moderately weak interference regime ($0.12<|g|^2\le 1$), the main regime of practical interest, where the proposed upper bound is tightest. Therefore, it suffices to investigate the behavior of the new sum-rate upper bound.



\begin{figure}
  \center \includegraphics[scale=.72]{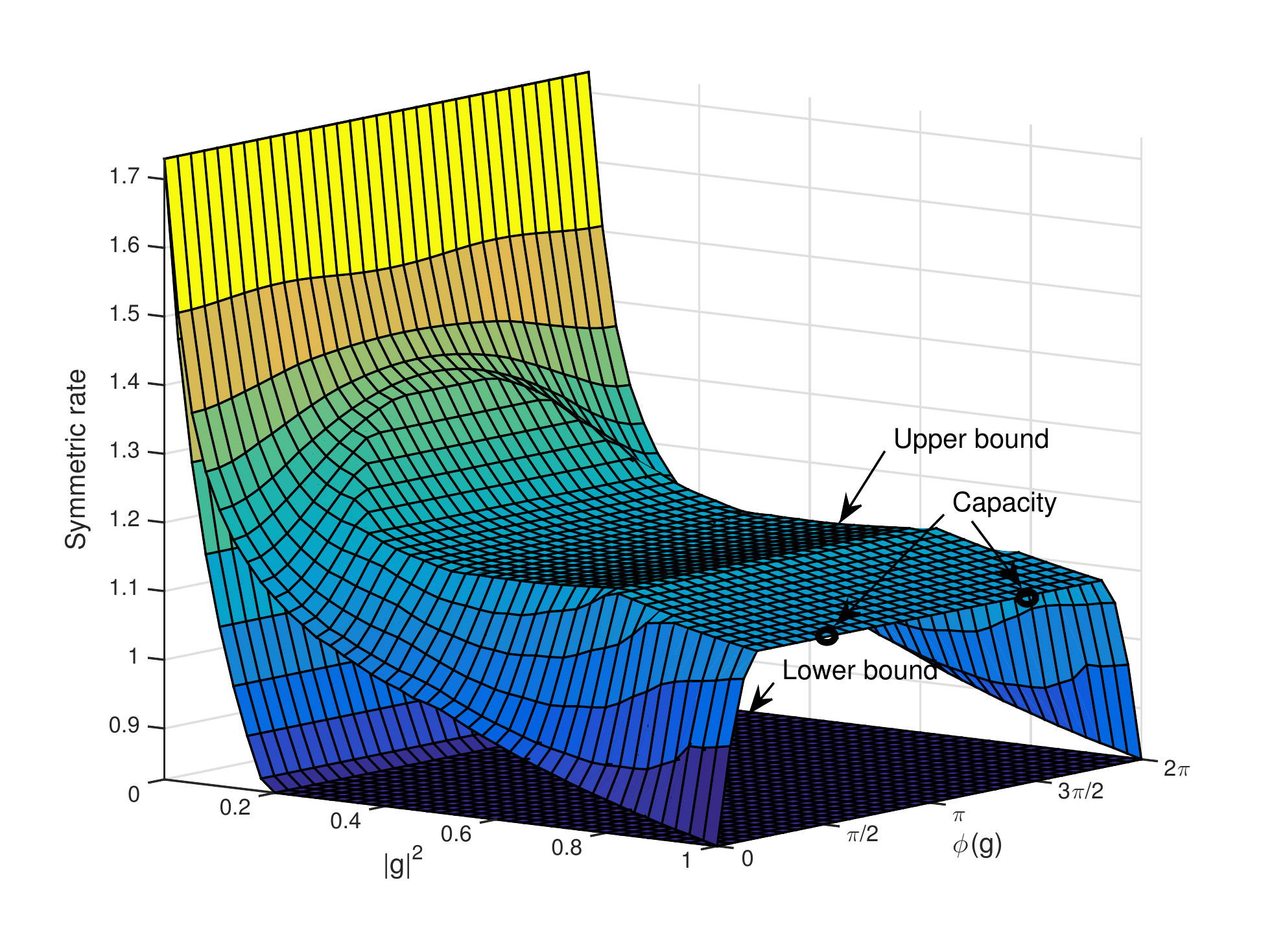}
  \caption{Bounds on the sum capacity of three-user symmetric GIC for different amplitudes $|g|^2$ and phases $\phi(g)$ of cross-channel coefficient $g$ when $P=10$ and $\phi(g)=0:\frac{\pi}{16}:2\pi$. {The points labeled by `capacity' are $\phi(g)=\frac{\pi}{2},\frac{3\pi}{2}$, respectively, with $|g|^2=1$, which follows from the well-known toy example in \cite{Cad08}.}}\label{fig-2}
\end{figure}

Fig. \ref{fig-2} shows the symmetric-rate behavior of the three-user symmetric {complex} GIC with the phase of the symmetric cross-channel coefficient $g$ varying between $0$ and $2\pi$. The four upper bounds derived in Section \ref{sec:3IC} have their own range of $|g|^2$, over which one is tighter than the other three, as the phase $\phi(g)$ varies. 
{Our capacity upper bound result improves upon understanding the potential impact of phase difference between the direct-channel and the cross-channel coefficients on the three-user complex GIC.}  This sum-rate behavior could not be predicted by the DoF results. For example, the known result in \cite{Cad10a} for the three-user complex GIC shows that DoF is discontinuous at a subset of channel coefficients with measure zero. In particular, the result therein proved that the phase alignment scheme achieves only $1$ DoF under certain amplitude and phase conditions, while it can achieve the same $1.2$ DoF for almost all values of channel coefficients. Moreover, another result in \cite{Cad08}  shows that the three-user constant complex GIC has 1 DoF for $g=1$ and 1.5 DoF for $g=\sqrt{-1}$. This does not provide much insight on the prediction of sum-rate behavior at finite SNR for particular values of $|g|$ and $\phi(g)$. 
{Hence we trace the trajectory of our upper bounds for different $\phi(g)$ with $|g|$ fixed.}




\begin{figure}
\vspace{-5mm}
\hspace{-2mm}
\ifdefined\TIT
   \center \includegraphics[scale=.72]{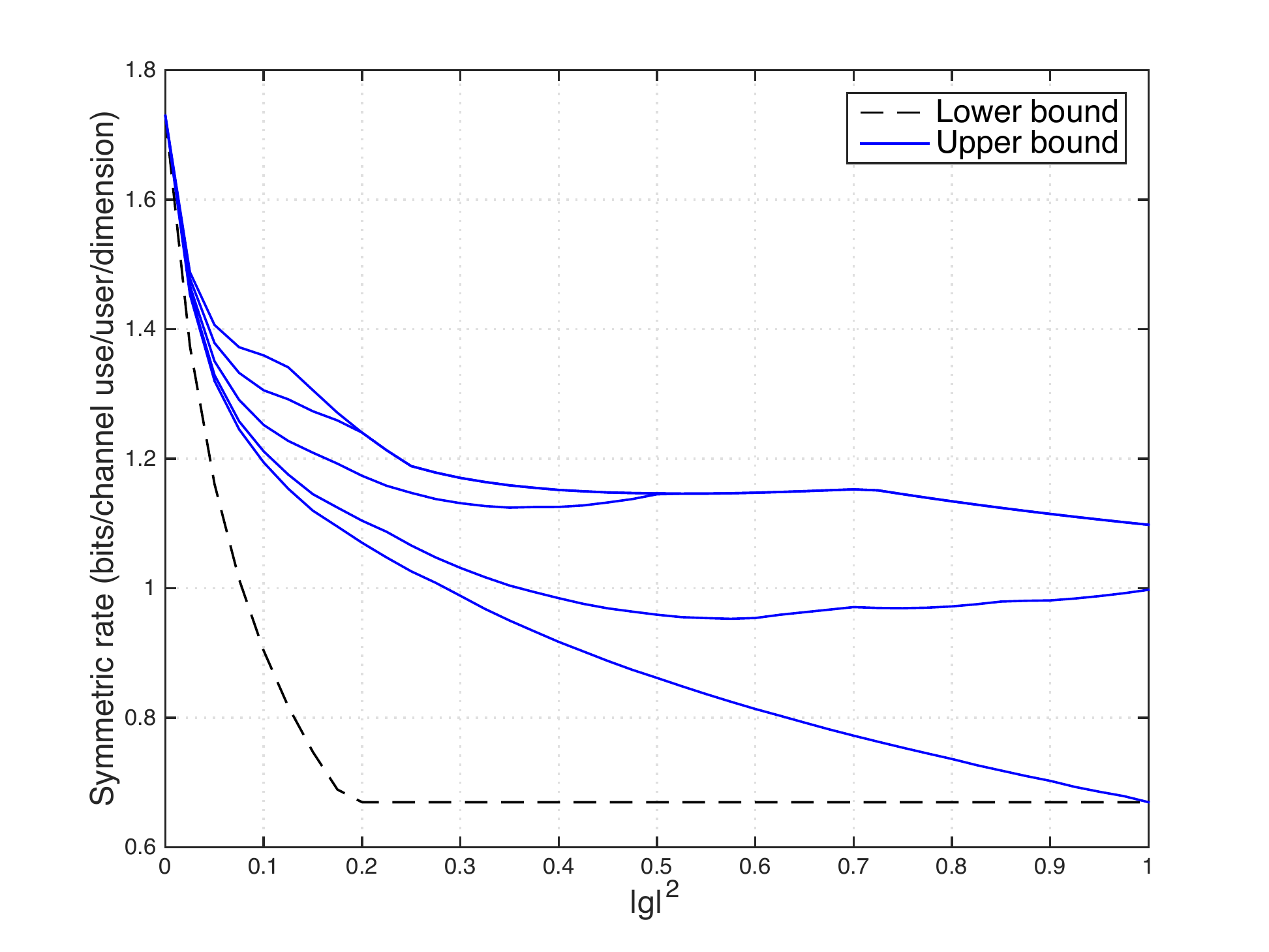}
\else
   \center \includegraphics[scale=.55]{./Fig/GIC_sum_rate_4K_pi}
\fi
\vspace{-6mm}
  \caption{Bounds on the sum capacity of four-user symmetric Gaussian interference channel for different amplitudes $|g|^2$ and phases $\phi(g)=0, \pi/8, \pi/4, 3\pi/8, \pi/2$ (blue curves from bottom to top): $P=10$. }\label{fig-5}
\end{figure}


Fig. \ref{fig-5} depicts the sum-rate behavior of the four-user symmetric {complex} GIC  at SNR $=10$ dB by varying the phase $\phi(g)$ such that $\phi(g)=0, \pi/16, \pi/8, 3\pi/16,  \ldots, \pi/2$. We can see that the impact of phase offset on the capacity upper bound is very analogous to the three-user case in (b) of Fig. \ref{fig-2}. We also verified that this sum-rate upper bound behavior remains similar at least for the five-user symmetric GIC, but the corresponding figure is omitted for compactness of this work.

In contrast to the symmetric real case, the capacity upper bound behavior for the complex GIC suggests  that  {one may improve the lower bound on the capacity of the complex GIC}. This might be done by existing sophisticated schemes exploiting the phase offset between direct-channel and cross-channel coefficients. 
However, we need to carefully interpret Figs. \ref{fig-2} and \ref{fig-5} where the phases of cross-channel coefficients are assumed to be already \emph{aligned}, which is a very special case in realistic systems.

\begin{rem}
{ For $\phi(g)=\frac{\pi}{2}$ or $\frac{3\pi}{2}$, the symmetric capacity (per dimension) is well known as $\frac{1}{4}\log(1+2P)$ \cite{Cad08}. As seen from Figs. \ref{fig-2} and \ref{fig-5}, the proposed upper bound is tight for those values of $\phi(g)$. This implies that our bound may be useful in the complex symmetric case. In fact, the Kramer upper bound is tight at those values of $\phi(g)$.}
\end{rem}

\ifdefined\TIT
\subsection{Semi-Symmetric Case} \label{sec:3IC-B}

Studying the $K$-user \emph{symmetric} real/complex GIC only is obviously insufficient to predict the sum-capacity behavior since the probability that all interfering links have the same channel coefficient (or even the same phase of channel coefficients) is quickly vanishing  as $K$ increases. Meanwhile, the fully asymmetric case where all channel coefficients $h_{ij}$ can be arbitrarily different is too difficult to evaluate the resulting upper bounds and to provide an insight. To compromise between the symmetric and the fully asymmetric case, we  introduce a $K$-user ``semi-symmetric" GIC, where complex cross-channel coefficients for each user are different but all users experience the same SNR and interference situation, ({e.g., the same INR defined by $\inr = \sum_{i=1}^{K-1}|g_{i}|^2\snr$}). The $K$-user semi-symmetric GIC can be formally written as
\begin{align} \label{eq:A-2c}
  Y_k&=X_k+\sum_{i=1}^{K-1}g_{i}X_{{k,i+1}}+Z_k 
\end{align}
where $P_k=P$ for all $k=1,\ldots,K$.
In particular, the three-user semi-symmetric GIC is given by 
\begin{align} \label{eq:A-2b}
  Y_1&=X_1+g_{1}X_2+g_{2}X_3+Z_1 \nonumber \\ 
  Y_2&=X_2+g_{1}X_3+g_{2}X_1+Z_2 \nonumber \\ 
  Y_3&=X_3+g_{1}X_1+g_{2}X_2+Z_3 
\end{align}
where $P_k=P$ for $k=1,2,3$. 

\begin{figure}
\centerline{\subfigure[$|g_1|^2=|g_2|^2=0.3$]{\includegraphics[width=4in]{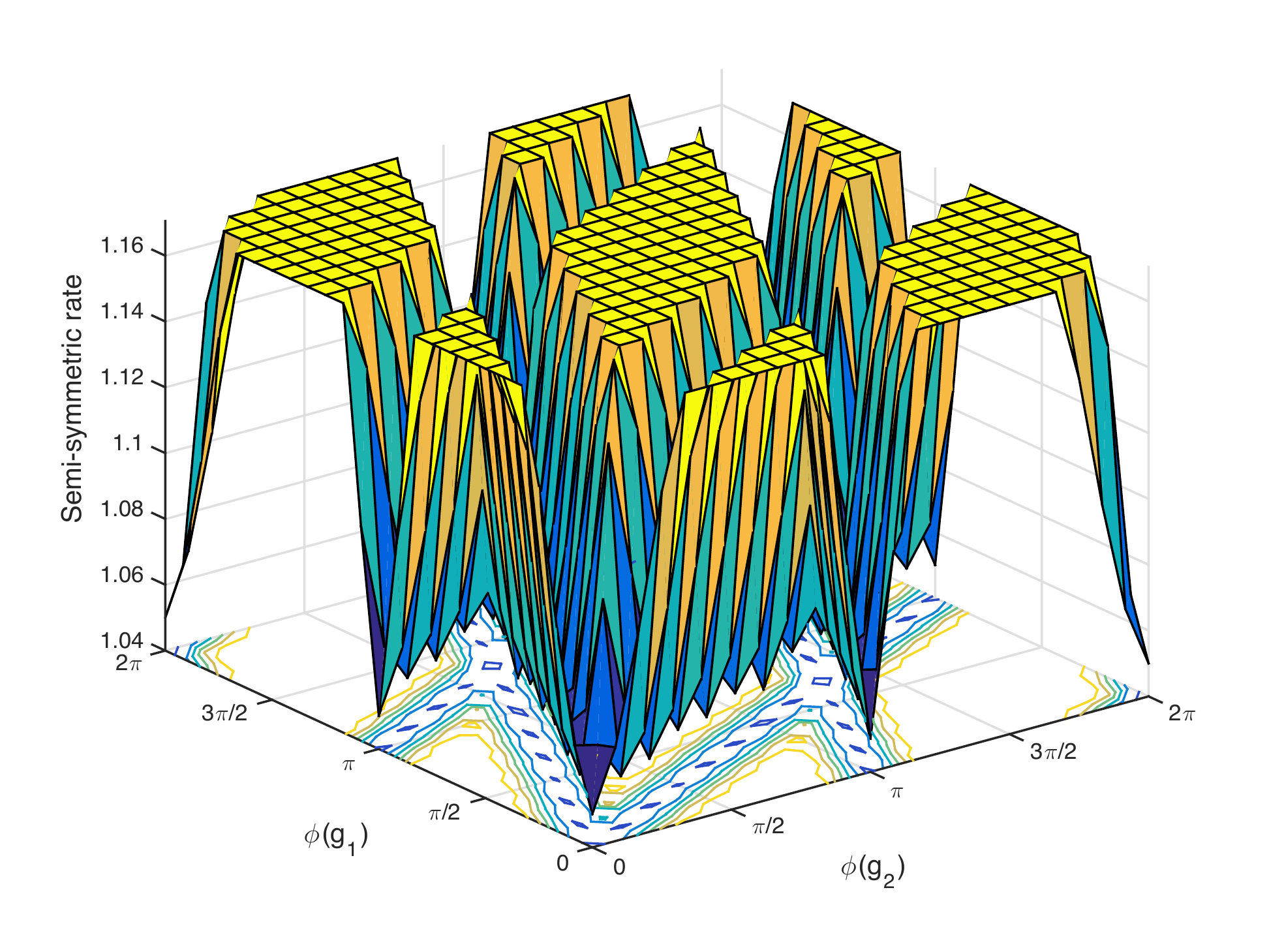}\label{fig_1st}}
\hfil  \subfigure[$|g_1|^2=|g_2|^2=0.5$]{\includegraphics[width=4in]{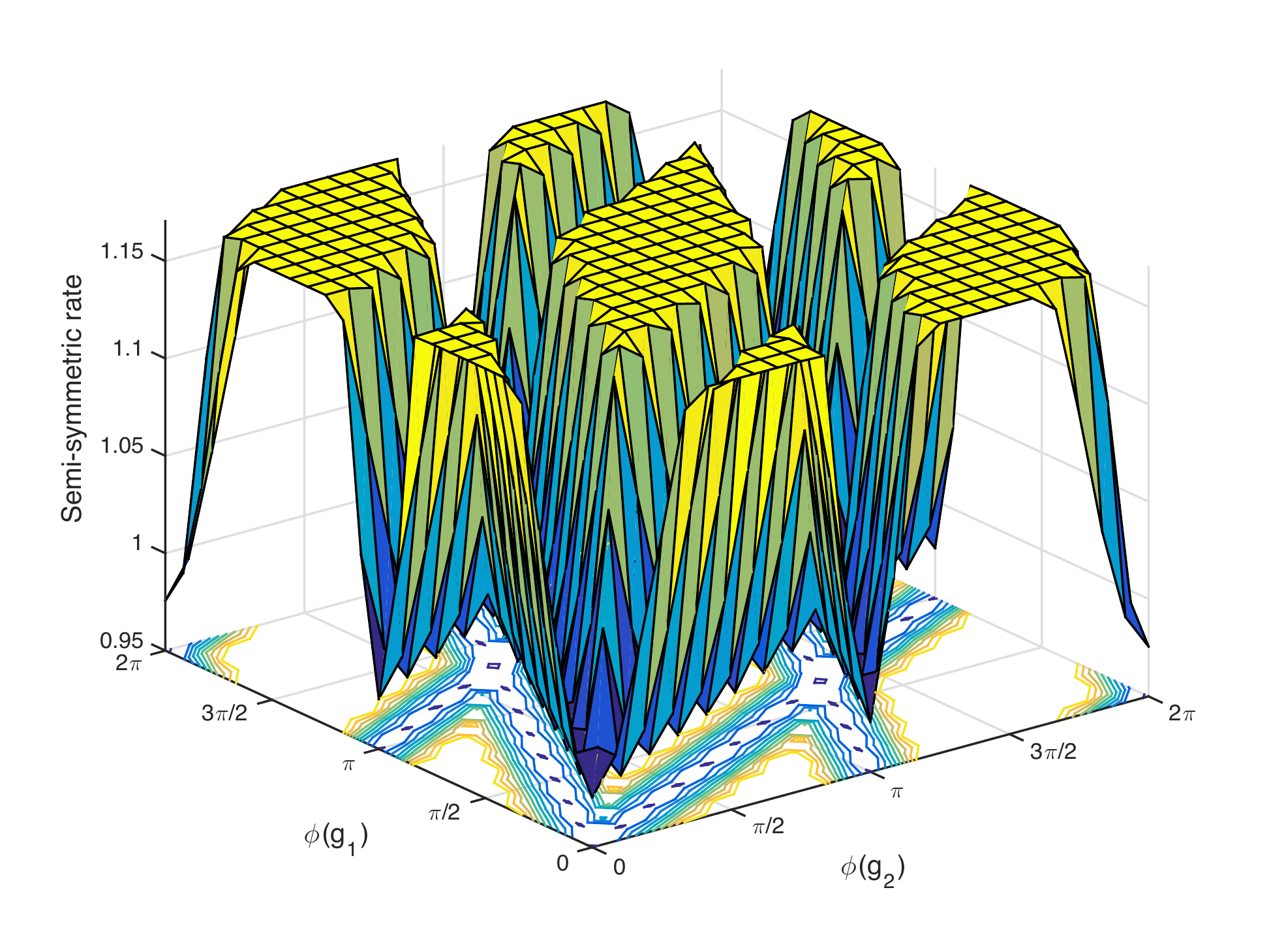}\label{fig_2nd}} }
\vfil 
\centerline{\subfigure[$|g_1|^2=|g_2|^2=0.7$]{\includegraphics[width=4in]{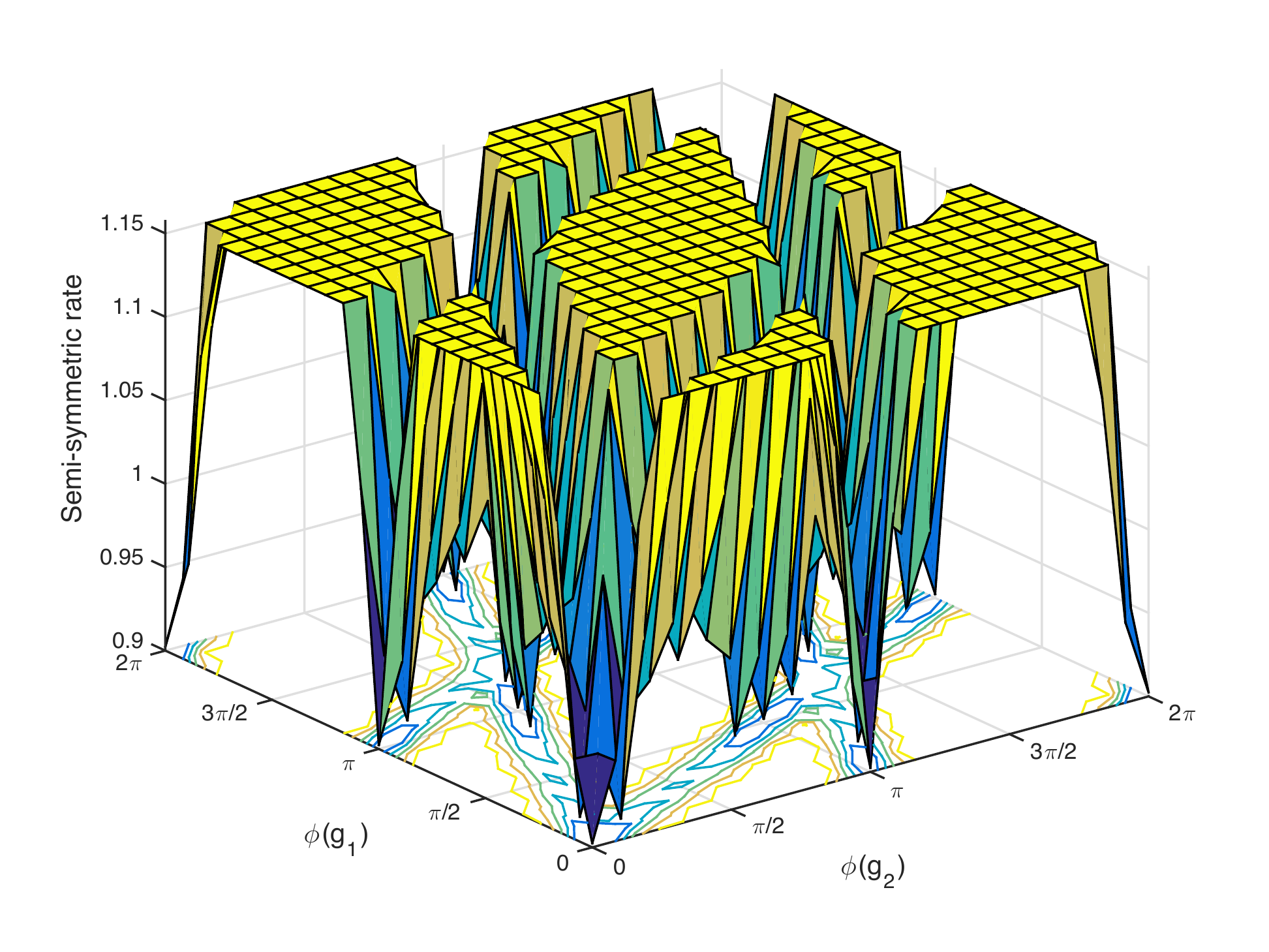}\label{fig_3rd}}
\hfil  \subfigure[$|g_1|^2=|g_2|^2=1$]{\includegraphics[width=4in]{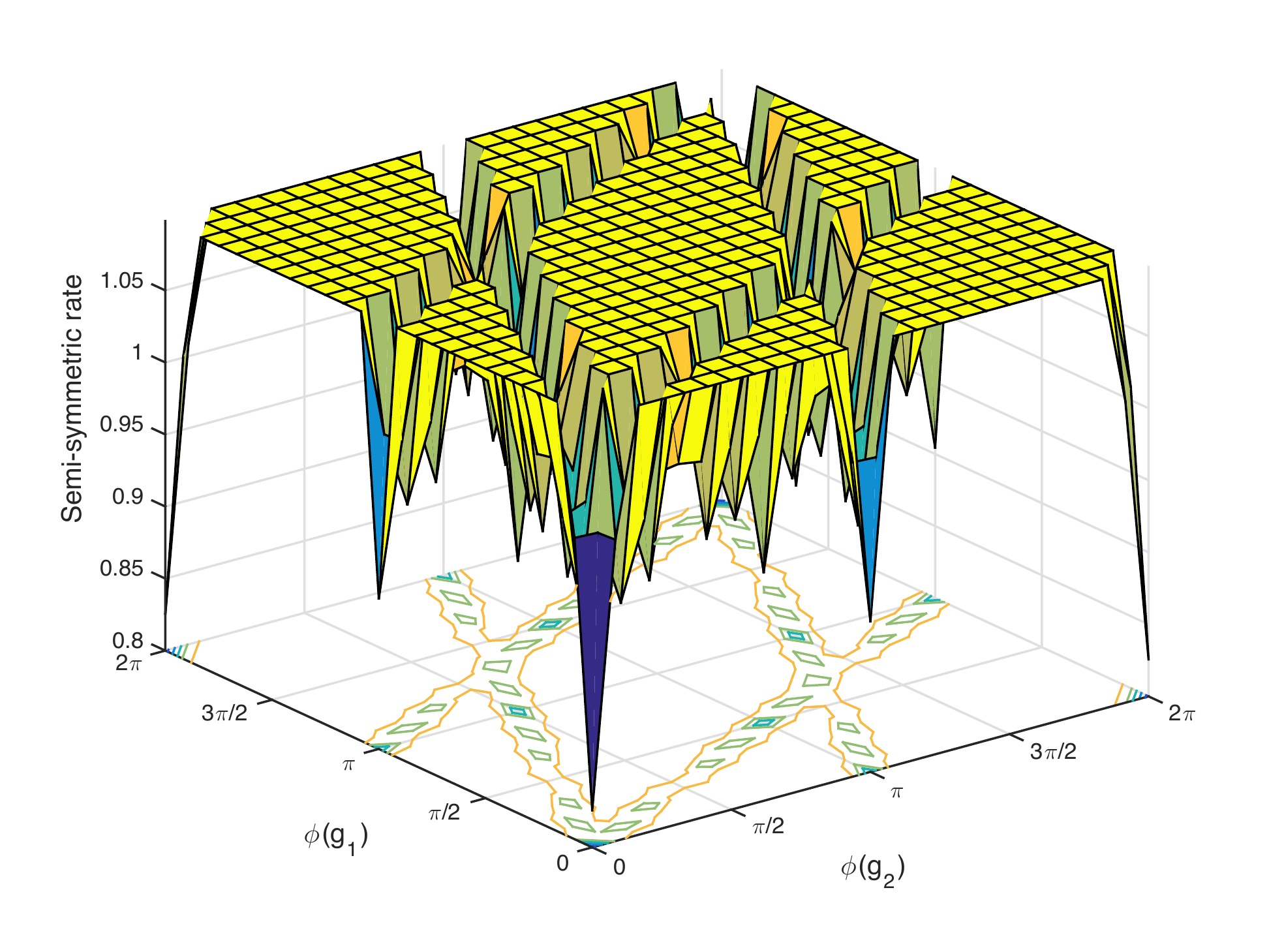}\label{fig_4th}}}  
\caption{Upper bound on the sum capacity of three-user \emph{semi-symmetric} GIC for different phases ($\phi(g_1), \phi(g_2)$) when $P=10$.} \label{fig-6}
\end{figure} 

Fig. \ref{fig-6} illustrates the new upper bound on the sum capacity of three-user {semi-symmetric} {complex} GIC for four different amplitudes ($|g_1|^2 , |g_2|^2$) with phases ($\phi(g_1), \phi(g_2)$) varying between $0$ and $2\pi$, where $|g_1|^2 = |g_2|^2$. For all cases, the  time division lower bound is $\frac{1}{6}\log(1+3P)=0.8257$ (bit/channel use/user/dimension). 
In Fig. \ref{fig-8}, we depicts the same upper bound for {$|g_1|^2 \neq |g_2|^2$}.  In this case, our upper bound depends on the order of users and hence it was obtained by the minimum of (\ref{eq:3IC-25}) and (\ref{eq:3IC-27}) in Theorem \ref{thm-3ub2}. {The upper bound behavior shows a large variation over the phase offset between $\phi(g_1)$ and $\phi(g_2)$. However, it should be pointed out that the behavior observed on an upper bound that is not matched by a lower bound may not be fundamental.
Potentially, the upper bound behavior reveals some interesting results.} 
In accordance with a consistent observation from Fig. \ref{fig-6}, Fig. \ref{fig-8}, and other numerical results with different settings not shown here due to compactness of this paper, we have the following conjecture.   

\begin{conj}
The sum capacity for the three-user semi-symmetric GIC {may} take its maximum values along the intersection of the following two types of lines  
\begin{align} \label{eq:3IC-15}
  2\phi(g_1) - \phi(g_2) +\pi &= 0 \mod 2\pi \nonumber \\
  2\phi(g_2) - \phi(g_1) +\pi &= 0 \mod 2\pi 
\end{align}
and it can be minimized along any of the following two types of lines 
\begin{align} \label{eq:3IC-16}
  2\phi(g_1) - \phi(g_2) &= 0 \mod 2\pi \nonumber \\
  2\phi(g_2) - \phi(g_1) &= 0 \mod 2\pi .
\end{align} 
\end{conj}

Notice that our conjecture does not necessarily imply that the points in the lines are always the extreme points of the upper bound, i.e., its extreme points are not always continuous as shown in Figs. \ref{fig-6} and \ref{fig-8}. 

\begin{figure}
  \center \includegraphics[scale=.72]{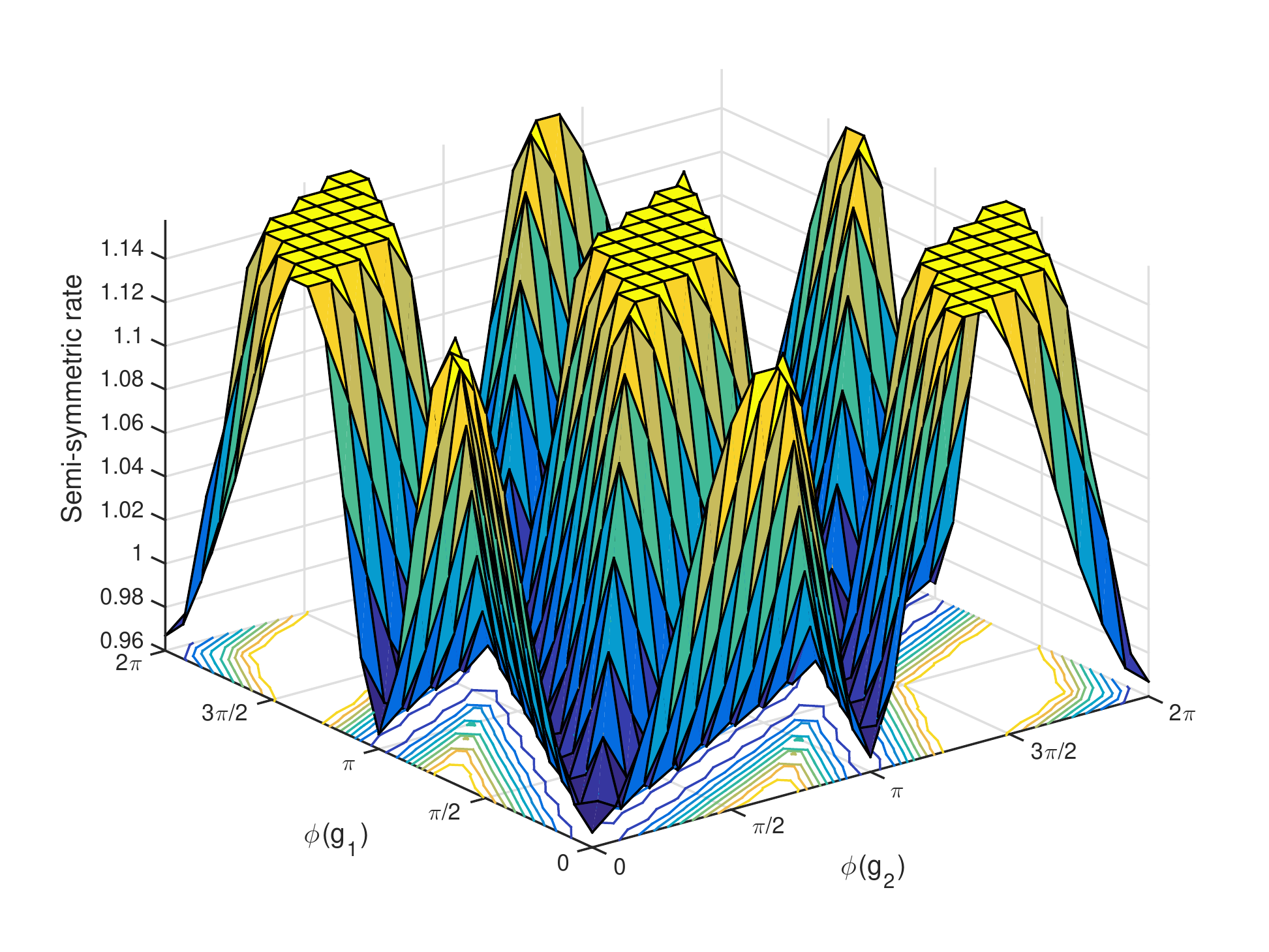} 
\caption{Upper bound on the sum capacity of three-user \emph{semi-symmetric} GIC for different phases ($\phi(g_1), \phi(g_2)$) when $P=10$, $|g_1|^2=0.3,$ and $|g_2|^2=0.7$. 
} \label{fig-8}
\end{figure}

\begin{rem}
As a special case, the above conjecture includes some discontinuous singularities shown in some previous results (e.g., \cite[Thm. 2 and 3]{Cad10a}). For our semi-symmetric case, the conditions in Theorem 3 of \cite{Cad10a} translate into 
\begin{align} 
  {|g_1|^2} = {|g_2|} \ &\text{ and } \ 2\phi(g_1) - \phi(g_2) = 0 \mod 2\pi \nonumber \\
  {|g_2|^2} = {|g_1|} \ &\text{ and } \ 2\phi(g_2) - \phi(g_1) = 0 \mod 2\pi . \nonumber 
\end{align}
If any of the above conditions is satisfied, the constant three-user complex GIC has only $1$ DoF.
Interestingly,  the conditions (\ref{eq:3IC-16}) in our conjecture and the above conditions from \cite[Thm.  3]{Cad10a} have the common phase conditions. 
\end{rem}


The semi-symmetric rate behavior provides an insight into the good (or bad) condition in terms of the phase offset between $g_1$ and $g_2$ for which a sophisticated scheme {might (or could not} in the bad condition) achieve a significant performance gain over the simple time division scheme. This is unpredictable by the existing DoF results. {While the bad condition on our upper bound is supported to some extent by the TDM lower bound, the good condition should be verified by a matching lower bound.} 
Furthermore, we can infer from the three-user semi-symmetric case that the capacity of the $K$-user complex GIC {may} significantly depend on the phase offsets of channel coefficients unlike the two-user case in \cite{Nam15a}, which shows that the phase offset between two cross-channel coefficients has a limited impact on the sum capacity.

Although the semi-symmetric GIC is shown to be very relevant for the three-user case, its sum-rate upper bounds for $K>3$ are given by Theorems \ref{thm-kub2b} and \ref{thm-kub3b} for the asymmetric case unless $g_1,g_2,\ldots,g_{K-1}$ in (\ref{eq:A-2c}) satisfy (\ref{eq:KIC-25}). Therefore, the value of the semi-symmetric GIC in (\ref{eq:A-2c}) may be undermined due to the penalty terms in the asymmetric case that incur some loss of tightness of our bounds.


\fi











\fi

\section{Conclusion}
\label{sec:con}

We have developed upper bounds on the capacity of the $K$-user complex GIC using a new type of genie-aided channels and a hybrid form of Etkin-type and change-of-interference bounding techniques. The resulting upper bounds were shown to be tighter than the existing bounds over a certain range of channel coefficients for the three-user case. We formulated closed-form expressions of the new upper bounds for the $K$-user symmetric GIC. Based on the analytical bounds, this paper has investigated $K/2$ DoF achievable by interference alignment for almost all constant GICs at realistic SNR.
In particular, we showed that {for large $K$, the performance benefit promised by the $K/2$ DoF results may not be realized even at high SNR over a certain range around $g^2=1$ for the symmetric real case. As a consequence, it has been argued that the DoF results should be carefully interpreted at finite SNR.} 
On the positive side, our result showed that the potential gain proposed by the existing DoF results may be realized at moderate SNR for the symmetric \emph{complex} GIC, depending on the phase offset between the direct-channel and the cross-channel coefficients. We have further introduced the semi-symmetric GIC and evaluated sum-rate upper bounds for the three-user semi-symmetric case, yielding a conjecture with respect to certain conditions on good and bad phase offsets between cross-channel coefficients. 

{We may leverage the conjecture with respect to potentially good and bad conditions, which suggests that an interference management scheme exploiting the phase offset conditions might be promising. For instance, one can perform phase rotation of multiple interference links to favorably align interfering signals according to the good phase condition (e.g., \cite{Cad10a}). Alternatively, one may exploit the opportunistic nature of slow-fading wireless channels so that only user pairs whose channel coefficients are near the desirable phase condition are opportunistically scheduled to communicate with each other. Therefore, the good condition on phase offset with realistic values of SNR may deserve  attention of a sophisticated scheme. On the contrary, we need to avoid trying to achieve an appreciable performance benefit while channel coefficients are near the bad condition.

An interesting future study would be finding an interference management scheme that provides a matching lower bound to the good phase-offset condition of the upper bound in this work. The conditions on the phase offsets of channel coefficients can be generalized to more than three-user cases by using the closed-form $K$-user upper bounds. Yet another item would be to improve the proposed upper bounds based on a technique in the companion paper for two-user GICs in \cite[Thm. 5]{Nam15a}.}

\vspace{2em}

\appendices

\section{Proof of Theorem \ref{thm-3ub1}}
\label{proof-1}

{We first present a lemma to be used in various proofs in this work. The proof of the above lemma is skipped.} 

\begin{lem} \label{lem-6} 
Let $Y^n_i=a_iX^n+Z^n_i,  i=1,2,\ldots,m$, where $X^n$ is a random sequence with the average power constraint such that $\sum_{j=1}^n\mathbb{E}[X_j^2]\le nP$, and for all $i$, $a_i$ is a complex number and $Z^n_i$ is i.i.d. $\mathcal{CN}(0,\sigma_{Z_i}^2)$, independent of $X^n$. For $i,k\in [1:m],$ the $j$th components of $Z_i$ and $Z_k$ are  correlated each other with $\rho_{ik}=E[Z_iZ_k]$. Let ${\mathcal S}$ be the set of $[1:m-1]$ and let $Y^n_{\mathcal S} =(Y^n_1,Y^n_2,\ldots,Y^n_{m-1})$. Then we have
\begin{align} \label{eq:B-17}
  h(Y^n_{\mathcal S}|Y^n_{m}) = h(a_{m}X^n+V^n)+nh(\overline{W})-h(Y^n_{m})
\end{align}
where $\overline{W} = (Z_{{\mathcal S}(1)}-a_{{\mathcal S}(1)}a_{m}^{-1}Z_{m},\ldots,Z_{{\mathcal S}(m-1)}-a_{{\mathcal S}(m-1)}a_{m}^{-1}Z_{m})$ is an $(m-1)$-dim zero-mean Gaussian random vector whose i.i.d. sequence is denoted by $\overline{W}^n$ and $V^n$ is i.i.d. $\mathcal{N}(0,\mathrm{Cov}(Z_{m}|\overline{W}))$.
\end{lem}

Using Fano's inequality and \cite[Lemma 1]{Tho87}, we have
\begin{align} \label{eq:3IC-22}
  n(R_1-\epsilon_n)&\le I(X^n_1; Y^n_1) \nonumber \\
  &\overset{(a)}{\le}   I(X^n_1; Y^n_1|X^n_3,U^n_1) \nonumber \\
   &= h(Y^n_1|X^n_3,U^n_1)-h(Y^n_1|X^n_1,X^n_3,U^n_1) \nonumber \\
   &= h(X_{1}^n+Z_1^n-W_1^n|h_{12}X^n_2+W^n_1) - h(h_{12}X^n_2+Z^n_1|h_{12}X^n_2+W^n_1) \nonumber \\
  &\overset{(b)}{=} h(X_{1}^n+Z_1^n-W_1^n|h_{12}X^n_2+W^n_1) -h(h_{12}X^n_2+V_{W_1}^n) \nonumber \\
   & \hspace{6mm} +h(h_{12}X^n_2+W^n_1) -nh(Z_1-W_1) 
\end{align}
where $(a)$ follows by definition of $U_1$ and by the independence of the inputs, and $(b)$ follows from Lemma \ref{lem-6}.
Similarly, we can get 
\begin{align} 
   n(R_1-\epsilon_n)&\le I(X^n_1; Y^n_1) \nonumber \\
  &\le I(X^n_1; Y^n_1|X^n_3) \nonumber \\
   &\le nh(X_{1G}+h_{12}X_{2G}+Z_1)-h(h_{12}X^n_2+Z^n_1) . \label{eq:3IC-23b} 
\end{align}
Similar to (\ref{eq:3IC-22}) and (\ref{eq:3IC-23b}), we have
\begin{align}  \label{eq:3IC-29}
  I(X^n_1; Y^n_1|X^n_3) &\le nh(X_{1G}+h_{12}X_{2G}+Z_1)-\underbrace{h(h_{12}X^n_2+Z^n_1)}_{\triangleq \;{B_1}}  \nonumber \\
  I(X^n_1; Y^n_1|X^n_3,U^n_1) & \le \underbrace{h(X_{1}^n+Z_1^n-W_1^n|h_{12}X^n_2+W^n_1)}_{\triangleq \;{A_6}} -\underbrace{h(h_{12}X^n_2+V_{W_1}^n)}_{\triangleq \;{B_2}} \nonumber \\
   & \hspace{6mm} +\underbrace{h(h_{12}X^n_2+W^n_1)}_{\triangleq \;{A_1}} -nh(Z_1-W_1)   \nonumber \\
  I(X^n_2; Y^n_2|X^n_1) &\le nh(X_{2G}+h_{23}X_{3G}+Z_2)-\underbrace{h(h_{23}X^n_3+Z^n_2)}_{\triangleq \;{B_3}}  \nonumber \\
  I(X^n_2; Y^n_2|X^n_1,U^n_2) &\le \underbrace{h(X_{2}^n+Z_2^n-W_2^n|h_{23}X^n_3+W^n_2)}_{\triangleq \;{A_2}} -\underbrace{h(h_{23}X^n_3+V_{W_2}^n)}_{\triangleq \;{B_4}} \nonumber \\
   & \ \ \  +\underbrace{h(h_{23}X^n_3+W^n_2)}_{\triangleq \;{A_3}} -nh(Z_2-W_2) \nonumber \\
  I(X^n_3; Y^n_3|X^n_2) &\le nh(X_{3G}+h_{31}X_{1G}+Z_3)-\underbrace{h(h_{31}X^n_1+Z^n_3)}_{\triangleq \;{B_5}}  \nonumber \\
  I(X^n_3; Y^n_3|X^n_2,U^n_3) &\le \underbrace{h(X_{3}^n+Z_3^n-W_3^n|h_{31}X^n_1+W^n_3)}_{\triangleq \;{A_4}} -\underbrace{h(h_{31}X^n_1+V_{W_3}^n)}_{\triangleq \;{B_6}} \nonumber \\
   & \ \ \  +\underbrace{h(h_{31}X^n_1+W^n_3)}_{\triangleq \;{A_5}} -nh(Z_3-W_3).
\end{align}


Applying the worst additive noise lemma in \cite{Dig01} to $A_1$ and $B_1$ in (\ref{eq:3IC-29}), and using the assumption of $\sigma_{W_1}^2\le 1$, we can bound $$h(h_{12}X^n_2+W^n_1)-h(h_{12}X^n_2+Z^n_1)\le nh(h_{12}X_{2G}+W_1)-nh(h_{12}X_{2G}+Z_1).$$ 
Noticing that $h_{23}X^n_3+W^n_2$ and $h_{12}X^n_2+V_{W_1}^n$ are independent and applying Lemma \ref{lem-8} in Appendix \ref{proof-1}   to $A_2$ and $B_2$, we can also bound 
\begin{align} \label{eq:3IC-12e}
  h(X_{2}^n&+Z_2^n-W_2^n| h_{23}X^n_3+W^n_2)-h(h_{12}X^n_2+V_{W_1}^n) \le  \nonumber \\   
  &nh(X_{2G}+Z_2-W_2|h_{23}X_{3G}+W_2)-nh(X_{2G}+Z_2-W_2+|h_{12}|^{-2}\tilde{V}_{W_1}|h_{23}X_{3G}+W_2) \nonumber \\ & -n\log |h_{12}|^2 
\end{align}
for $|h_{12}|^2\le 1$ and $\sigma^2_{V_{W_1}}\ge |h_{12}|^2 \sigma^2_{Z_2-W_2}$. 
Then, the second term in the right hand side of (\ref{eq:3IC-12e}) can be rewritten as 
\begin{align}  
   h&(X_{2G}+Z_2-W_2+\tilde{V}_{W_1}|h_{23}X_{3G}+W_2) \nonumber \\
   &= h(X_{2G}+Z_2-W_2+\tilde{V}_{W_1}|h_{23}X_{3G}+W_2) -h(X_{2G}+h_{12}^{-1}V_{W_1})+h(X_{2G}+h_{12}^{-1}V_{W_1}) \nonumber \\
   &= h(X_{2G}+Z_2-W_2+\tilde{V}_{W_1}|h_{23}X_{3G}+W_2) -h(X_{2G}+Z_2-W_2+\tilde{V}_{W_1})+h(X_{2G}+h_{12}^{-1}V_{W_1}) \label{eq:3IC-12d} \\
   &= h(Y_{2G}+\tilde{V}_{W_1}|X_{1G},U_{2G}) -h(Y_{2G}+\tilde{V}_{W_1})+h(X_{2G}+h_{12}^{-1}V_{W_1}) \nonumber \\
   &= -I(U_{2G};Y_{2G}+\tilde{V}_{W_1}|X_{1G}) +h(X_{2G}+h_{12}^{-1}V_{W_1}) \label{eq:3IC-12c}
\end{align} 
where we used the fact that the Gaussian random variables $h_{12}^{-1}V_{W_1}$ and $Z_2-W_2+\tilde{V}_{W_1}$ are statistically equivalent, and the condition in (\ref{eq:3ub-30b}) is intended to guarantee $\sigma^2_{\tilde{V}_{W_1}}\ge 0$. 

Substituting (\ref{eq:3IC-12c}) into (\ref{eq:3IC-12e}), we have 
\begin{align}  \label{eq:3IC-13b}
   \frac{1}{n}\big(A_2-B_2\big) & \le h(X_{2G}+Z_2-W_2|h_{23}X_{3G}+W_2)\nonumber \\
   &\hspace{-10mm} -h(h_{12}X_{2G}+V_{W_1}) +I(U_{2G};Y_{2G}+\tilde{V}_{W_1}|X_{1G}). 
\end{align}  
In fact, $I(U_{2G};Y_{2G}+\tilde{V}_{W_1}|X_{1G})$ is a {penalty term due to the conditional worst additive noise lemma. However, it can be seen from (\ref{eq:3IC-12d}) that this penalty would be marginal in general. This is because assuming $\sigma^2_{\tilde{V}_{W_1}}=0$,\footnote{It will be later shown in Sec. III.D that this condition is used to tighten the bound in Theorem \ref{thm-3ub1}.} we have 
\begin{align}
  I(U_{2G};Y_{2G}|X_{1G}) &=h(X_{2G}+Z_2-W_2)-h(X_{2G}+Z_2-W_2|h_{23}X_{3G}+W_2) \nonumber \\
&=(???) h(X_{2G}+Z_2-W_2)-h(X_{2G}+Z_2-W_2|h_{23}X_{3G}+W_2,X_{3G})  \nonumber \\
  &\le h(X_{2G}+Z_2-W_2)-h(X_{2G}+Z_2|W_2) \nonumber \\
  &=\log\frac{P_2+\sigma^2_{Z_2-W_2}}{P_2+\sigma^2_{Z_2|W_2}}.
\end{align}
Repeating the same techniques to the remaining $A_3$ through $B_6$, we can upper-bound $2(R_1 +R_2+R_3-3\epsilon_n)$ as (\ref{eq:3ub-1c}).}

\section{Proof of Theorem \ref{thm-3ub2}}
\label{proof-2}

We first bound $R_1$ and $R_3$ as
\begin{align} 
  n(R_1-\epsilon_n)&\le I(X^n_1; Y^n_1) \nonumber \\
 &\le nh(Y_{1G})-h(Y_1^n|X_1^n)  \label{eq:3IC-20} \\
  n(R_3-\epsilon_n) &\le I(X_3^n; Y_3^n |  X_1^n,X_2^n) \nonumber \\ 
  &= h(X_3^n+Z_3^n) -nh(Z_3). \label{eq:3IC-4}
\end{align}
Combining (\ref{eq:3IC-20}) and (\ref{eq:3IC-3}) and using the worst additive noise lemma and the definition of $\sigma_{N_k}^2\le 1$, we get 
\ifdefined\TIT
\begin{align} \label{eq:3IC-5}
  h(S_{2}^n) -h(Y_1^n|X_1^n) &=  h(h_{12}X_{2}^n+h_{13}X_{3}^n+N_2^n) -h(h_{12}X_{2}^n+h_{13}X_{3}^n+Z_1^n) \nonumber \\ 
  &\le  nh(S_{2G}) -nh(Y_{1G}|X_{1G}). 
\end{align}
\else
\begin{align} \label{eq:3IC-5}
  h(S_{2}^n) -h(Y_1^n|X_1^n) \le  nh(S_{2G}) -nh(Y_{1G}|X_{1G}). 
\end{align}
\fi
Combining (\ref{eq:3IC-4}) and (\ref{eq:3IC-3}), we similarly have
\ifdefined\TIT
\begin{align} \label{eq:3IC-6}
  h(X_3^n+Z_3^n)-h(h_{13}X_3^n+V_{N_2}^n) &= h(X_3^n+Z_3^n)-h(X_3^n+h_{13}^{-1}V_{N_2}^n) -n\log|h_{23}|^2  \nonumber \\
  &\le nh(X_{3G}+Z_3)-nh(h_{13}X_{3G}+V_{N_2}) 
\end{align}
\else
\begin{align} \label{eq:3IC-6}
  h(X_3^n+Z_3^n)&-h(h_{13}X_3^n+V_{N_2}^n) \nonumber \\
  & \le  nh(X_{3G}+Z_3) -nh(h_{13}X_{3G}+V_{N_2}) 
\end{align}
\fi
for $V_{N_2}$ satisfying $\sigma^2_{V_{N_2}}\ge |h_{13}|^2$, the first condition in (\ref{eq:3ub-2d}), which also implies that $|h_{13}|^2\le 1$ should be satisfied due to $\sigma^2_{V_{N_2}}\le 1$.
Substituting (\ref{eq:3IC-5}) and (\ref{eq:3IC-6}) into the sum of (\ref{eq:3IC-20}), (\ref{eq:3IC-4}), and (\ref{eq:3IC-3}), we get 
\ifdefined\TIT
\begin{align}  \label{eq:3IC-7}
    R_1 +R_2+R_3-3\epsilon_n &\le  I(X_{1G};Y_{1G}) +h(S_{2G}) -h(Z_2-h_{23}h_{13}^{-1}N_2) +h(Y_{2G}|X_{1G},S_{2G})\nonumber \\ &-h(h_{13}X_{3G}+V_{N_2}) +I(X_{3G}; Y_{3G} |  X_{1G},X_{2G}).
\end{align}
\else
\begin{align}  \label{eq:3IC-7}
    R_1 &+R_2+R_3-3\epsilon_n \le  I(X_{1G};Y_{1G}) +h(S_{2G})\nonumber \\ &-h(Z_2-h_{23}h_{13}^{-1}N_2) +h(Y_{2G}|X_{1G},S_{2G})\nonumber \\ &-h(h_{23}X_{3G}+V_{N_2}) +I(X_{3G}; Y_{3G} |  X_{1G},X_{2G}).
\end{align}
\fi
We can then immediately translate (\ref{eq:3IC-7}) into (\ref{eq:3ub-2a}) by reversely using Lemma \ref{lem-6}. 

As for the second condition in (\ref{eq:3ub-2d}), it suffices to replace $-h(h_{13}X_{3}^n+N_2^n)-h(h_{23}X_3^n+Z_2^n|h_{13}X_{3}^n+N_2^n)$ in (\ref{eq:3IC-3c}) with $-h(h_{23}X_3^n+Z_2^n)-h(h_{13}X_{3}^n+N_2^n|h_{23}X_3^n+Z_2^n)$ and to follow the same steps. This completes the proof.

\section{Proof of Theorem \ref{thm-kub2}}
\label{proof-4}

We begin with
\begin{align}  
  I(X^n_1; Y^n_1) &\le nh(Y_{1G})-\underbrace{\textstyle h(g\sum_{i=2}^{K}X_i^n+Z_1^n)}_{\triangleq \;\beta_{1B}}  \nonumber \\
  I(X^n_2;Y^n_2,S_2^n| X_1^n) &\le  {\textstyle h(g\sum_{i=2}^{K}X_i^n+N_2^n)} -{\textstyle h(g\sum_{i=3}^{K}X_i^n+N_2^n)}  \nonumber \\ &\ \ \ +nh(Y_{2G}| X_{1G},S_{2G})  -{\textstyle h(g\sum_{i=3}^{K}X_i^n+Z_2^n|g\sum_{i=3}^{K}X_i^n+N_2^n)} \nonumber \\
  &=  \underbrace{\textstyle h(g\sum_{i=2}^{K}X_i^n+N_2^n)}_{\triangleq \;\beta_{1A}}-{n h(Z_2-N_2)} \nonumber \\ &\ \ \  +nh(Y_{2G}| X_{1G},S_{2G})  -\underbrace{\textstyle h(g\sum_{i=3}^{K}X_i^n+V_{N_2}^n)}_{\triangleq \;\beta_{2B}} \label{eq:KIC-1c} \\
  I(X^n_3;Y^n_3,S_3^n| X_1^n,X_2^n) &\le  \underbrace{\textstyle h(g\sum_{i=3}^{K}X_i^n+N_3^n)}_{\triangleq \;\beta_{2A}}-{n h(Z_3-N_3)} \nonumber \\ &\ \ \  +nh(Y_{3G}| X_{1G},X_{2G},S_{3G})  -{\textstyle h(g\sum_{i=4}^{K}X_i^n+V_{N_3}^n)} \nonumber \\ & \ \ \ \vdots \nonumber \\
  I(X^n_{K-1};Y^n_{K-1},S_{K-1}^n| X_1^n,\ldots,X_{K-2}^n) &\le  {\textstyle h(g\sum_{i=K-1}^{K}X_i^n+N_{K-1}^n)}-{n h(Z_{K-1}-N_{K-1})} \nonumber \\ &\ \ \  +nh(Y_{(K-1)G}| X_{1G},\ldots,X_{(K-2)G},S_{(K-1)G})  -\underbrace{\textstyle h(gX_K^n+V_{N_{K-1}}^n)}_{\triangleq \;\beta_{3B}} \nonumber \\ 
  I(X^n_K; Y^n_K|X_1^n,\ldots,X_{K-1}^n) &= \underbrace{ h(X_K^n+Z_K^n)}_{\triangleq \;\beta_{3A}}  -nh(Z_K). \label{eq:KIC-1}
\end{align}
Notice that the cross-channel coefficient $g$ in $V_{N_i}^n$ should be canceled out as shown in (\ref{eq:3IC-14c}). 
Applying the worst additive noise lemma to $\beta_{1A}-\beta_{1B}$, $\beta_{2A}-\beta_{2B}$ for the condition in (\ref{eq:Kub-2a}), and $\beta_{3A}-\beta_{3B}$ for (\ref{eq:Kub-2b}), respectively, we get (\ref{eq:kub-2}).

\section{Proof of Theorem \ref{thm-kub3}}
\label{proof-5}

\begin{figure}
\hspace{-7mm}
  \center 
  \includegraphics[scale=.95]{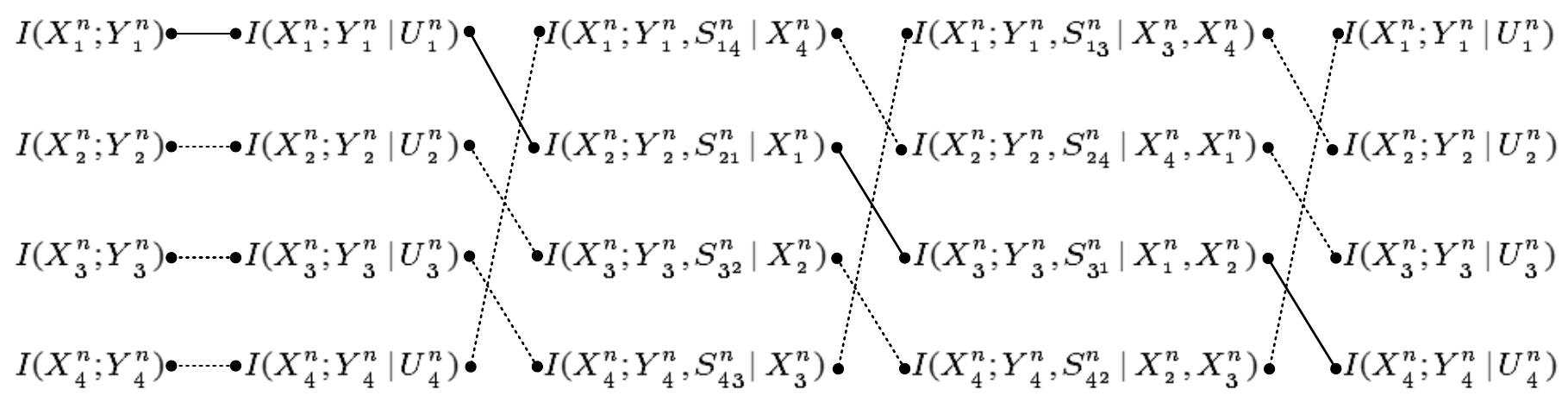}
  \caption{Graphical representation of the relation among mutual informations for the four-user case, where the solid lines corresponds to one of the four components in (\ref{eq:kub-3}) for $K=4$ and the last column is redundant and just given for illustrative convenience. {This shows how positive and negative entropies should be paired with each other in (\ref{eq:3IC-8b}) to derive a useful upper bound.}}\label{fig-0b}
\end{figure}

By symmetry, we can write
\begin{align}  
  I(X^n_1; Y^n_1) &\le nh(Y_{1G})-\underbrace{\textstyle h(g\sum_{i=2}^{K}X_i^n+Z_1^n)}_{\triangleq \;\gamma_{1B}}  \nonumber \\
  I(X^n_2;Y^n_2,S_{2}^n| X_1^n) &\le  \underbrace{\textstyle h(g\sum_{i=2}^{K}X_i^n+N_{2}^n)}_{\triangleq \;\gamma_{2A}}-nh(Z_2-N_{2})  \nonumber \\ &\ \ \ +nh(Y_{2G}| X_{1G},S_{2G})  -\underbrace{\textstyle h(g\sum_{i=3}^{K}X_i^n+V_{N_{2}}^n)}_{\triangleq \;\gamma_{3B}} \nonumber \end{align}
  \vspace{-7mm}
\begin{align}  \label{eq:3IC-8b}
  I(X^n_3;Y^n_3,S_{3}^n| X_1^n,X_2^n) &\le  \underbrace{\textstyle h(g\sum_{i=3}^{K}X_i^n+N_{3}^n)}_{\triangleq \;\gamma_{3A}}-nh(Z_3-N_{3})  \nonumber \\ &\ \ \ +nh(Y_{3G}| X_{1G},X_{2G},S_{3G})  -{\textstyle h(g\sum_{i=4}^{K}X_i^n+V_{N_{3}}^n)} \nonumber \\
  & \ \ \ \vdots \nonumber \\
  I(X^n_{K-1};Y^n_{K-1},S_{K-1}^n| X_1^n,\ldots,X_{K-2}^n) &\le  {\textstyle h(g\sum_{i=K-1}^{K}X_i^n+N_{K-1,1}^n)}-nh(Z_{K-1}-N_{K-1})  \nonumber \\ &\ \ \ +nh(Y_{(K-1)G}| X_{1G},\ldots,X_{(K-2)G},S_{K-1)G})  -\underbrace{\textstyle h(gX_K^n+V_{N_{K-1}}^n)}_{\triangleq \;\gamma_{4B}} \nonumber \\  
  I(X^n_K;Y^n_K|U^n_K) &= \underbrace{h(X^n_K+Z_K^n-W_K^n|U^n_K)}_{\triangleq \;\gamma_{4A}} - \underbrace{\textstyle h(g\sum_{i=1}^{K-1}X_i^n+V^n_{W_K})}_{\triangleq \;\gamma_{2B}} \nonumber \\ & \ \ \ +\underbrace{\textstyle h(g\sum_{i=1}^{K-1}X_i^n+W_K^n)}_{\triangleq \;\gamma_{1A}} -nh(Z_K-W_K).
\end{align}
Noticing that $h(g\sum_{i=1}^{K-1}X_i^n+V^n_{W_K})$ and $h(g\sum_{i=1}^{K-1}X_i^n+W_K^n)$ are statistically equivalent to $ h(g\sum_{i=2}^{K}X_i^n+V^n_{W_1})$ and $h(g\sum_{i=2}^{K}X_i^n+W_1^n)$, respectively,  by symmetry and applying the worst additive noise lemma to $\gamma_{1A}-\gamma_{1B}$ for the condition in (\ref{eq:Kub-4a}) and to $\gamma_{2A}-\gamma_{2B}$ and $\gamma_{3A}-\gamma_{3B}$ for (\ref{eq:Kub-4b}), respectively, and applying the conditional worst additive noise lemma to $\gamma_{4A}-\gamma_{4B}$ for (\ref{eq:Kub-4c}), we get (\ref{eq:kub-3}).

\section{Proof of Theorem \ref{thm-kub2b}}
\label{proof-7}

Here, we cannot use the technique in (\ref{eq:3IC-3b}) and (\ref{eq:KIC-1c}). Therefore, rather assuming $N_k$ and $Z_k$ are independent, for $3\le k\le K-1$, we can get the following upper bound:
\begin{align}  
   I(X^n_{k-1}&;Y^n_{k-1},S_{2}^n| X_1^n,\ldots,X_{k-2}^n) \nonumber \\&= I(X^n_{k-1};S_{2}^n| X_1^n,\ldots,X_{k-2}^n) +I(X^n_{k-1};Y^n_{k-1}| X_1^n,\ldots,X_{k-2}^n,S_{2}^n) \nonumber \\
   &\overset{(a)}{\le} I(X^n_{k-1};S_{2}^n| X_1^n,\ldots,X_{k-2}^n) +I(X^n_{k-1};Y^n_{k-1}| X_1^n,\ldots,X_{k-2}^n,X_{k}^n,\ldots,X_{K}^n,S_{2}^n) \nonumber \\
   &\overset{(b)}{=} {\textstyle h(\sum_{i=k-1}^{K}h_{1i}X_i^n+N_{2}^n)}-{\textstyle h(\sum_{i=k}^{K}h_{1i}X_i^n+N_{2}^n)} \nonumber \\ 
   &\ \ \ +h(X^n_{k-1}+Z^n_{k-1}| h_{1,k-1}X_{k-1}^n+N_{2}^n) -{\textstyle h(Z_{k-1}^n)} \nonumber \\
   &\le {\textstyle h(\sum_{i=k-1}^{K}h_{1i}X_i^n+N_{2}^n)}-{\textstyle h(\sum_{i=k}^{K}h_{1i}X_i^n+N_{2}^n)} \nonumber \\ 
   &\ \ \ +nh(X_{(k-1)G}+Z_{k-1}| h_{1,k-1}X_{(k-1)G}+N_{2}) -nh(Z_{k-1}) \label{eq:KIC-4a} \\
   I(X^n_{k};&\;Y^n_{k},S_{2}^n| X_1^n,\ldots,X_{k-1}^n) \nonumber \\ &\le {\textstyle h(\sum_{i=k}^{K}h_{1i}X_i^n+N_{2}^n)}-{\textstyle h(\sum_{i=k+1}^{K}h_{1i}X_i^n+N_{2}^n)} \nonumber \\ 
   &\ \ \ +nh(X_{kG}+Z_{k}| h_{1k}X_{kG}+N_{2})-nh(Z_{k}) \label{eq:KIC-4b}
\end{align}
where $(a)$ follows from the independence assumption between $X_1^n,\ldots,X_{k-2}^n,X_{k}^n,\ldots,X_{K}^n$ and $S_{2}^n$, and in $(b)$ we used the assumption that $X_k^n$, $Z_k^n$, and $N_k^n$ are mutually independent for all $k$. We can see that the second term in (\ref{eq:KIC-4a}) and the first term in (\ref{eq:KIC-4b}) are canceled out.
As before, using the worst additive noise lemma, we get (\ref{eq:kub-2b}). 

\section{Proof of Theorem \ref{thm-kub3b}}
\label{proof-8}

We rewrite some mutual information terms that are different from the symmetric case in Appendix \ref{proof-5} as follow:
\begin{align} 
  I(X^n_1;Y^n_1|U^n_1) &= h(X^n_1+Z_1^n-W_1^n|U^n_1) - \underbrace{\textstyle h(\sum_{i=2}^{K}h_{1i}X_i^n+V^n_{W_1})}_{\triangleq \;\delta_{1B}} \nonumber \\ & \ \ \ +{\textstyle h(\sum_{i=2}^{K}h_{1i}X_i^n+W_1^n)} -nh(Z_1-W_1)  \nonumber \\
  I(X^n_2;Y^n_2,S_{2}^n| X_1^n) &\le \underbrace{\textstyle h(\sum_{i=2}^{K}h_{1i}X_i^n+N_{2}^n)}_{\triangleq \;\delta_{1A}}  -\underbrace{\textstyle h(\sum_{i=3}^{K}h_{1i}X_i^n+N_{2}^n)}_{\triangleq \;\delta_{2B}} \nonumber  \\ 
   &\ \ \ +nh(X_{2G}+Z_{2}| h_{12}X_{2G}+N_{2}) -nh(Z_{2}) \label{eq:3IC-17a}  \\
  I(X^n_3;Y^n_3,S_{2}^n| X_1^n,X_2^n) &\le \underbrace{\textstyle h(\sum_{i=3}^{K}h_{1i}X_i^n+N_{2}^n)}_{\triangleq \;\delta_{2A}}  -{\textstyle h(\sum_{i=4}^{K}h_{1i}X_i^n+N_{2}^n)} \nonumber \\ 
   &\ \ \ +nh(X_{3G}+Z_{3}| h_{13}X_{3G}+N_{2}) -nh(Z_{3})  \nonumber \\
     I(X^n_{K-1};\;Y^n_{K-1},S_{2}^n| X_1^n,\ldots,X_{K-2}^n) &\le {\textstyle h(\sum_{i=K-1}^{K}h_{1i}X_i^n+N_{2}^n)} -\underbrace{\textstyle h(h_{1K}X_K^n+N_{2}^n)}_{\triangleq \;\delta_{3B}} \nonumber \\ 
   &\ \ \ +nh(X_{(K-1)G}+Z_{K-1}| h_{1,K-1}X_{(K-1)G}+N_{2}) -nh(Z_{K-1}) \nonumber \\ 
   I(X^n_K;Y^n_K|U^n_K) &= \underbrace{h(X^n_K+Z_K^n-W_K^n|U^n_K)}_{\triangleq \;\delta_{3A}} - \textstyle h(\sum_{i=1}^{K-1}h_{Ki}X_i^n+V^n_{W_K}) \nonumber \\ & \ \ \ +{\textstyle h(\sum_{i=1}^{K-1}h_{Ki}X_i^n+W_K^n)} -nh(Z_K-W_K)  \label{eq:3IC-17b}
\end{align}
where  (\ref{eq:3IC-17a}) comes from (\ref{eq:KIC-4a}).
We can then apply the worst additive noise lemma to $\delta_{1A}-\delta_{1B}$ for $ \sigma^2_{V_{W_1}}\ge \sigma^2_{N_{2}}$, apply the conditional worst additive noise lemma to $\delta_{3A}-\delta_{3B}$ for $\sigma^2_{{N_{2}}}\ge |h_{1K}|^2\sigma^2_{Z_{K}-W_{K}} $, and notice $\delta_{2A}-\delta_{2B}=0$. 
Rewriting $\delta_{3A}-\delta_{3B}$ in a similar fashion to (\ref{eq:3IC-12}), we can get (\ref{eq:kub-3b}).

\bibliographystyle{IEEEtran}
\bibliography{OB_Kuser_GIC}

\begin{thebibliography}{10}
\providecommand{\url}[1]{#1}
\csname url@samestyle\endcsname
\providecommand{\newblock}{\relax}
\providecommand{\bibinfo}[2]{#2}
\providecommand{\BIBentrySTDinterwordspacing}{\spaceskip=0pt\relax}
\providecommand{\BIBentryALTinterwordstretchfactor}{4}
\providecommand{\BIBentryALTinterwordspacing}{\spaceskip=\fontdimen2\font plus
\BIBentryALTinterwordstretchfactor\fontdimen3\font minus
  \fontdimen4\font\relax}
\providecommand{\BIBforeignlanguage}[2]{{%
\expandafter\ifx\csname l@#1\endcsname\relax
\typeout{** WARNING: IEEEtran.bst: No hyphenation pattern has been}%
\typeout{** loaded for the language `#1'. Using the pattern for}%
\typeout{** the default language instead.}%
\else
\language=\csname l@#1\endcsname
\fi
#2}}
\providecommand{\BIBdecl}{\relax}
\BIBdecl

\bibitem{Car78}
A.~B. Carleial, ``Interference channels,'' \emph{{IEEE} Trans. on Inform.
  Theory}, vol.~24, no.~1, pp. 60--70, Jan. 1978.

\bibitem{Etk08}
R.~H. Etkin, D.~N.~C. Tse, and H.~Wang, ``Gaussian interference channel
  capacity to within one bit,'' \emph{{IEEE} Trans. on Inform. Theory},
  vol.~54, no.~12, pp. 5534--5562, Dec. 2008.

\bibitem{Ann09}
V.~S. Annapureddy and V.~V. Veeravalli, ``{Gaussian interference networks: Sum
  capacity in the low-interference regime and new outer bounds on the capacity
  region},'' \emph{{IEEE} Trans. on Inform. Theory}, vol.~55, no.~7, pp.
  3032--3050, Jul. 2009.

\bibitem{Sha09}
X.~Shang, G.~Kramer, and B.~Chen, ``A new outer bound and the
  noisy-interference sum-rate capacity for the {G}aussian interference
  channels,'' \emph{{IEEE} Trans. on Inform. Theory}, vol.~55, no.~2, pp.
  689--699, Feb. 2009.

\bibitem{Mot09}
A.~S. Motahari and A.~K. Khandani, ``{Capacity bounds for the Gaussian
  interference channel},'' \emph{{IEEE} Trans. on Inform. Theory}, vol.~55,
  no.~2, pp. 620--643, Feb. 2009.

\bibitem{Etk09}
R.~H. Etkin, ``{New sum-rate upper bound for the two-user Gaussian interference
  channel},'' in \emph{Proc. {IEEE} Int. Symp. on Inform. Theory (ISIT)}.\hskip
  1em plus 0.5em minus 0.4em\relax Seoul, Korea, Jun. 2009.

\bibitem{Cha11}
A.~Chaaban and A.~Sezgin, ``{An extended Etkin-type outer bound on the capacity
  of the Gaussian interference channel},'' in \emph{Proc. {IEEE} Asilomar Conf.
  on Signals, Systems, and Computers (ACSSC)}, 2011, pp. 1860 -- 1864.

\bibitem{Nam12}
J.~Nam and G.~Caire, ``A new outer bound on the capacity region of {G}aussian
  interference channels,'' in \emph{Proc. {IEEE} Int. Symp. on Inform. Theory
  (ISIT)}.\hskip 1em plus 0.5em minus 0.4em\relax Boston, MA, Jun/Jul. 2012.

\bibitem{Nam15a}
\BIBentryALTinterwordspacing
J.~Nam, ``{New outer bounds on the capacity of the two-user {G}aussian
  interference channel},'' \emph{submitted to {IEEE} Trans. on Inform. Theory},
  2015. [Online]. Available: \url{http://arxiv.org/abs/1506.03324v2}
\BIBentrySTDinterwordspacing

\bibitem{Kra04}
G.~Kramer, ``Outer bounds on the capacity region of gaussian interference
  channels,'' \emph{{IEEE} Trans. on Inform. Theory}, vol.~50, no.~3, pp.
  581--586, Mar. 2004.

\bibitem{Sri08}
S.~Sridharan, A.~Jafarian, S.~Vishwanath, and S.~A. Jafar, ``Capacity of
  symmetric {$K$}-user {G}aussian very strong interference channels,'' in
  \emph{Proc. {IEEE} Global Commun. Conf. (GLOBECOM)}, Dec. 2008.

\bibitem{Zho13}
L.~Zhou and W.~Yu, ``On the capacity of the k-user cyclic gaussian interference
  channel,'' \emph{{IEEE} Trans. on Inform. Theory}, vol.~59, no.~1, pp.
  154--165, Jan. 2013.

\bibitem{Tun11}
D.~Tuninetti, ``{A new sum-rate outer bound for interference channels with
  three source-destination pairs},'' \emph{in Proc. Inf. Theory and App.
  (ITA)}, pp. 1--8, 2011.

\bibitem{Tun11b}
------, ``{$K$}-user interference channels: {G}eneral outer bound and
  sum-capacity for certain {G}aussian channels,'' \emph{Proc. {IEEE} Int. Symp.
  on Inform. Theory (ISIT)}, pp. 1--6, 2011.

\bibitem{Mad08}
M.~A. Maddah-Ali, A.~S. Motahari, and A.~K. Khandani, ``Communication over
  {MIMO X} channels: {I}nterference alignment, decomposition, and performance
  analysis,'' \emph{{IEEE} Trans. on Inform. Theory}, vol.~54, no.~8, pp.
  3457--3470, 2008.

\bibitem{Cad08}
V.~R. Cadambe and S.~A. Jafar, ``Interference alignment and degrees of freedom
  of the {$K$}-user interference channel,'' \emph{{IEEE} Trans. on Inform.
  Theory}, vol.~54, no.~8, pp. 3425--3441, Aug. 2008.

\bibitem{Ave11}
S.~Avestimehr, S.~Diggavi, and D.~Tse, ``Wireless network information flow: {A}
  deterministic approach,'' \emph{{IEEE} Trans. on Inform. Theory}, vol.~57,
  no.~4, pp. 1872--1905, 2011.

\bibitem{Bre10}
G.~Bresler, A.~Parekh, and D.~Tse, ``The approximate capacity of the
  many-to-one and one-to-many {G}aussian interference channels,'' \emph{{IEEE}
  Trans. on Inform. Theory}, vol.~56, no.~9, pp. 4566--4592, Sep. 2010.

\bibitem{Naz11}
B.~Nazer and M.~Gastpar, ``{Compute-and-forward: Harnessing interference
  through structured codes},'' \emph{{IEEE} Trans. on Inform. Theory}, vol.~57,
  no.~10, pp. 6463--6486, Oct. 2011.

\bibitem{Sat77}
H.~Sato, ``Two-user communication channels,'' \emph{{IEEE} Trans. on Inform.
  Theory}, vol.~23, pp. 295--304, May 1977.

\bibitem{Cad09}
V.~R. Cadambe and S.~A. Jafar, ``Parallel {G}aussian interference channels are
  not always separable,'' \emph{{IEEE} Trans. on Inform. Theory}, vol.~55,
  no.~9, pp. 3983--3990, 2009.

\bibitem{Jos10}
J.~Jose and S.~Vishwanath, ``{Sum capacity of $K$ user Gaussian degraded
  interference channels},'' in \emph{Proc. IEEE Info. Theory Workshop
  (ITW)}.\hskip 1em plus 0.5em minus 0.4em\relax Dublin, Ireland, Sep. 2010.

\bibitem{Naz12}
B.~Nazer, M.~Gastpar, S.~Jafar, and S.~Vishwanath, ``Ergodic interference
  alignment,'' \emph{{IEEE} Trans. on Inform. Theory}, vol.~58, no.~10, pp.
  6355--6371, Oct. 2012.

\bibitem{Mot09b}
A.~S. Motahari, S.~O. Gharan, M.-A. Maddah-Ali, and A.~K. Khandani, ``{Real
  interference alignment: Exploiting the potential of single antenna
  systems},'' \emph{{IEEE} Trans. on Inform. Theory}, vol.~60, no.~8, pp.
  4799--4810, 2014.

\bibitem{Wu15}
Y.~Wu, S.~{Shamai (Shitz)}, and S.~Verd{\'u}, ``A formula for the degrees of
  freedom of the interference channel,'' \emph{{IEEE} Trans. on Inform.
  Theory}, vol.~61, no.~1, pp. 256--279, Jan. 2015.

\bibitem{Cad10a}
V.~R. Cadambe, S.~A. Jafar, and C.~Wang, ``{Interference alignment with
  asymmetric complex signaling - Settling the Host-Madsen-Nosratinia
  conjecture},'' \emph{{IEEE} Trans. on Inform. Theory}, vol.~56, no.~9, pp.
  4552--4565, 2010.

\bibitem{Bre11}
\BIBentryALTinterwordspacing
G.~Bresler, D.~Cartwright, and D.~Tse, ``{Settling the feasibility of
  interference alignment for the MIMO interference channel: The symmetric
  square case},'' 2011. [Online]. Available: \url{http://arXiv:1104.0888v1}
\BIBentrySTDinterwordspacing

\bibitem{Etk09b}
R.~Etkin and E.~Ordentlich, ``{The degrees-of-freedom of the $K$-user Gaussian
  interference channel is discontinuous at rational channel coefficients},''
  \emph{{IEEE} Trans. on Inform. Theory}, vol.~55, no.~11, pp. 4932--4946,
  2009.

\bibitem{Nie13}
U.~Niesen and M.~A. Maddah-Ali, ``Interference alignment: {F}rom degrees of
  freedom to constant-gap capacity approximations,'' \emph{{IEEE} Trans. on
  Inform. Theory}, vol.~59, no.~8, pp. 4855--4888, Aug. 2013.

\bibitem{Jaf10}
S.~A. Jafar and S.~Vishwanath, ``{Generalized degrees of freedom of the
  symmetric Gaussian $K$-user interference channel},'' \emph{{IEEE} Trans. on
  Inform. Theory}, vol.~56, no.~7, pp. 3297--3303, 2010.

\bibitem{Ord14}
\BIBentryALTinterwordspacing
O.~Ordentlich, U.~Erez, and B.~Nazer, ``The approximate sum capacity of the
  symmetric {G}aussian {$K$}-user interference channel,'' Mar. 2014. [Online].
  Available: \url{http://arxiv.org/abs/1206.0197v2}
\BIBentrySTDinterwordspacing

\bibitem{Tul05}
A.~Tulino, A.~Lozano, and S.~Verd{\'u}, ``Impact of antenna correlation on the
  capacity of multiantenna channels,'' \emph{{IEEE} Trans. on Inform. Theory},
  vol.~51, no.~7, pp. 2491--2509, 2005.

\bibitem{Sha01}
S.~Shamai and S.~Verd\'u, ``The impact of frequency-flat fading on the spectral
  efficiency of {CDMA},'' \emph{{IEEE} Trans. on Inform. Theory}, vol.~47,
  no.~4, pp. 1302--1327, May 2001.

\bibitem{Dig01}
S.~Diggavi and T.~M. Cover, ``Worst additive noise under covariance
  constraints,'' \emph{{IEEE} Trans. on Inform. Theory}, vol.~47, no.~7, pp.
  3072--3081, Nov. 2001.

\bibitem{Tel07}
E.~Telatar and D.~Tse, ``Bounds on the capacity region of a class of
  interference channels,'' in \emph{Proc. {IEEE} Int. Symp. on Inform. Theory
  (ISIT)}.\hskip 1em plus 0.5em minus 0.4em\relax Nice, France, Jun. 2007.

\bibitem{She12}
\BIBentryALTinterwordspacing
M.~Shen and A.~Host-Madsen, ``The wideband slope of interference channels: The
  small bandwidth case,'' 2012. [Online]. Available:
  \url{http://arxiv.org/abs/1506.03324v2}
\BIBentrySTDinterwordspacing

\bibitem{Tho87}
J.~A. Thomas, ``{Feedback can at most double Gaussian multiple access channel
  capacity},'' \emph{{IEEE} Trans. on Inform. Theory}, vol.~33, no.~5, pp.
  711--716, Sep. 1987.

\bibitem{Han81}
T.~S. Han and K.~Kobayashi, ``A new achievable rate region for the interference
  channel,'' \emph{{IEEE} Trans. on Inform. Theory}, vol.~27, no.~1, pp.
  49--60, Jan. 1981.

\end{thebibliography}

\end{document}